%% file: ucd_fcci_rob_j_meijer.tex
\documentclass[a4paper,12pt,oneside]{book}
\usepackage[toc,page]{appendix}
\usepackage{lmodern}
\usepackage[T1]{fontenc}
\usepackage[procnames]{listings}
\usepackage{color}
\usepackage{graphicx}
\usepackage{subfig}
\usepackage{mathtools}
\usepackage{rail}
\usepackage[nottoc,numbib]{tocbibind}
\usepackage{multibib}
\begin{document}
\definecolor{keywords}{RGB}{255,0,90}
\definecolor{comments}{RGB}{0,0,113}
\definecolor{red}{RGB}{160,0,0}
\definecolor{green}{RGB}{0,150,0}
\lstset{language=Python, 
        basicstyle=\ttfamily\small, 
        keywordstyle=\color{keywords},
        commentstyle=\color{comments},
        stringstyle=\color{red},
        showstringspaces=false,
        identifierstyle=\color{green},
        procnamekeys={def,class}}

\begin{titlepage}
\include{conf}
\newcommand{\HRule}{\rule{\linewidth}{0.5mm}}
\center
\HRule \\[0.4cm]
{ \Large \bfseries \mymaintitle}\\[0.4cm]
\emph{\large{\mysubtitle}}
\\[20pt]
{ \large \bfseries \myauthor}\\[0.3cm]
\HRule \\[1.5cm]
A minor thesis submitted in part fulfilment of the degree of \mytitle in \myfield \\
under the supervision of \mysupervisor.
\vfill
\vfill
\vfill
\center
\vfill
\vfill
\vfill
\includegraphics[width=6cm]{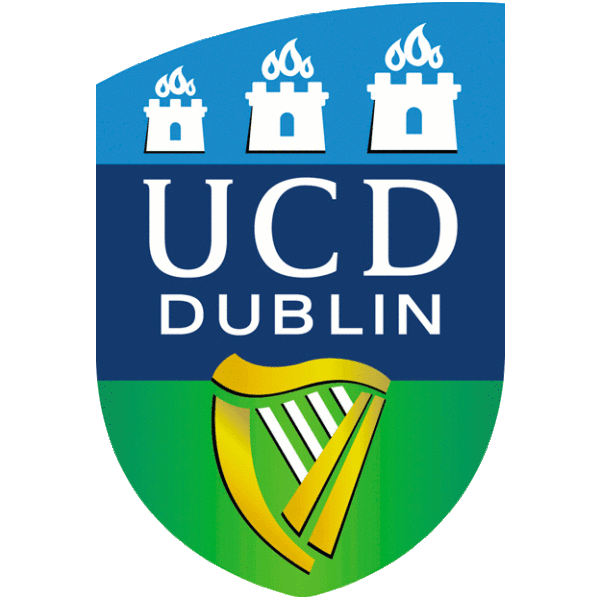}
\vfill
{\large \myschool}\\[12pt]
{\large \myuniversity}\\[12pt]
{\large June 16, 2016}
\vfill
\end{titlepage}
\include{mainpaper-abstract}

\tableofcontents
\include{mainpaper-introduction}

\include{mainpaper-literature}
\include{mainpaper-problem}

\include{mainpaper-approach}
\include{mainpaper-results}

\include{ocfa-step2-results}
\include{ocfa-step3-results}
\include{ocfa-step4-results}
\include{ocfa-step5-results}
\include{ocfa-conclussions}

\include{mainpaper-fuse}
\include{mainpaper-results2}

\include{mainpaper-archive}
\include{mainpaper-ohash}
\include{mainpaper-bus}
\include{mainpaper-integrity}
\include{mainpaper-poc}
\include{mainpaper-evaluation}
\include{mainpaper-references}

\begin{appendices}
\include{ocfa-scriptsreference}

\include{linux-linux}
\include{mattock-hash}
\include{mattock-ocfa}
\include{mattock-mattock}

\include{MattockFS-fsapi}
\end{appendices}

\end{document}

%% file: conf.tex
\def \mymaintitle {MattockFS}
\def \mysubtitle {Page-cache and access-control concerns in asynchronous message-based forensic frameworks on the Linux platform.}
\def \myauthor {Rob J Meijer}
\def \mysupervisor {Dr Pavel Gladishev} 
\def \mytitle {M.Sc.}
\def \myfield {Forensic Computing and Cyber Crime Investigations}
\def \myschool {School of Computer Science and Informatics}
\def \myuniversity {University College Dublin}

%% file: mainpaper-abstract.tex
\chapter*{\centering \begin{normalsize}Abstract\end{normalsize}}
\begin{quotation}
\noindent 
In this dissertation the feasibility of creating a page-cache efficient storage- and messaging solution with integrity geared access control for a scalable forensic framework is researched. The Open Computer Forensics Architecture (OCFA),a lab-side scalable computer forensics framework, introduced the concept of a message passing concurrency based forensic framework. Since then, the amount of per-investigation data to be processed in a lab environment has continued to grow significantly while available RAM and CPU processing power combined with prohibitive cost and limited capacity of SSD solutions have shifted processing from being largely CPU constrained to being much more IO constrained. OCFA suffered from several page-cache-miss related performance issues that have grown more significant as a result of this shift. In the light of anti-forensics and general issues related to process integrity, OCFA did not leverage the power of its message passing based design to address integrity concerns.

The main purpose of this dissertation is to analyze and evaluate a number of page-cache friendly technologies that could contribute to the creation of a computer forensics lab-geared scalable message-passing-concurrency based forensic framework with a significantly reduced quantity of page-cache-miss induced spurious IO operations, taking into account integrity related issues.

Provenance logs from historic investigations conducted using the Open Computer Forensics Architecture were thoroughly analyzed in this study, during which several bottlenecks and design flaws in OCFA were identified. A number of strategies were devised to address these bottlenecks in future computer forensic frameworks. Finally, the most prominently page-cache related strategies were consolidated with access-control measures into a user-space file-system and low-level API prototype. 
\end{quotation}
\clearpage

%% file: mainpaper-introduction.tex
\chapter{Introduction}
\noindent 
In 2006, the Dutch National Police open-sourced a scalable message-passing concurrency based digital forensic framework that had already been used internally for several years. Development of The Open Computer Forensics Architecture (OCFA) was first prompted by a large scale investigation involving fraud in the real estate sector in 2001, where digital evidence from 80 distinct search and seizures led to serious scalability issues in the forensic process when using the then standard commercial forensic tools available at that time.  Development on and support of OCFA by the Dutch National Police was discontinued in 2012, leaving a void that is yet to be filled by an alternative open source solution.     

In order to address scalability and robustness concerns, OCFA introduced the concept of a message passing concurrency based forensic framework, and due to storage concerns, early versions of OCFA made use of Content Addressed  Storage (CAS) technology. OCFA, as most forensic solutions of that time was geared towards completeness and less so on IO efficiency.

One of the main architectural components in the Open Computer Forensics Architecture was a server named the Anycast Relay. This Anycast Relay was what today would be called a message bus or an enterprise-messaging solution. The Anycast Relay allowed a module process in the OCFA set-up to register as a worker instance. Any given OCFA set-up could have multiple instances running simultaneously for any given module functionality. Submitting a message to the Anycast Relay directed at a given module name would result in it (eventually) being delivered to exactly one, or to phrase it differently, to \emph{any} of the worker instances. Hence the name \emph{Anycast}. The Anycast message bus worked with persistent priority queues that for practical intents and purposes were infinite in size.

While somewhat mitigated by advances in computer forensic triage, leading to large amounts of triaged data not finding its way to the lab, the amount of per-investigation data to be processed in a lab environment has continued to grow significantly. At the same time, while Solid State Disk (SSD) technology has shifted bottlenecks for small-scale investigations from IO-constrained to CPU-constrained, the combination of increased available RAM and CPU processing power combined with prohibitive cost and limited capacity of SSD solutions have created a situation where for a full-scale investigation the opposite has become a reality. 
The old completeness strategy should no longer be considered in sync with today's reality in digital forensics, and some of the design decisions made for OCFA have proven to be seriously IO constrained on modern server hardware. The IO performance of OCFA is greatly limited by naive implementation of infinite size queue based message passing and other fundamental design choices that end up resulting in several page-cache-miss related performance issues. While OCFA development has been discontinued, there is a continued need for a computer forensics architecture that combines some of the features that OCFA offered such as scalability, a unified forensic API and the ability to assist in cross house-search data analysis. Many of the lessons learned from a decade of OCFA usage should prove useful in the development of the next generation message-passing concurrency based digital forensic framework. This is assuming that IO efficiency issues can be addressed.

Also in 2006, the Dutch National Police released a computer forensic file-system for \emph{zero-storage} file carving named CarvFS, while at that same time Richards et al \cite{inplace} were independently working on a paper on \emph{In-place} file carving and a relatively similar file-system named ScalpelFS. Both \emph{zero-storage} and \emph{In-place} file carving refer to the idea that instead of creating a copy of the data of a carved file and storing that copy in a file-system using its own allocated disk-space, a user-space file-system can be used to present a read-only file abstraction of the carved file without the need to allocate any additional disk storage on a regular file-system.
CarvFS introduced the concept of using pseudo file-names as designation for potentially fragmented data files within an archive. Later versions of OCFA used CarvFS not just for regular carving, but also in conjunction with regular file-system forensics, using the Sleuth-kit library for different modules. A way to look at ScalpelFS and CarvFS is as a file-system abstraction for a flat secondary storage address space.

OCFA and CarvFS use technologies, user-space file-systems and an actor-model message passing approach for concurrency, that are fundamentally well suited for implementing privilege separation and access control. The way these were used and combined by the Dutch National Police however, while reasonable within the reality of that day, should no longer be considered to be in sync with the reality of anti-forensic technology. There is need for a computer forensic process that is able to maintain data integrity even when the integrity of individual tools in the tool-chain end up being compromised.

The research described in this paper leverages the use of flat secondary storage address space for the purpose of page-cache miss induced spurious IO event reduction. This within the context of an OCFA inspired sub system that aims to be a main component in a next generation computer forensic framework. Aside from this primary concern, it also tries to take important steps regarding the provisioning of measures towards fulfilling integrity needs of the modern day computer forensic process. 

In this research, a set of provenance log records from real-life investigations is used to identify the most important IO bottlenecks in the OCFA system, and to give direction to the rest of the research. Using these findings, a general outline for a possible future computer forensic framework is created, and the main potential IO optimization techniques are explored and later integrated in a user-space file-system that combines most of the the old CarvFS and Anycast Relay functionality with measures and strategies for IO efficiency. In the interface and functionality provided by this file-system, extra care is given to address privilege separation access control hooks and integrity provisions regarding the logging of provenance information. 

The results from the provenance log analysis proved very useful in further research. Four possible IO efficiency strategies were identified:
\begin{itemize}
\item Opportunistic hashing: Calculating hashes of open pseudo-files and in-queue files opportunistically from data accessed in low-level read and write actions.
\item Linux system calls on open archive file: The Linux OS provides specific system calls for communicating intent regarding future reads and writes that are used by the kernel for page-cache strategies.
\item Throttling: A good way for avoiding page-cache misses is throttling new data input so the page-cache over-commitment is kept in check.
\item Message bus job picking strategies: Implement different \emph{sane} job picking strategies for workers, either aimed at optimized page-cache outflow or at better opportunistic hashing outcome. 
\end{itemize}
These strategies were all implemented in a user-space file-system prototype.

%% file: mainpaper-literature.tex
\chapter{Literature Survey}
\noindent 
This study brings together a number of subjects with aspects that were found in academic literature. The combination of specific subjects brought together into a coherent digital forensic setting around one central paradigm appears to currently be a unique approach. The central paradigm of using a single flat secondary storage address space for digital forensic data has found relatively little traction since the first publications of the core technique as measure for storage effective file carving. No subsequent publications have been found on the subject. Much of this study focusses on the effective integration of our central paradigm into the message passing concurrency model. A concurrency model first introduced in digital forensics by the now orphaned open source Open Computer Forensics Architecture project. A concurrency model that is also linked tightly to the Actor model of computation. A model of computation widely covered in academic literature. Given the goal of applying the research subject to an envisioned digital forensics framework, the survey also looked into papers for clues on developments that could prove relevant for the research topic. This part of the survey identified a number of desired properties that proved partially relevant for our research and should prove relevant to the eventual deployment of the results of our research into a future computer forensics framework. It also deepened the link with the Actor model of computation and introduced essential access control concerns that while not the central subject of this study proved important enough to include in the implementation stage of this research project. The final part of our survey looked into the concept of opportunistic hashing. While the opportunistic part of this subject was not found in our survey, recent development in the cryptography sub-field of secure hashing ended up proving to be crucial to digital forensics as a whole and to the topic of this study in particular.

Richard et al \cite{inplace} Introduced the concept of a single flat secondary storage address space as a tool for doing In place file-carving. Unaware of this ongoing research, the Dutch National Police developed the user space file-system CarvFS \cite{carvfs} for exactly the same purpose, and dubbed the same concept zero-storage file-carving. The Open Computer Forensics Architecture as described by Vermaas et al \cite{vermaas} made limited use of this zero-storage carving concept. Not only for file carving in the narrow sense, but also for designating files that were found using file-system forensics.

S.L. Garfinkel \cite{garfinkel} identified a number of concerns that are relevant to any next generation computer forensics environment and some of these are also relevant to our research topic. Garfinkel states that to \emph{emphasize on completeness without concern for speed} has created a situation where very few tools can perform a five minute analysis. While a five minute analysis would be pressing in the field rather than in the lab, it is important for our research to allow for a less complete analysis and better support for a \emph{light} analysis than what OCFA provided for. Garfinkel also highlights the need to move away from monolithic applications, an approach that OCFA already introduced for robustness and software reuse purposes. A more pressing issue that Garfinkel raises is the problem with the transition of academia to end users when it comes to the results of academic research, and the problem of applying forensic frameworks primarily targeted at an investigation setting to an academic research setting. We take these problems as an important indicator towards the need for computer forensics frameworks and their components that are suitable for both an academic setting and for a full scale digital forensic investigation.  
An idea that is further underlined in the call for a stable and scale independent API for computer forensics frameworks. 

Wood et al \cite{wood} showed the role of data containers in the chain of custody recording and made a case for the use of AFF4 and/or DFXML within the context of multi disk image digital forensic data repositories. Access control is also looked at in detail with respect to access by individuals. They take the approach that image provenance and integrity are to be guarded by means of container format technology. Chatz and Clark \cite{sdeb} from the prospect of container files introduce the concept of Sealed Digital Evidence Bags. A non-monolithic variant of the Digital Evidence Bag concept. They describe how data and meta-data, once written should be sealed by means of signatures. They define a taxonomy that further highlights the importance of solid provenance information recording and integrity.

It is in provenance information that we reconnect with the Actor model and with aspects that most touch on the research at hand. Hewitt \cite{actor} amongst other principles lists Persistence, Concurrency and Provenance as core information system principles for information integration. He defines the actor model, to be based on one-way asynchronous communication. While in OCFA message ordering was strictly maintained, Hewitt explains how the Actor model puts no restriction on message reception order. With respect to performance and the Actor model, Hewitt defines as most important tool the minimization of latency along critical paths. It is exactly the lack of such minimization that was found at the core of the OCFA performance issues. Finally Hewitt defines a set of three Actor address acquisition possibilities that should prove crucial in addressing security concerns for the implementation. 
A Spiessens \cite{spiessens} provides us with an overview of patterns for the use of \emph{capabilities} for the implementation of privilege separation and cooperating objects or processes that prove applicable to part of our computer forensics access control requirements. Meijer \cite{minorfs} implemented a user space file system based on \emph{sparse-capabilities} especially for privilege separation and access control purposes in a high-integrity setting.  

With respect to scalability and modern scalability technology, Carrier et al \cite{sleuth} worked on a Sleuth-kit Hadoop framework. Work on this project seems to have discontinued. Some of the findings however were shared with us and should prove useful in future work on multi-node set-ups. An other scalability insight was presented within the FIVES Architecture \cite{fives} in its introduction of an alternate router for OCFA. 

The last subject of this study concerns opportunistic hashing. Until today most computer forensic tools make use of SHA1 or MD5 as algorithm for creating digital signatures. NIST \cite{nist} defines SHA1 as being deprecated for signing purposes, while use of SHA1 for the digital forensic file matching scenario is still considered \emph{acceptable use}. In 2015 Stevens et al \cite{sha1} presented a working free-start collision against SHA-1. While a free-start collision isn't a full SHA-1 collision yet, it does present an important milestone on the road to a full SHA-1 collision. From a cryptographic point of view this means that SHA-1 isn't a solid choice to base new technological developments on. On the other hand, while NIST has been providing SHA256 hashes with their hash sets that address security issues, there are serious performance concerns with SHA256 and its use in digital forensic frameworks. As such, the recent advances regarding SHA1 collision resistance make us face the additional challenge of finding a suitable core hashing algorithm to implement our opportunistic hashing study with. In the selection of SHA3 candidates by NIST \cite{sha3}, different aspects were looked at, including security and performance aspects of the different candidates. This assessment made clear that two former SHA3 candidates, Skein and BLAKE would be particularly suitable for usage within computer forensics due to their software performance characteristics. These candidates set us on the path of BLAKE2 \cite{aumasson}. Guo et al \cite{guo} showed an analysis of BLAKE2 that led to the conclusion that BLAKE2 would be the best candidate for our opportunistic hashing purposes, specifically for digital forensics in general.

While our literature survey yielded a wide range of relevant research, the idea of using a flat secondary storage address space as core of a page-cache friendly Actor based set-up for digital forensic frameworks, as well as the idea of opportunistic hashing within such a framework, appear to be a completely new  research topic.

%% file: mainpaper-problem.tex
\chapter{Problem statement \& adapted approach}
\noindent 
Lab-side semi-automated processing of large amounts of digital forensic evidence data within the context of medium to large criminal investigations requires tooling that goes beyond the scalability limitations of common commercial tools. In the past, OCFA provided an open source framework that has proven useful in both a law-enforcement and, to some extent, in an academic setting. Next to the fact that development and support of OCFA has discontinued, relative changes in the speed and latency between CPU, RAM, Network IO and disk IO of computer forensic lab servers, and new insights regarding anti-forensic threats and the requirements for system robustness have made some design choices made at OCFA inception, choices that in today's reality should be considered sub-optimal. 

At the time of inception, CPU, next to disk-IO, was a major bottleneck in most investigations. Avoiding page-cache misses wasn't that promising a goal as RAM in a typical server wasn't anywhere close to the size where high amounts of page-cache misses could have been expected to be avoided.
Looking at developments since OCFA inception we see that while high throughput low latency Solid State Disk (SSD) solutions have become available, their size and pricing should still be considered prohibitive for the deployment of the triple digit, and bigger, terabyte storage solutions required in modern day medium to large scale digital forensic investigations. RAM sizes and RAM speed have gone up considerably as has the number of cores available in a typical computer forensics lab server. Non SSD storage however, while having continued to grow cheaper, has not kept up with the rest of the hardware improvements. As a result of these changes, the shortcomings of the architecture used in OCFA with regards to inefficient use of secondary storage IO have become the dominating performance bottleneck. In today's reality the scalability properties of the OCFA architecture should be considered mostly deprecated.

Our literature survey identified a continued need for an open source scalable computer forensic framework suitable for both academic research and criminal investigations, that could fill the gap that the discontinuation of OCFA development has created. A framework that could only be realized if secondary storage IO efficiency and system robustness requirements can both be met.

\section{Goal}
We wish to investigate the viability of creating the foundation for an OCFA inspired next generation scalable message passing concurrency computer forensics framework suitable for both academic research and full scale criminal investigations that addresses the page-cache and system integrity concerns identified. 
\section{Research questions}
\begin{itemize}
\item What were the historic page-cache related bottlenecks within the OCFA architecture?
\item How could these page-cache related bottlenecks be addressed in a future framework?
\item What role could the flat secondary-storage address-space model, introduced by CarvFS and ScalpelFS, play in solving page-cache related bottlenecks?
\item What role could an opportunistic form of hashing play in solving page-cache related bottlenecks?
\item What were the anti-forensic weaknesses within the OCFA architecture?
\item How could these anti-forensic weaknesses be addressed in a future framework?
\item What role could the sparse capability based access-control model, introduced by MinorFS, play in solving anti-forensic weaknesses?
\item Can the above concerns be fully or partially implemented within a proof-of-concept computer-forensics framework component?
\end{itemize}

%% file: mainpaper-approach.tex
\section{Adopted Approach}
\noindent 
Before attempting to solve any problem, the first step in the approach is identifying and quantifying page-cache related bottlenecks and design shortcomings of the existing OCFA framework. In order to do this, we make use of profiling/provenance logs from a small set of real investigations that were made available for this research by the Dutch national police. 

Using the log files as extracted from archived historic database dumps, we look to quantify the problem as it existed in the past, and look at the effectiveness of the limited measures implemented by OCFA that should in theory mitigate some of the perceived problems. Using the same data, we map the typical tool-chain path, both quantified on a per-file basis and quantified on a per data-volume basis. 

From our findings in this first step we identify the main bottlenecks and look at finding ways to avoid or attenuate these problems in a future computer forensics framework. 
From this broader perspective, we then take the concept of a flat secondary-storage address-space, as simultaneously introduced by both CarvFS and ScalpelFS, and look at how that concept could facilitate a more page-cache friendly storage and/or messaging system that would be usable in the future implementation of such a framework. The potential flat secondary-storage address-space related concepts we investigate are:

\begin{itemize}
\item Page-cache friendly archive interaction. 
\item Page-cache friendly message-bus interaction.
\item Opportunistic hashing.
\end{itemize}

Special attention is given for opportunistic hashing to take the results from the literature survey into account regarding the suitability of hashing algorithms.

From our findings from the literature study, we look at fitting in those controls needed to guard the forensic integrity of the forensic data processing. We investigate the possibilities to guard the tool-chain and system integrity, even in the case of compromised or buggy code in a particular framework module. Robustness and process integrity concepts we investigated are:

\begin{itemize}
\item A lab-side privilege-separation equivalent of the Sealed Digital Evidence Bags concept.
\item Trusted provenance logs.
\end{itemize}

With the results of these two assessments, a proof of concept sub-system was implemented that could serve as the foundation of a future high-integrity page-cache friendly computer forensic framework.

To conclude the research, the results are evaluated and advice is given as to how future research could aid in the completion of a full successor to the now deprecated OCFA framework. 

%% file: mainpaper-results.tex
\chapter{Analysis of OCFA provenance logs}
\noindent 
In this chapter, the results of the first steps of the adopted approach are discussed in detail. This is done by describing the outcome of the first two main phases of the research:
\begin{enumerate}
\item Analyze the OCFA timing information.
\item Identify the main bottlenecks within the OCFA framework.
\end{enumerate}
\section{Analysis of the OCFA timing information}
In this section we look at identifying the main bottlenecks through an analysis of OCFA timing information.
\subsection{Extracting the timing information}
The Dutch National Police made a set of database dumps from old OCFA runs available for use in this study. One important prerequisite however was that the sensitive investigative data would under no circumstances become part of the scientific research. In order to adhere to these preconditions, a Python script was created to parse the database dumps and to extract only those parts of the data that were essential for \emph{technical} evaluation related to timing. In theory this information should be sufficient to base our research and evaluations on. It could be possible that a better foundation of our findings would be possible when including non-directly timing related information. The python script as referenced in a following appendix did the following:
\begin{itemize}
\item Extract the evidence XML blobs from the SQL dump.
\item Extract the oldest timestamp from the XML blobs and use that time as offset for the successive steps.
\item Process each individual XML blob and extract only the relative start and stop times and module types for each job, as well as the sizes of the job-bound data entities.
\item Write the extracted information to a JSON based log file.
\end{itemize}
The true analysis of the OCFA timing information took place on the resulting JSON based log files. It is important to note that due to a bug in the OCFA Java library at the time these investigations were active, there are some (detectable/reversible) offset bugs in the results. These needed to be corrected during the actual analysis.

%% file: ocfa-step2-results.tex
\subsection{A fictitious perfect cache}
\begin{figure}
\centering
\subfloat[case 1]{
  \includegraphics[width=70mm]{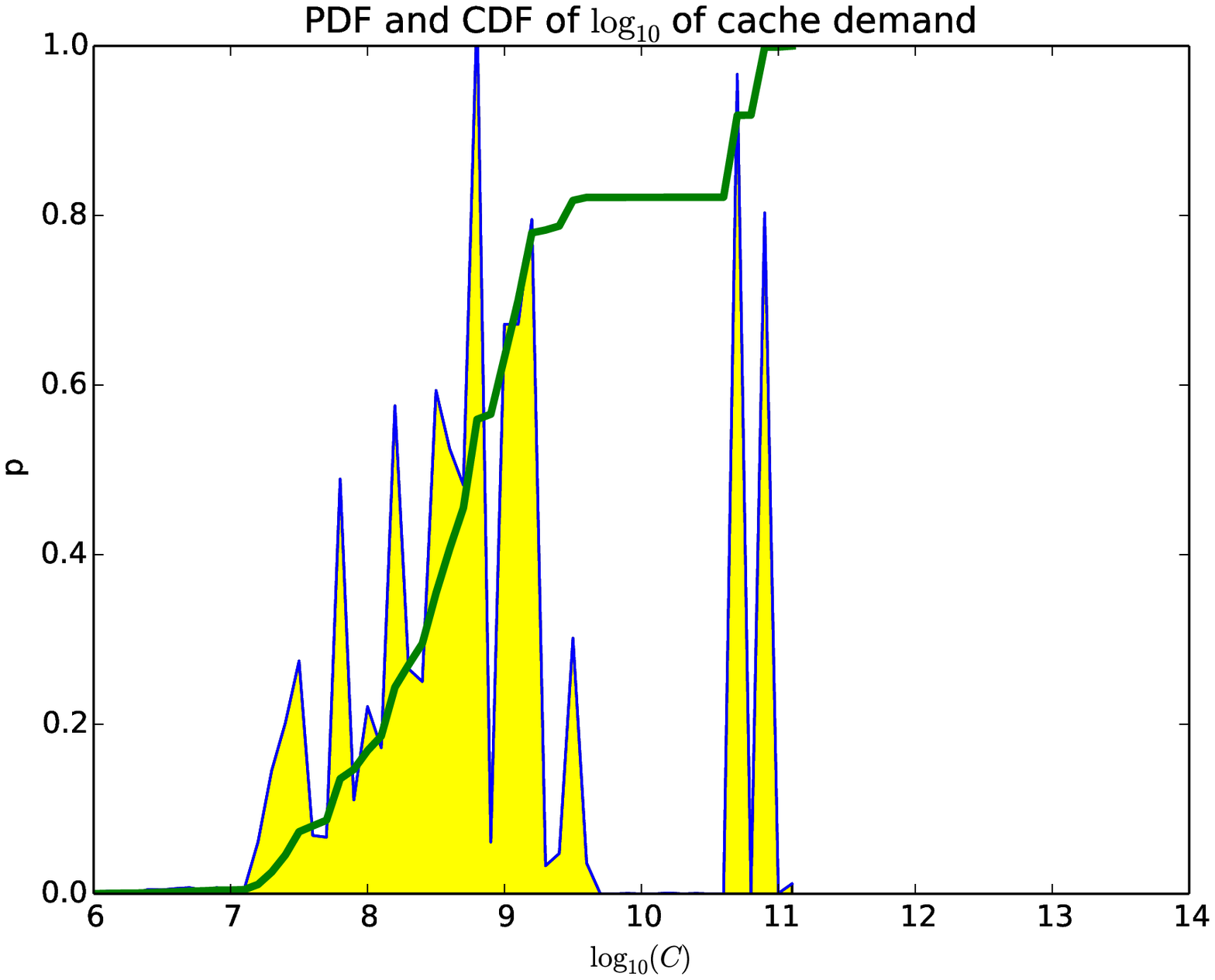}
}
\subfloat[case 2]{
  \includegraphics[width=70mm]{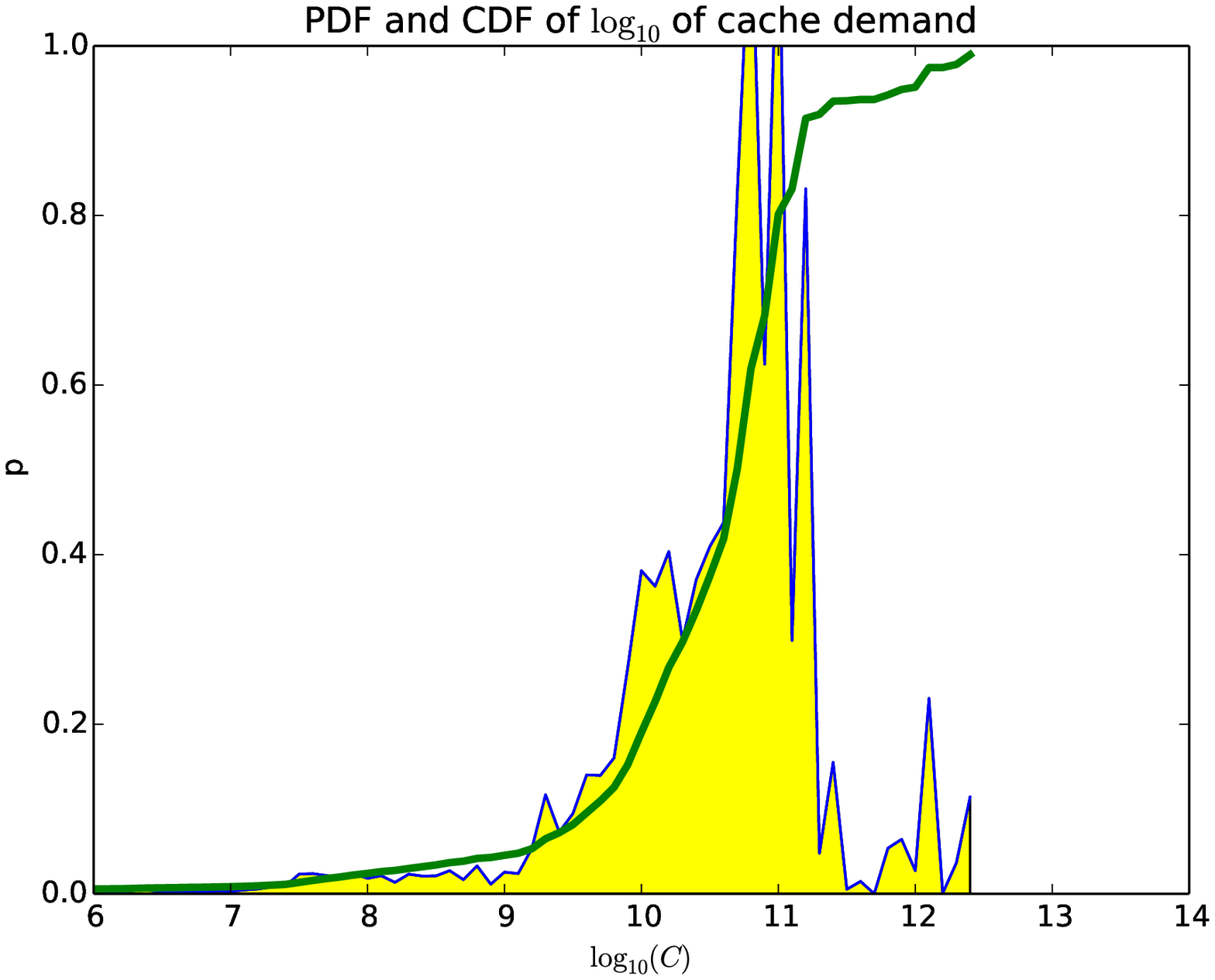}
}
\hspace{0mm}
\subfloat[case 3]{
  \includegraphics[width=70mm]{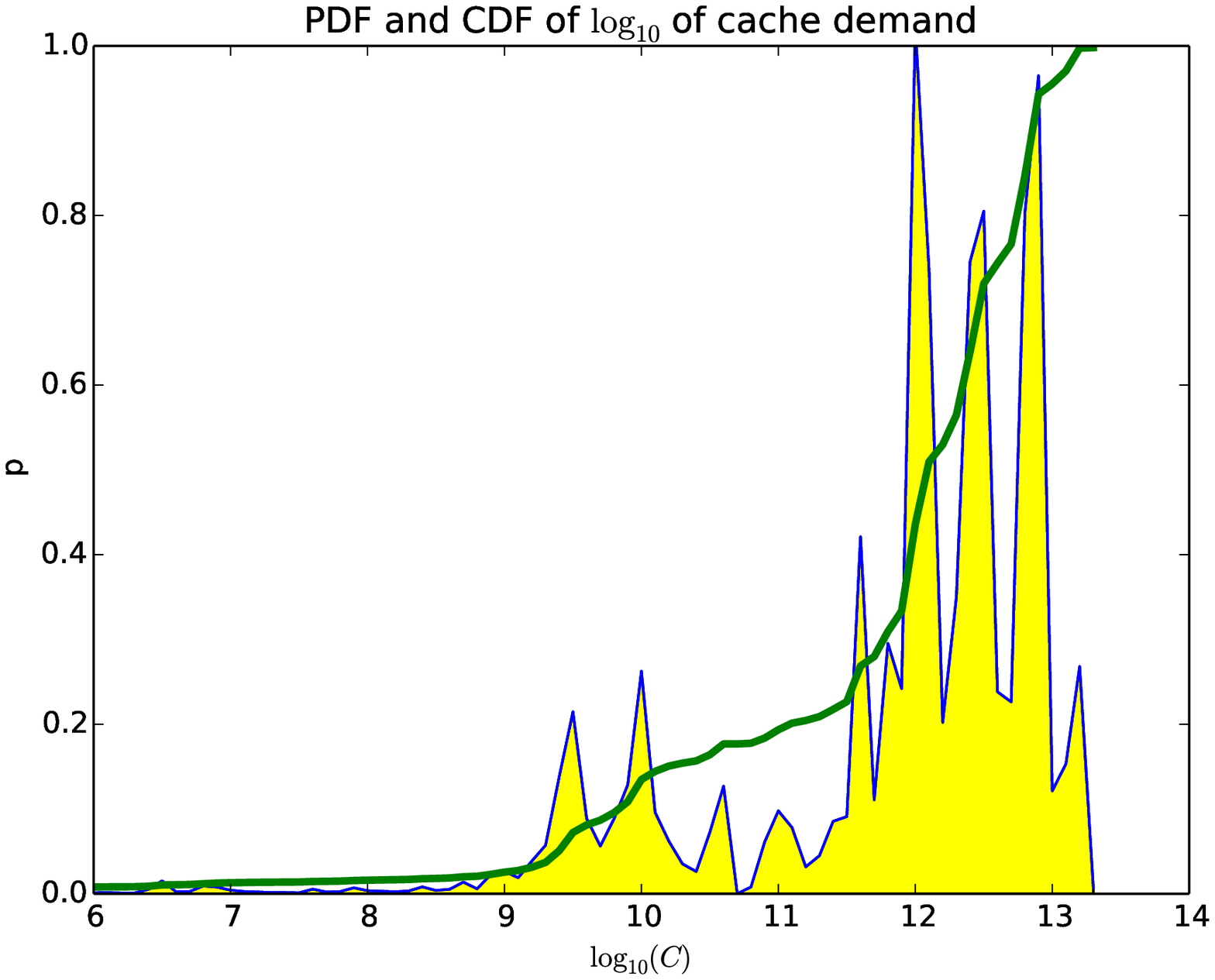}
}
\subfloat[case 4]{
  \includegraphics[width=70mm]{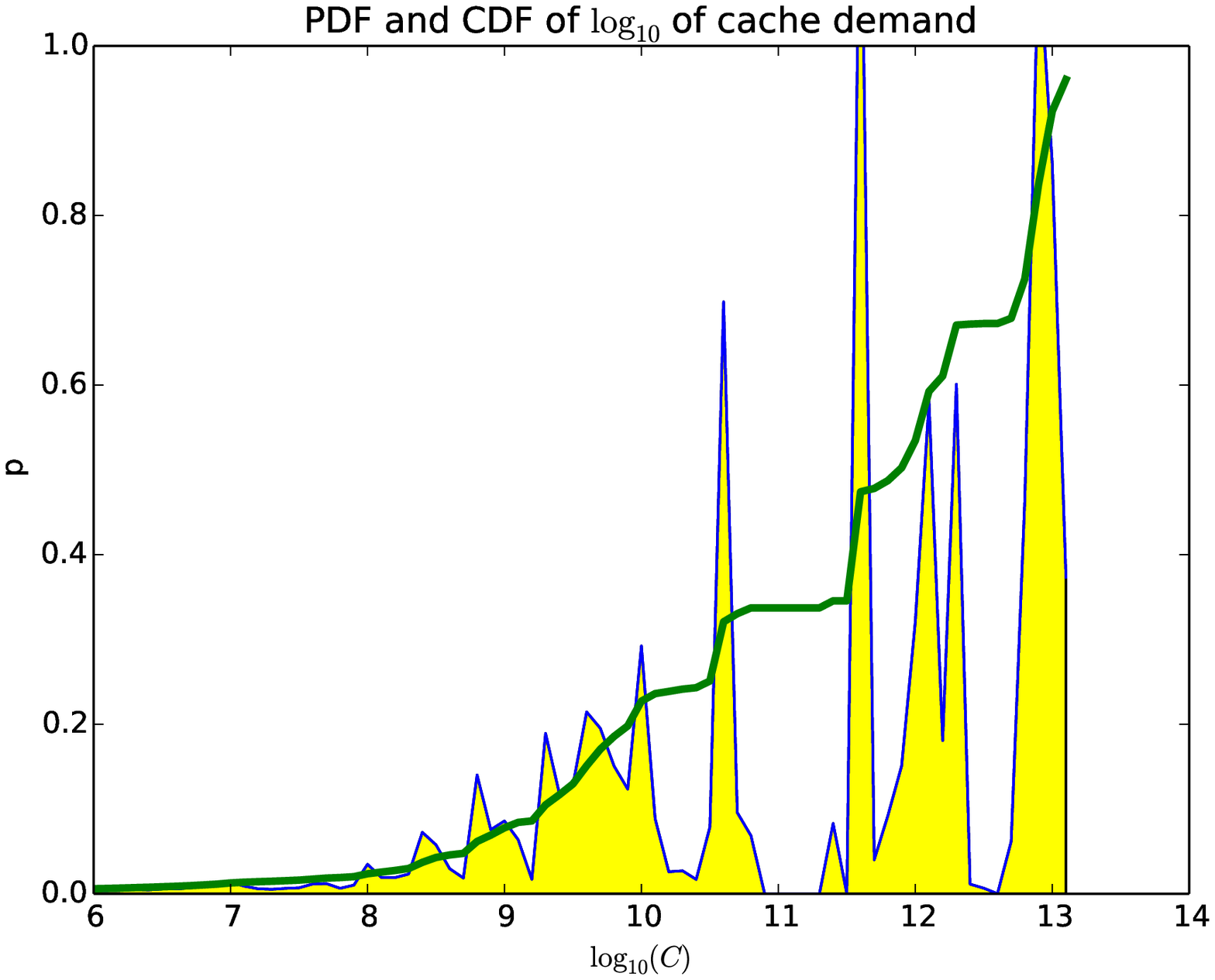}
}
\caption{Virtual cache size probability density}
\label{fig:VirtCacheSize}
\end{figure}

In our first analysis we look at data entities from our four investigations from a first seen / last seen event perspective. We define a model where a fictitious perfect infinite-size cache is used. Every time a piece of data is first seen, its data size is added to the virtual cache. Every time a piece of data is last seen, it is removed from the cache. Figure ~\ref{fig:VirtCacheSize} on page ~\pageref{fig:VirtCacheSize} shows the probability density function of the required amount of RAM to implement such a perfect disk cache. The X axis is the $\log_{10}{C}$ where $C$ is the required capacity for the fictitious cache size at any random point in time during an active OCFA run. When we look at figure ~\ref{fig:VirtCacheSize}, depending on the investigation, we see some different results depending on the investigation. While for example a fictitious cache size of $10^9$ (1 GB) would fit all active data for about 65\% for investigation one, the same amount would yield a fit for only one or two percent of the time in investigation three. One main conclusion that we can draw from three out of four investigations is that most still active data could not possibly stay contained in any reasonable amount of physical RAM memory that the OCFA machines might have had. In the following sub sections we discuss why this is the case.

%% file: ocfa-step3-results.tex
\subsection{Inter-job timing}
\begin{figure}
\centering
\subfloat[case 1]{
  \includegraphics[width=70mm]{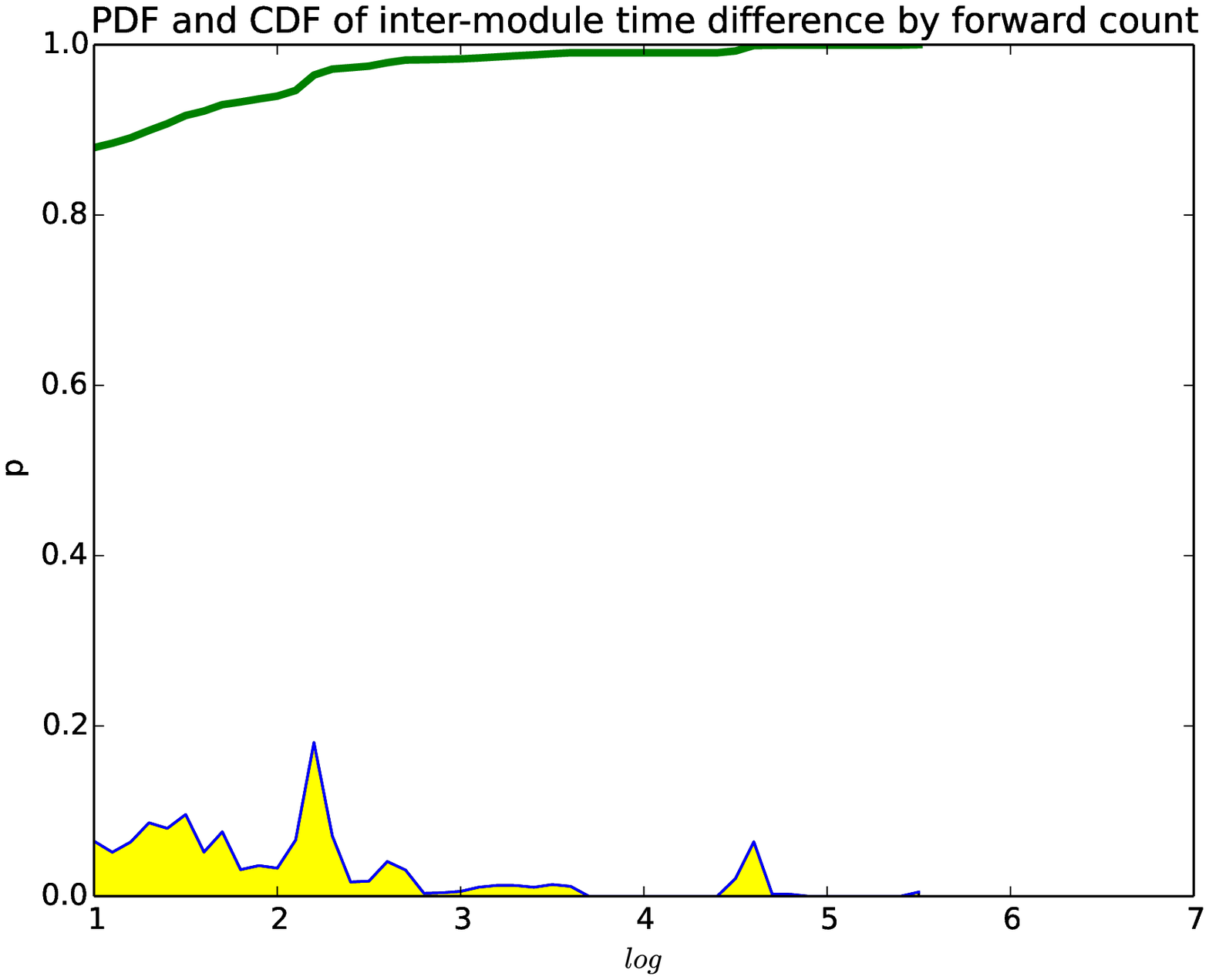}
}
\subfloat[case 2]{
  \includegraphics[width=70mm]{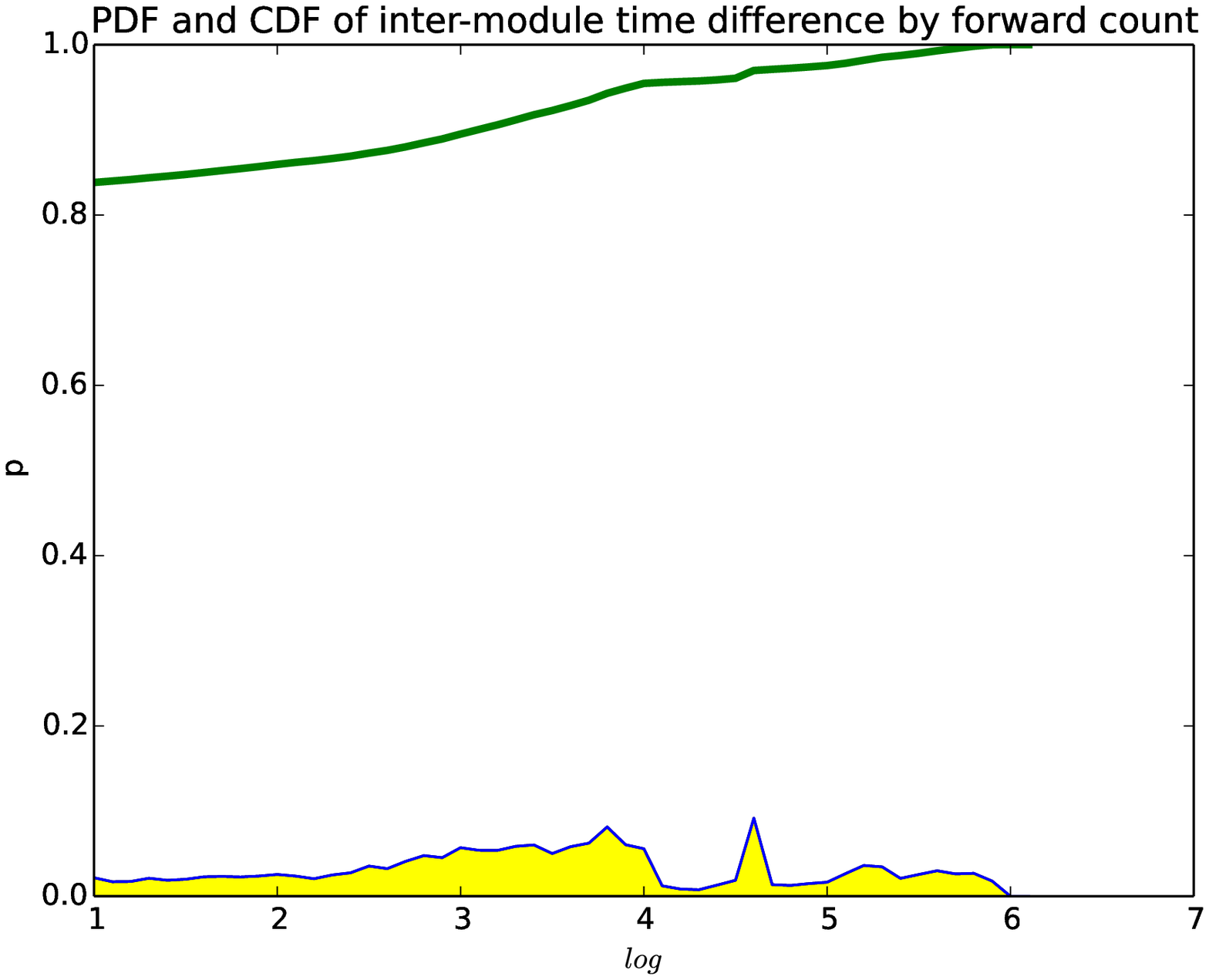}
}
\hspace{0mm}
\subfloat[case 3]{
  \includegraphics[width=70mm]{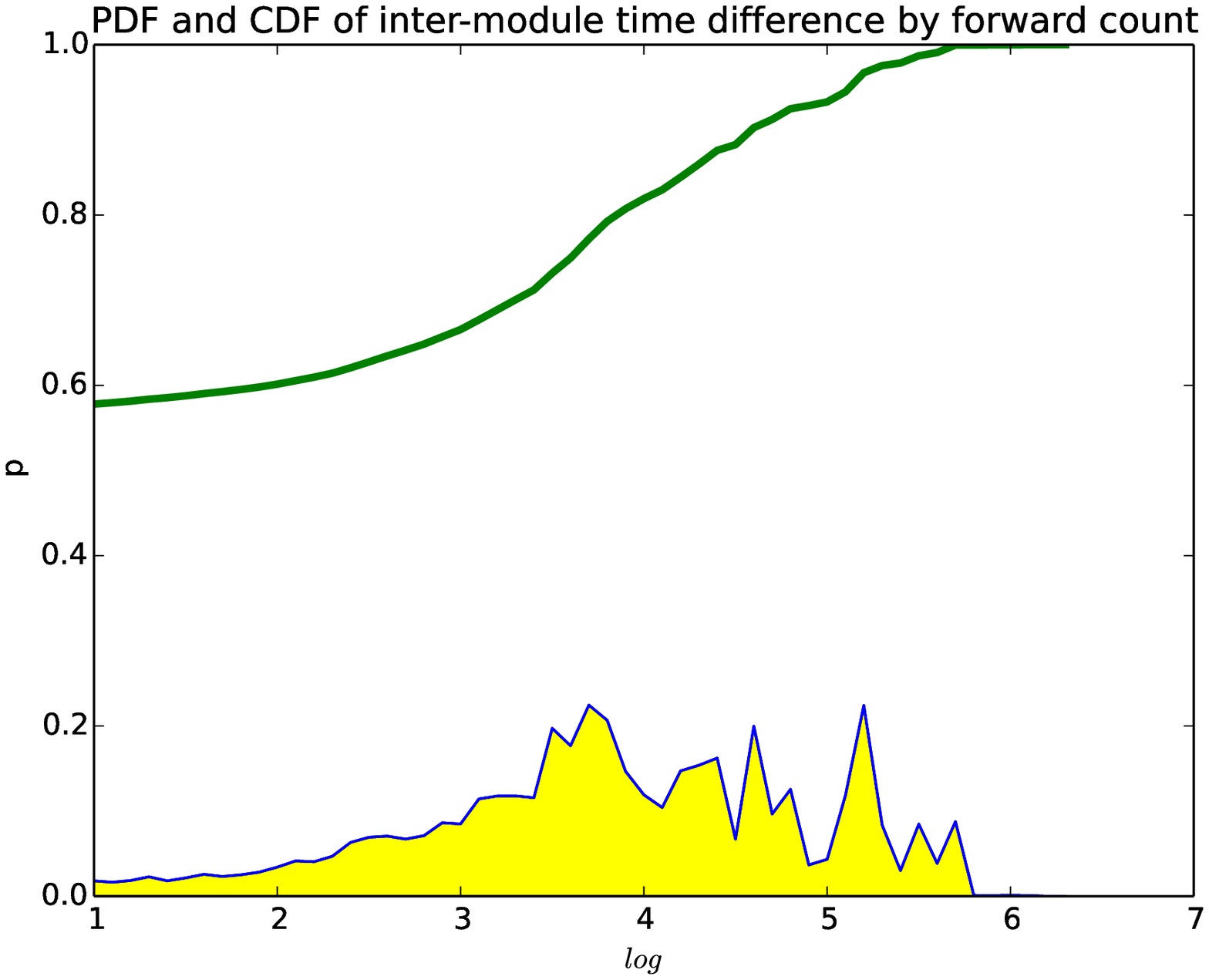}
}
\subfloat[case 4]{
  \includegraphics[width=70mm]{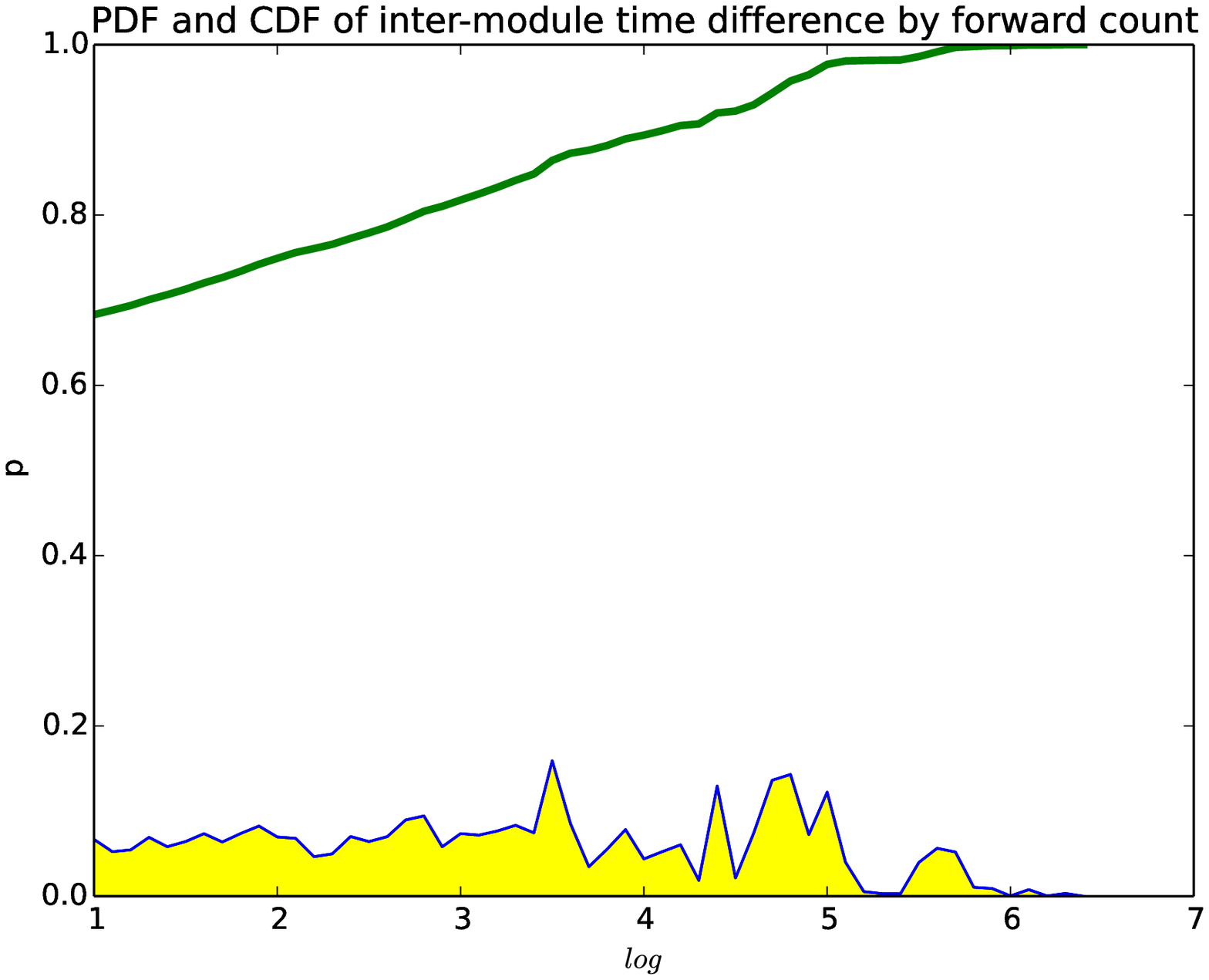}
}
\caption{Inter-job time probability density}
\label{fig:InterJob}
\end{figure}
Figure ~\ref{fig:InterJob} on page ~\pageref{fig:InterJob} shows the probability density function of the inter-module time for modules processing the same evidence.
In the previous subsection we discovered the fact that most of the active data would not fit in the physical memory of an OCFA server. When we look however at the probability density function of the time between two jobs, we discover that between about 60\% and 90\% of all inter-job times falls below $10^1$ seconds mark. While there are times extending up to and exceeding the $10^6$ seconds visible in some of the graphs, most of the inter-module times are so small in comparison that disk cache hits would seem almost inevitable. These results are quite inconclusive so far, so we need to look at other results.
\subsection{Inter-job timing by content size}
\begin{figure}
\centering
\subfloat[case 1]{
  \includegraphics[width=70mm]{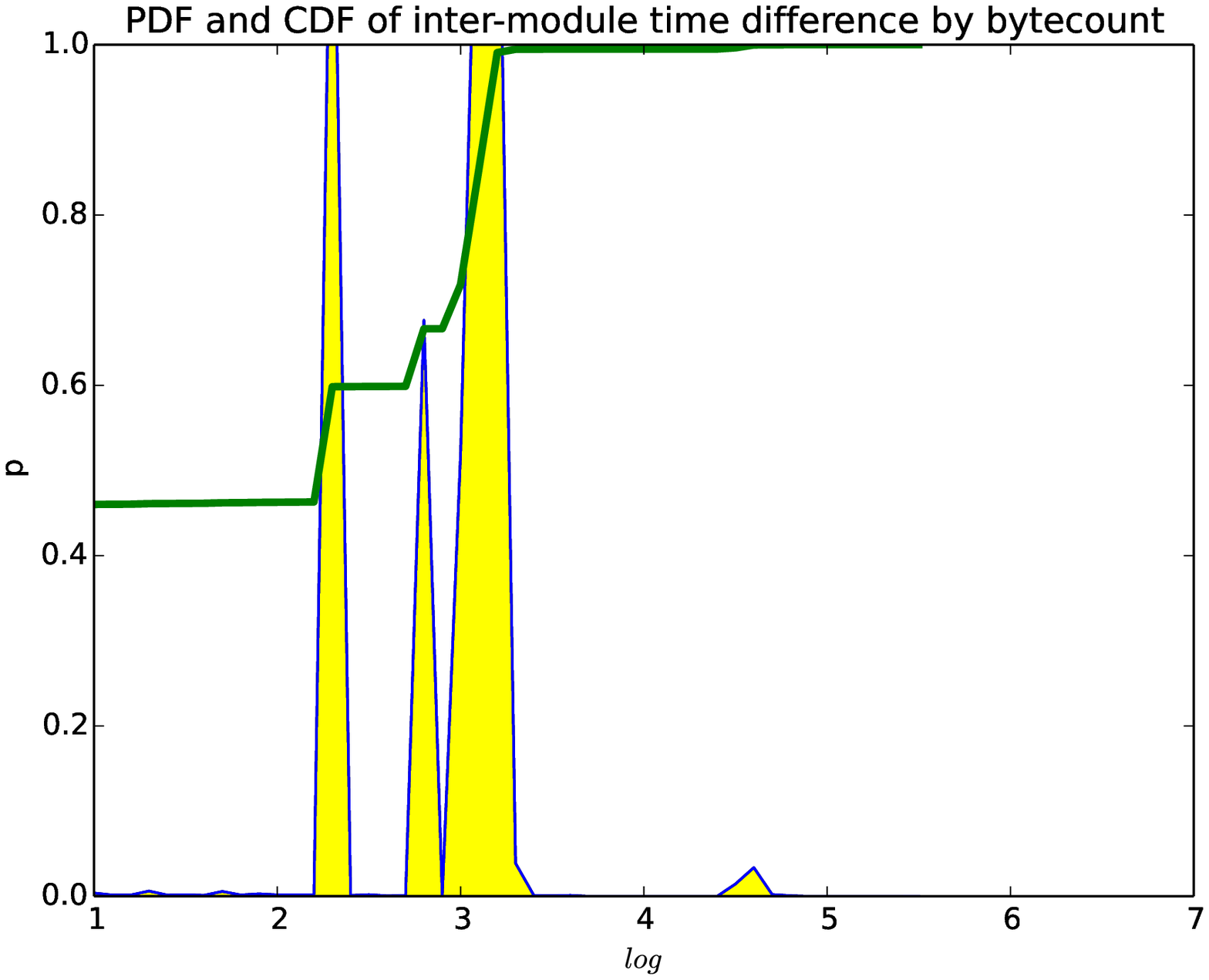}
}
\subfloat[case 2]{
  \includegraphics[width=70mm]{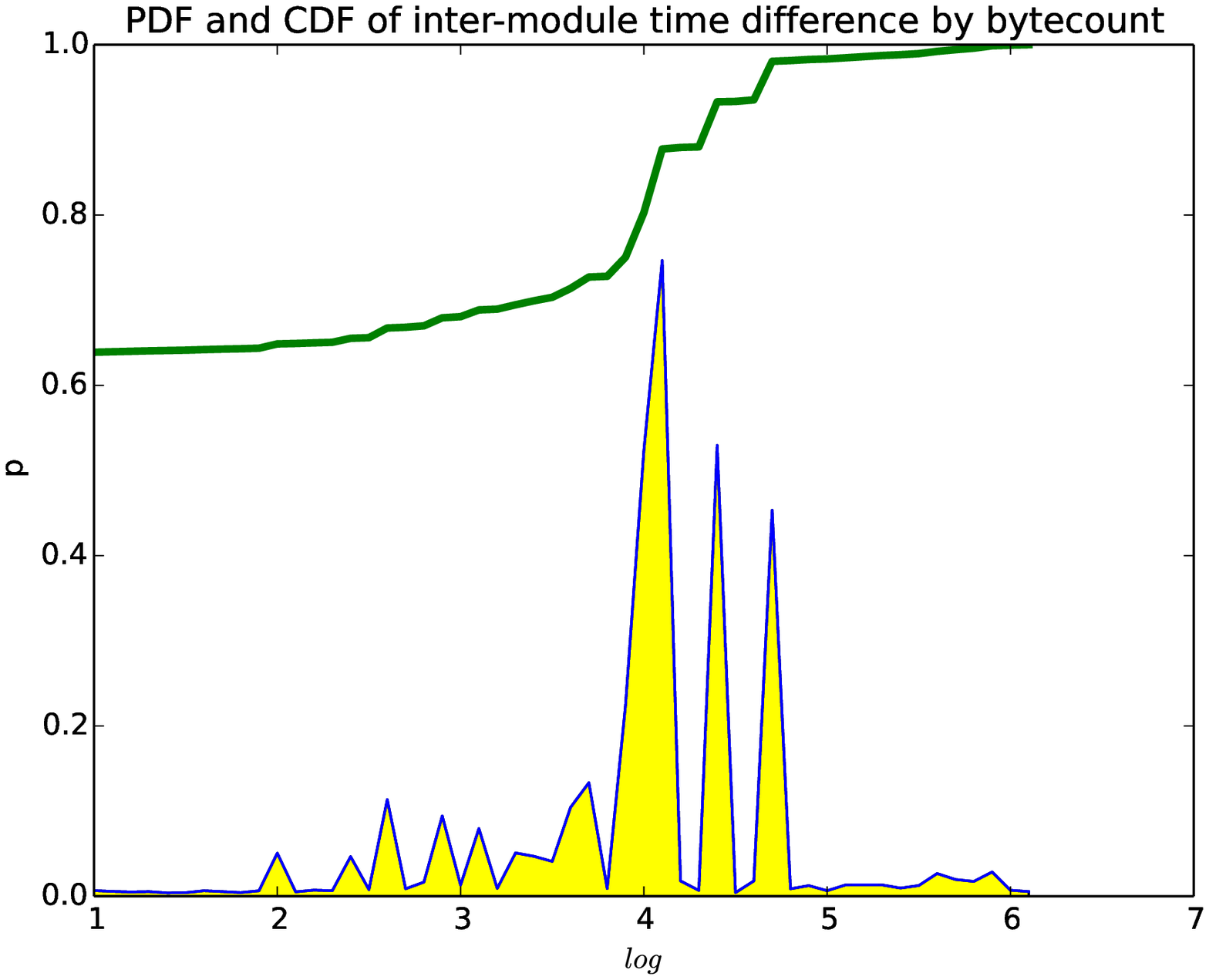}
}
\hspace{0mm}
\subfloat[case 3]{
  \includegraphics[width=70mm]{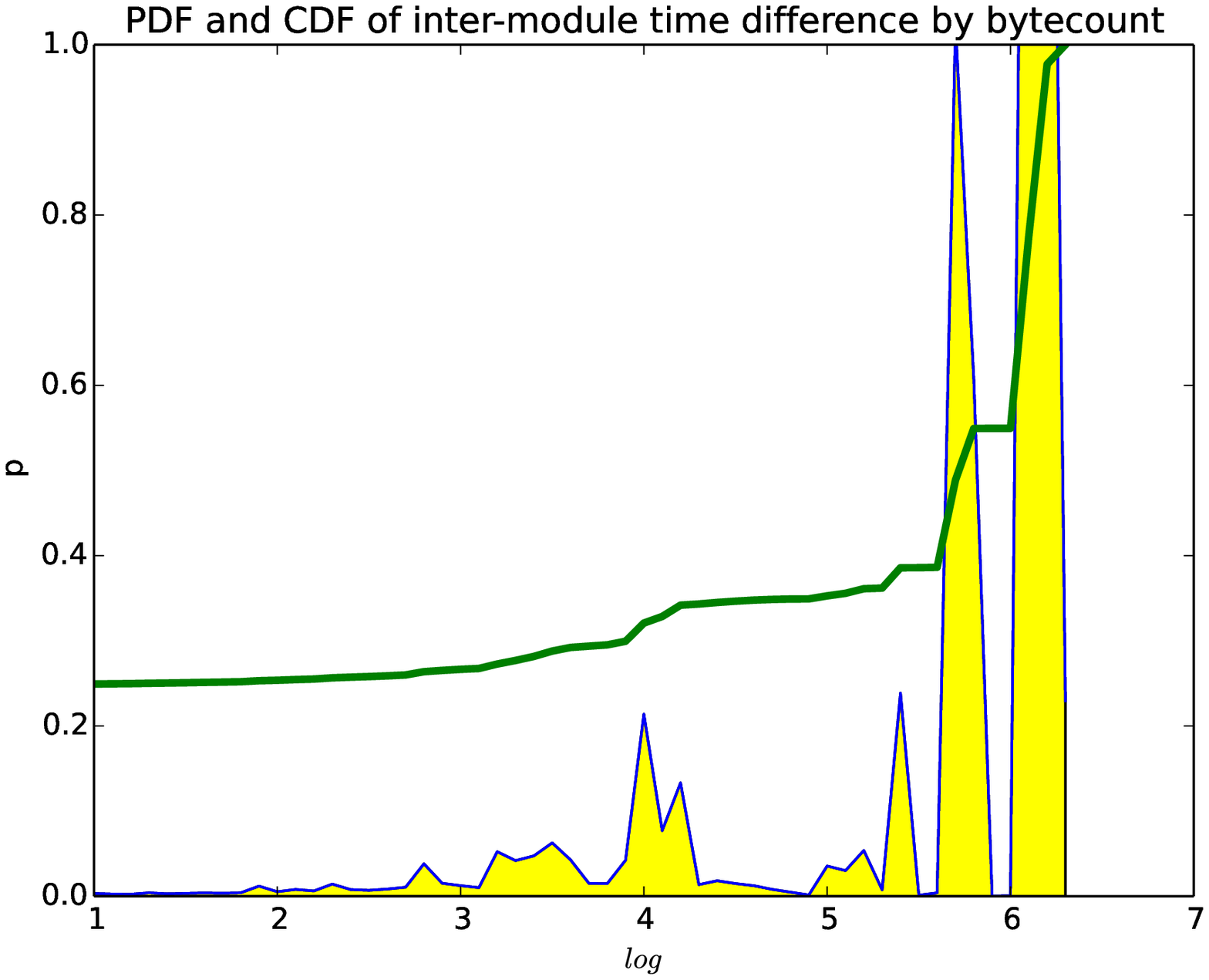}
}
\subfloat[case 4]{
  \includegraphics[width=70mm]{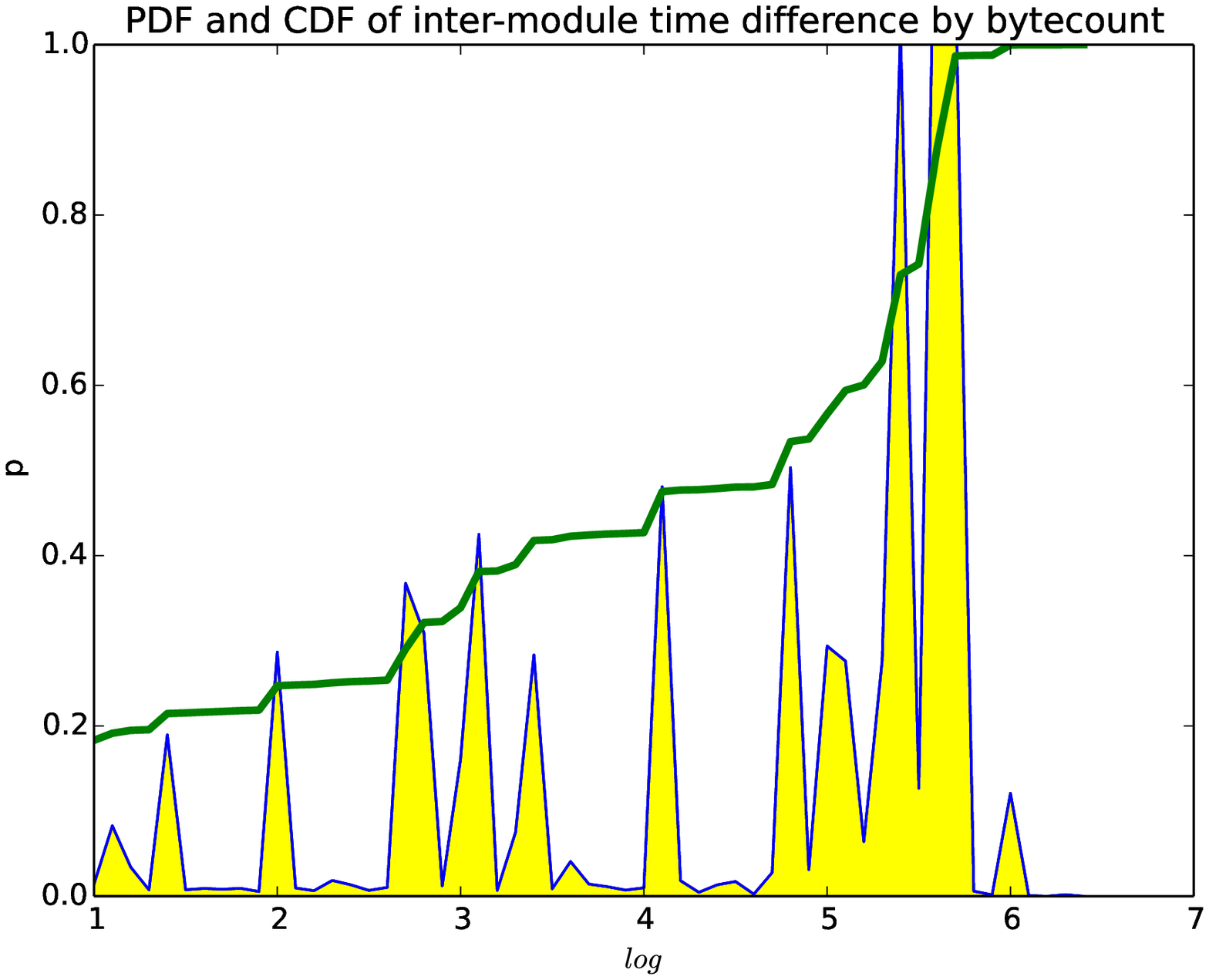}
}
\caption{Inter-job time probability density by volume}
\label{fig:InterJobBySize}
\end{figure}
If instead of weighing the probability density function by the number of data events, we weigh it by the amount of data, the picture that we saw in the previous subsection shifts significantly. Figure ~\ref{fig:InterJobBySize} on page ~\pageref{fig:InterJobBySize} shows this shift. A relatively large portion of the data now shows to have a job time interval of well over $10^4$ seconds. That means multiple hours, long enough to expect any disk caching to long have been overwritten so that a consecutive read would surely lead to a disk-cache miss.
\subsection{First-last timing}
\begin{figure}
\centering
\subfloat[case 1]{
  \includegraphics[width=70mm]{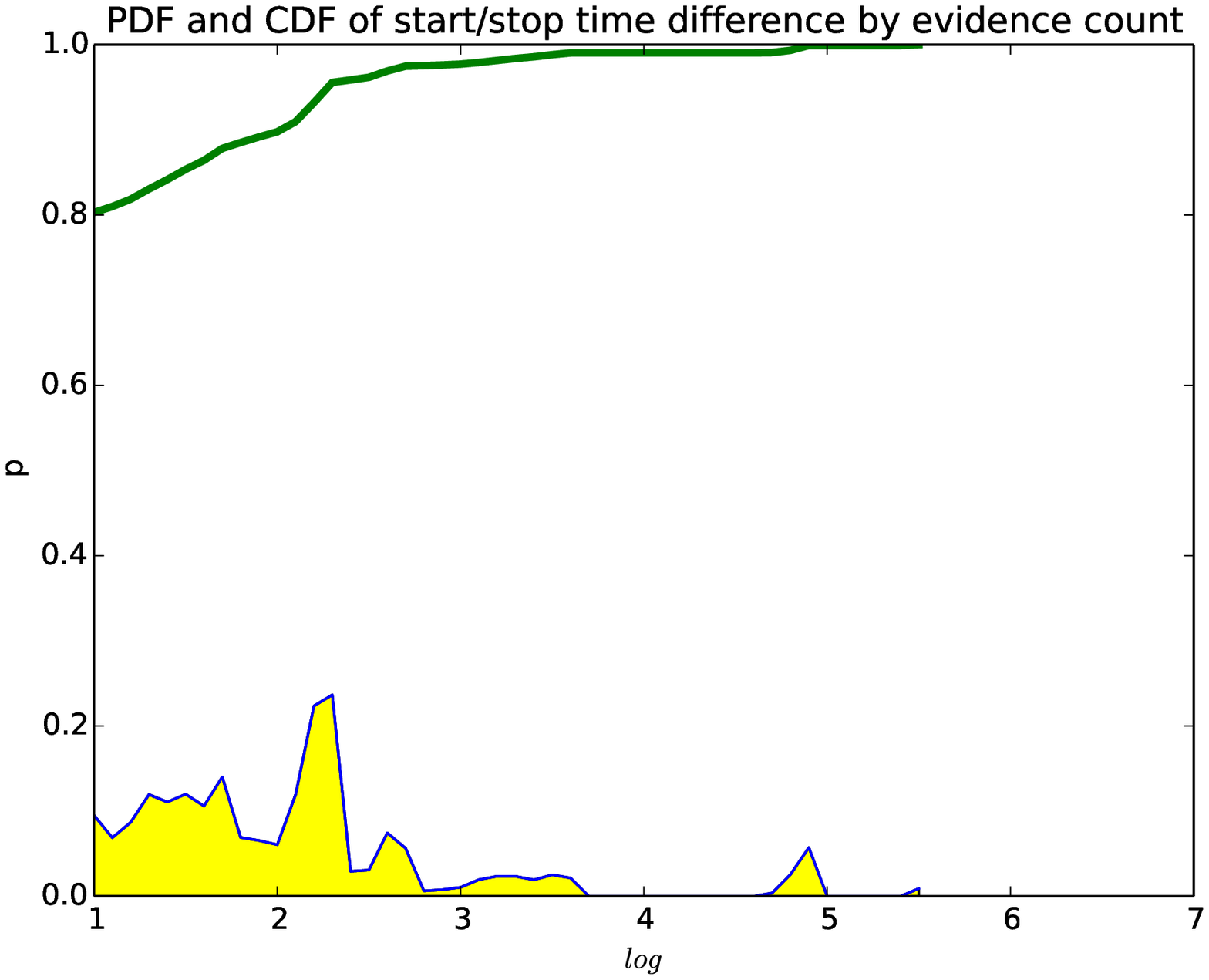}
}
\subfloat[case 2]{
  \includegraphics[width=70mm]{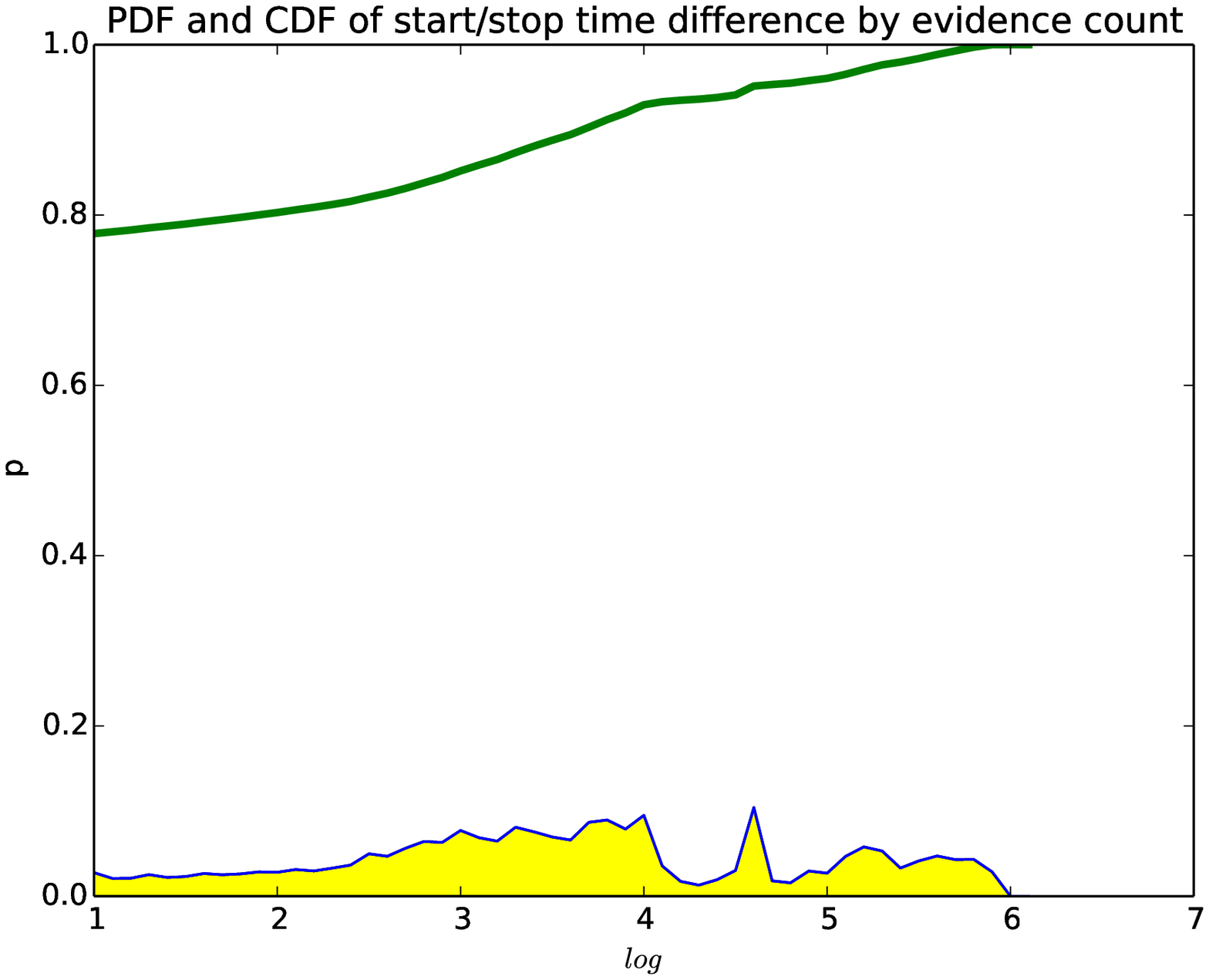}
}
\hspace{0mm}
\subfloat[case 3]{
  \includegraphics[width=70mm]{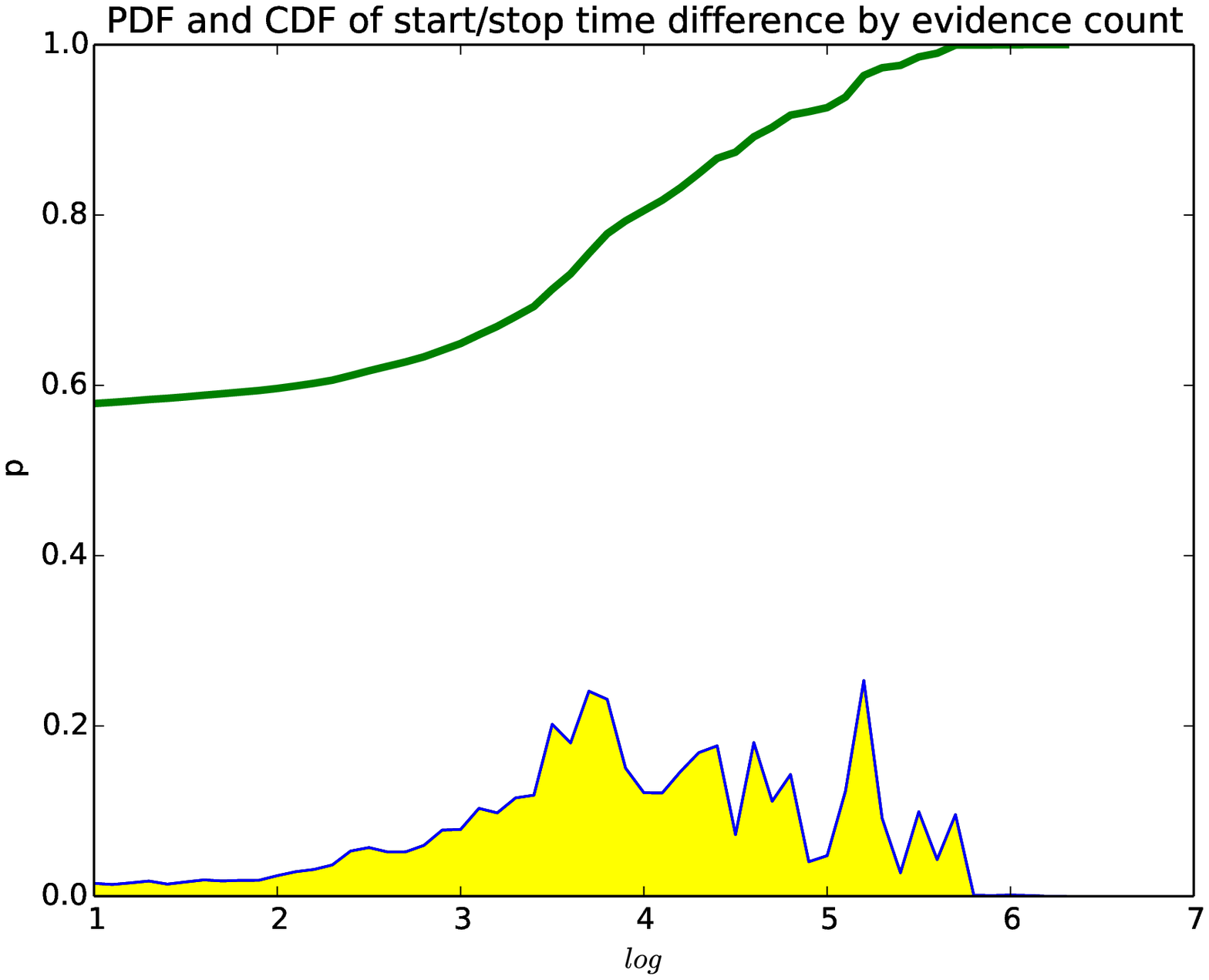}
}
\subfloat[case 4]{
  \includegraphics[width=70mm]{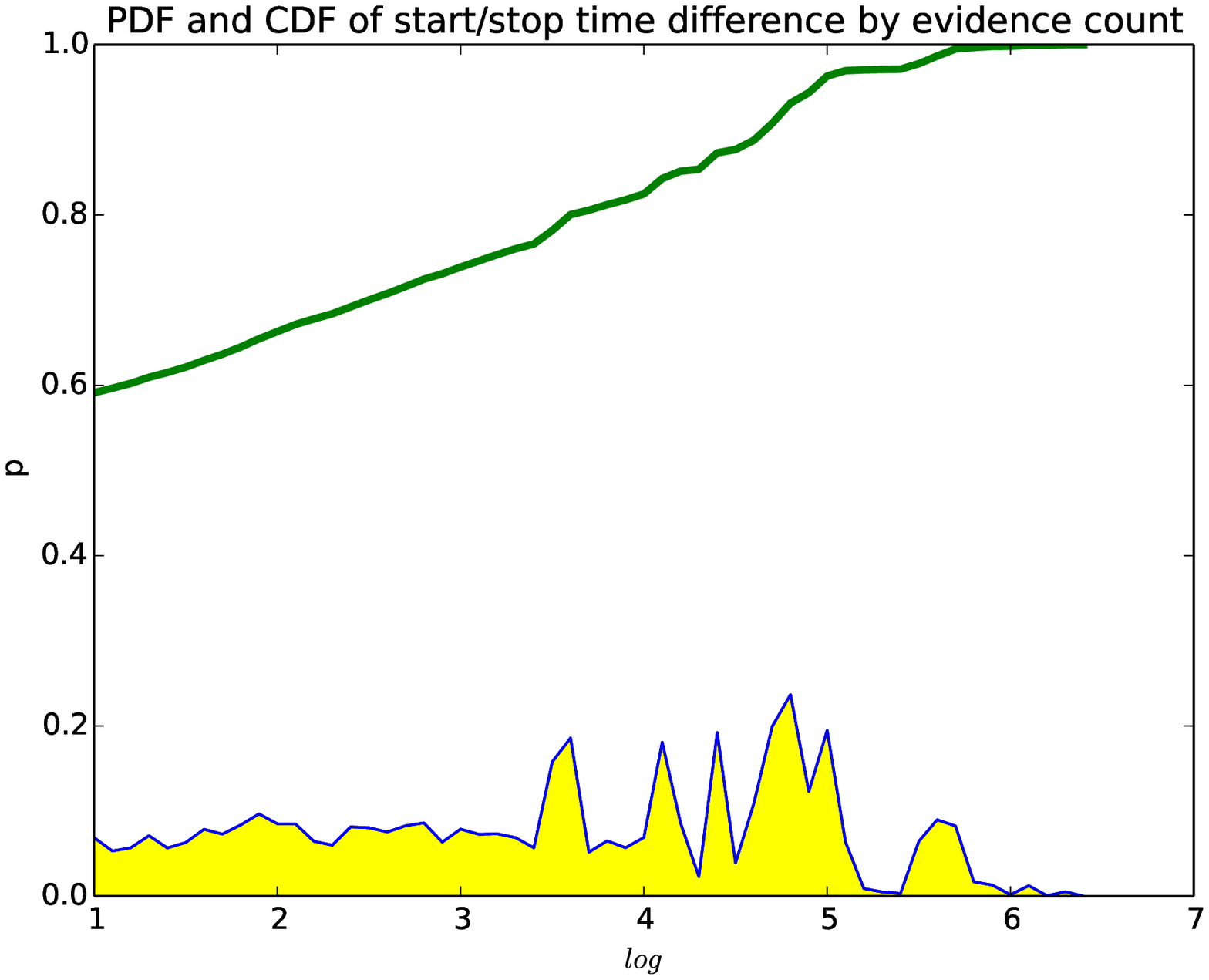}
}
\caption{First-last time probability density}
\label{fig:FirstLast}
\end{figure}
Given that our inter-job timing stats were inconclusive so far, we look at the timing interval between the first module creating the data entity and the last data processing module being done with it. In ~\ref{fig:FirstLast} on page ~\pageref{fig:FirstLast} we see that while less pronounced than for ~\ref{fig:Interjob} on page ~\pageref{fig:InterJob}, it is still a majority of about 60\% to 80\% of evidence entities that will be completely done in less than ten seconds. These results while being in line with our expectations for the OCFA priority system, still don't match the other results we have seen so far.
\pagebreak
\subsection{First-last timing by content size}
\begin{figure}
\centering
\subfloat[case 1]{
  \includegraphics[width=70mm]{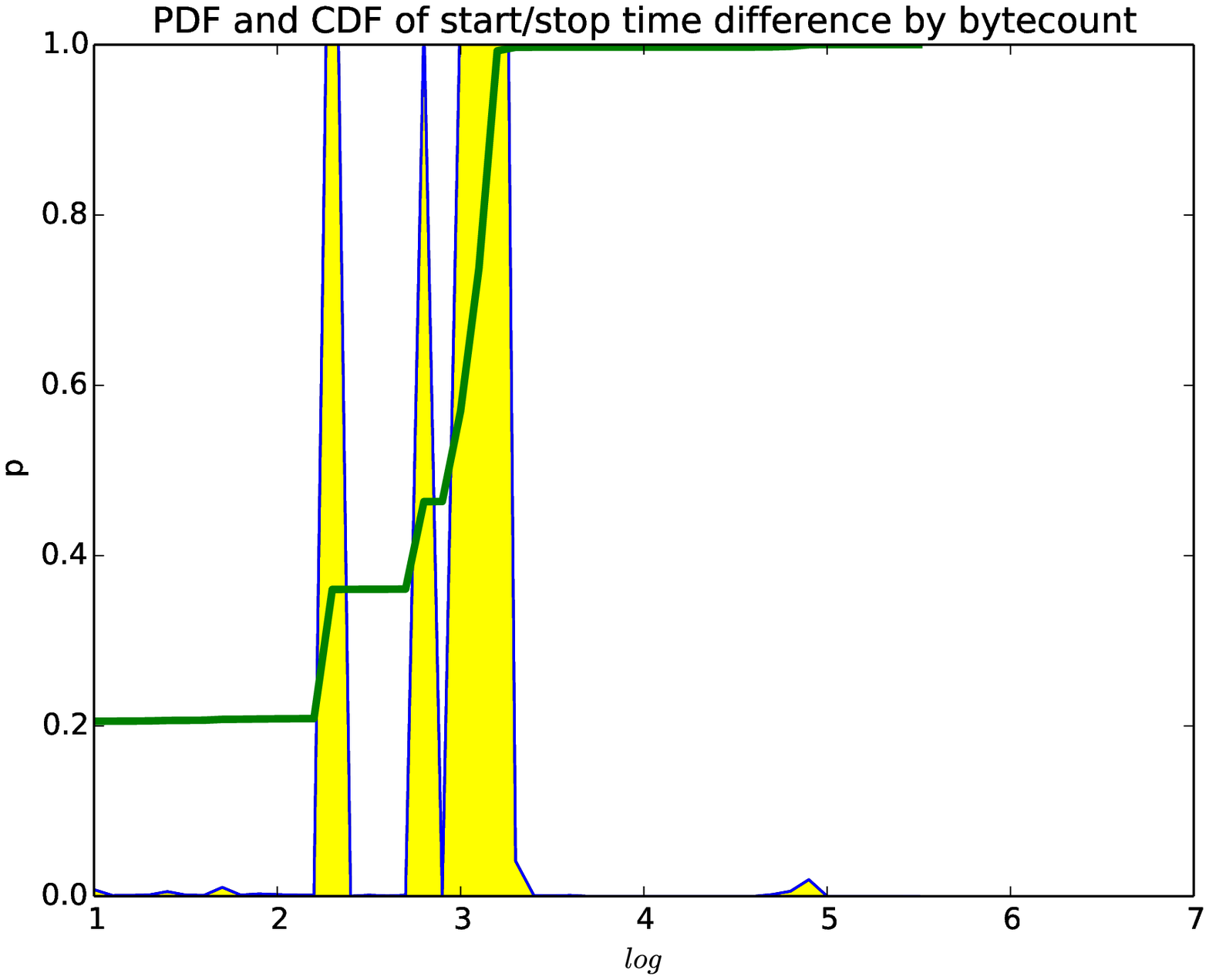}
}
\subfloat[case 2]{
  \includegraphics[width=70mm]{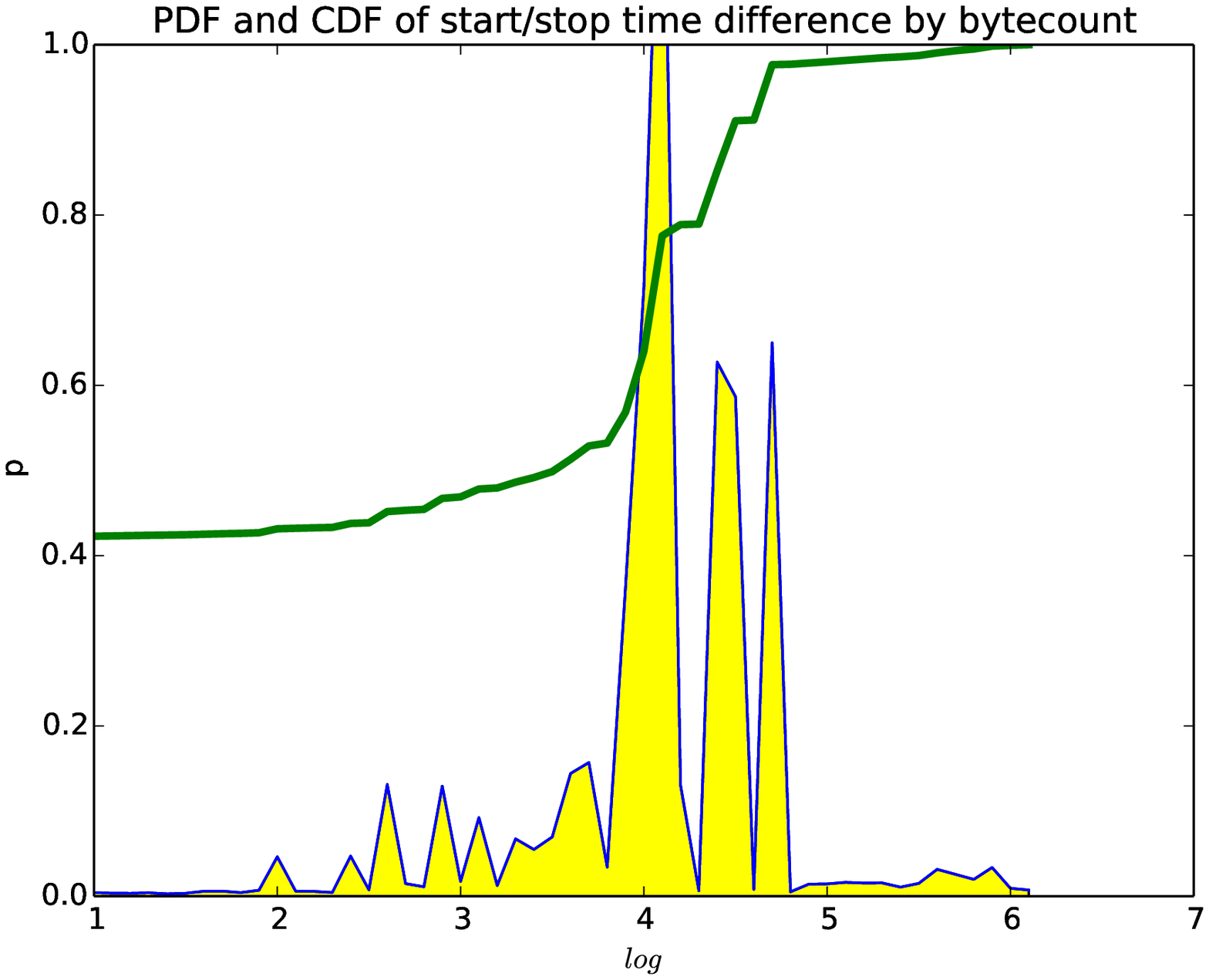}
}
\hspace{0mm}
\subfloat[case 3]{
  \includegraphics[width=70mm]{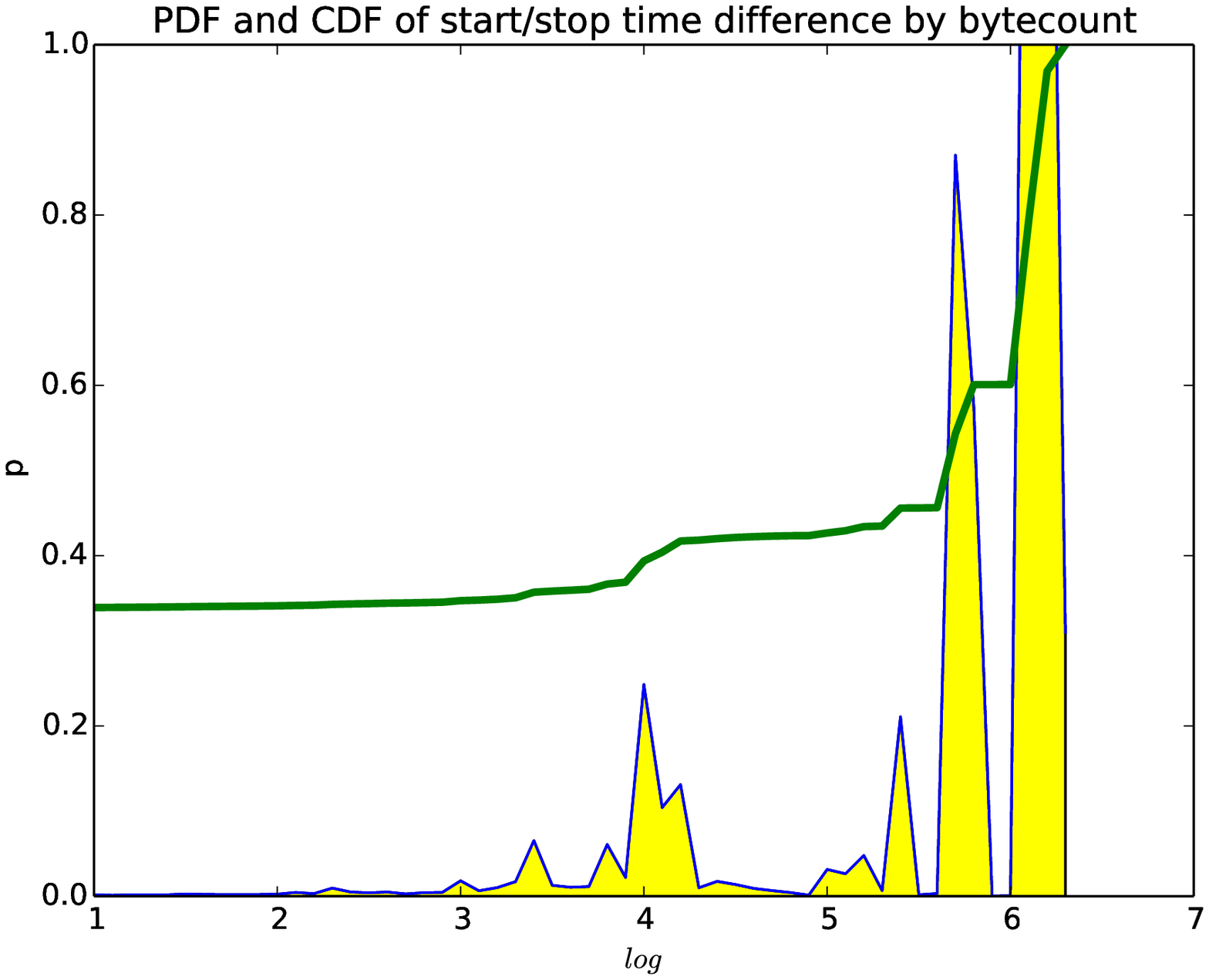}
}
\subfloat[case 4]{
  \includegraphics[width=70mm]{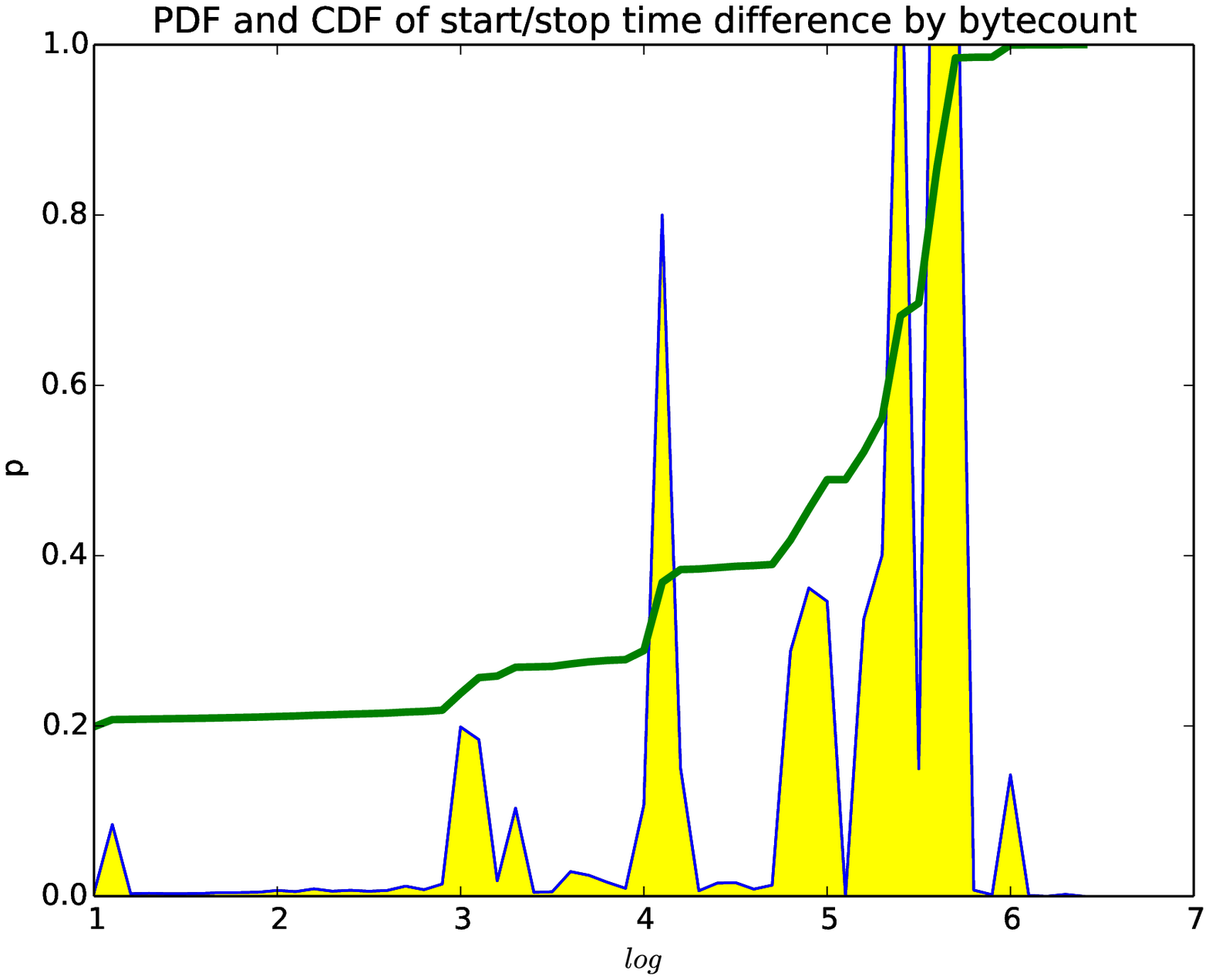}
}
\caption{First-last time probability density by volume}
\label{fig:FirstLastBySize}
\end{figure}

In ~\ref{fig:FirstLastBySize} on page ~\pageref{fig:FirstLastBySize} we see the pattern we saw earlier in figure ~\ref{fig:InterJobBySize}, but much more pronounced. In most of the four investigations, a majority of the data will take multiple hours to get from the first module to touch/produce it, to the last module to process its data. This is again a clear sign that we can expect a large number of disk cache misses during the lifetime of a piece of larger data within the OCFA system.

%% file: ocfa-step4-results.tex
\pagebreak
\subsection{Inflow and outflow for our fictitious cache}
\begin{figure}
\centering
\subfloat[case 1]{
  \includegraphics[width=70mm]{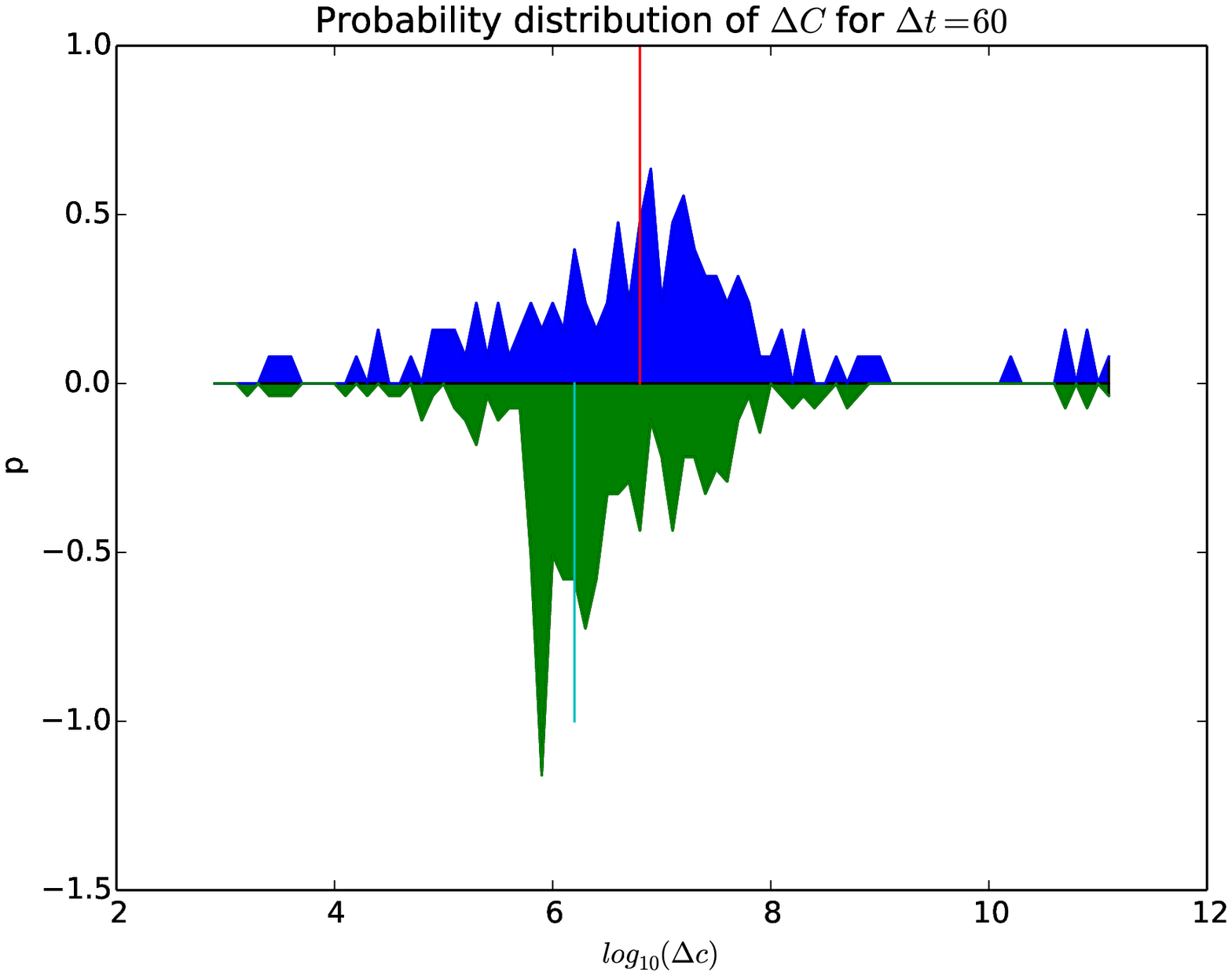}
}
\subfloat[case 2]{
  \includegraphics[width=70mm]{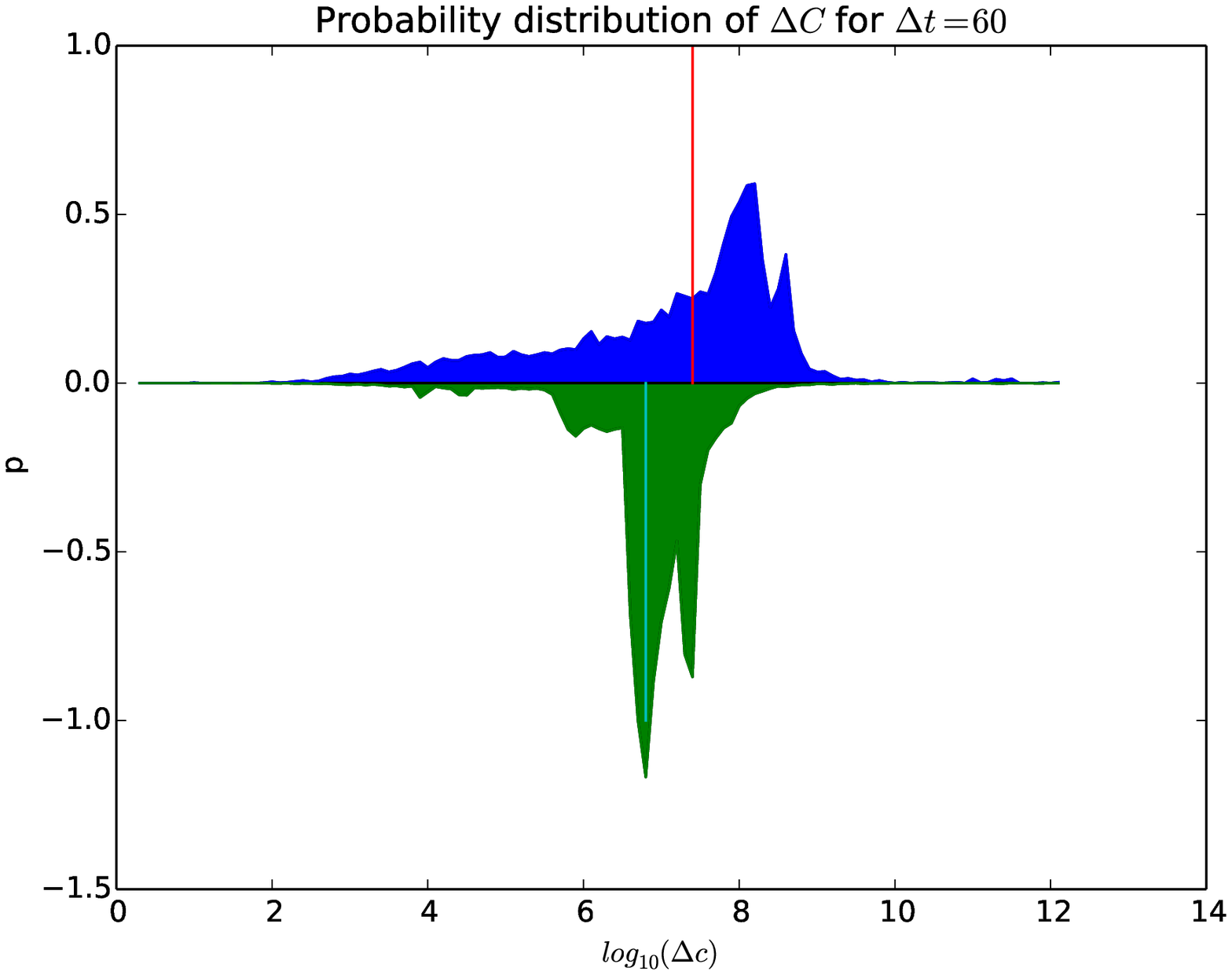}
}
\hspace{0mm}
\subfloat[case 3]{
  \includegraphics[width=70mm]{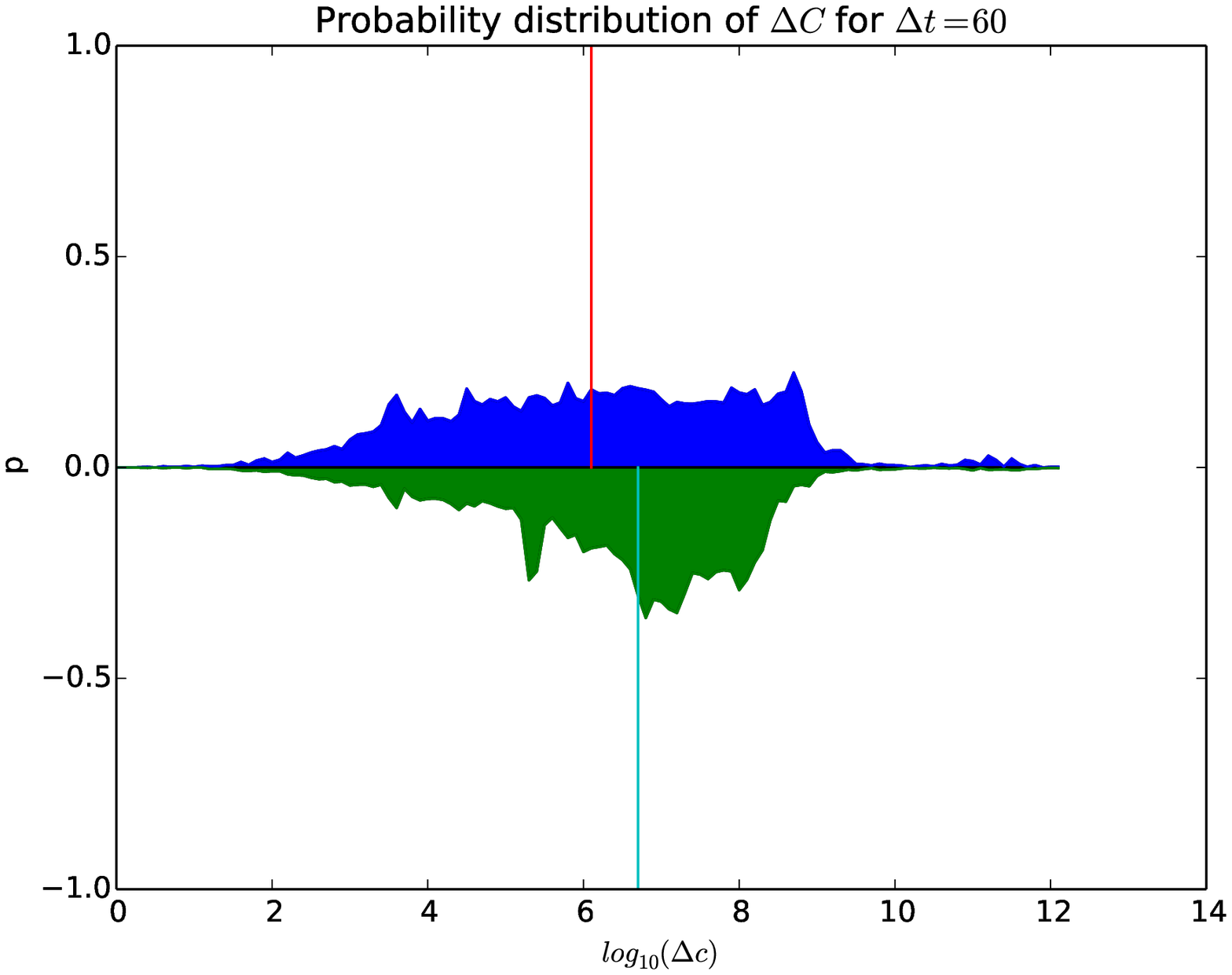}
}
\subfloat[case 4]{
  \includegraphics[width=70mm]{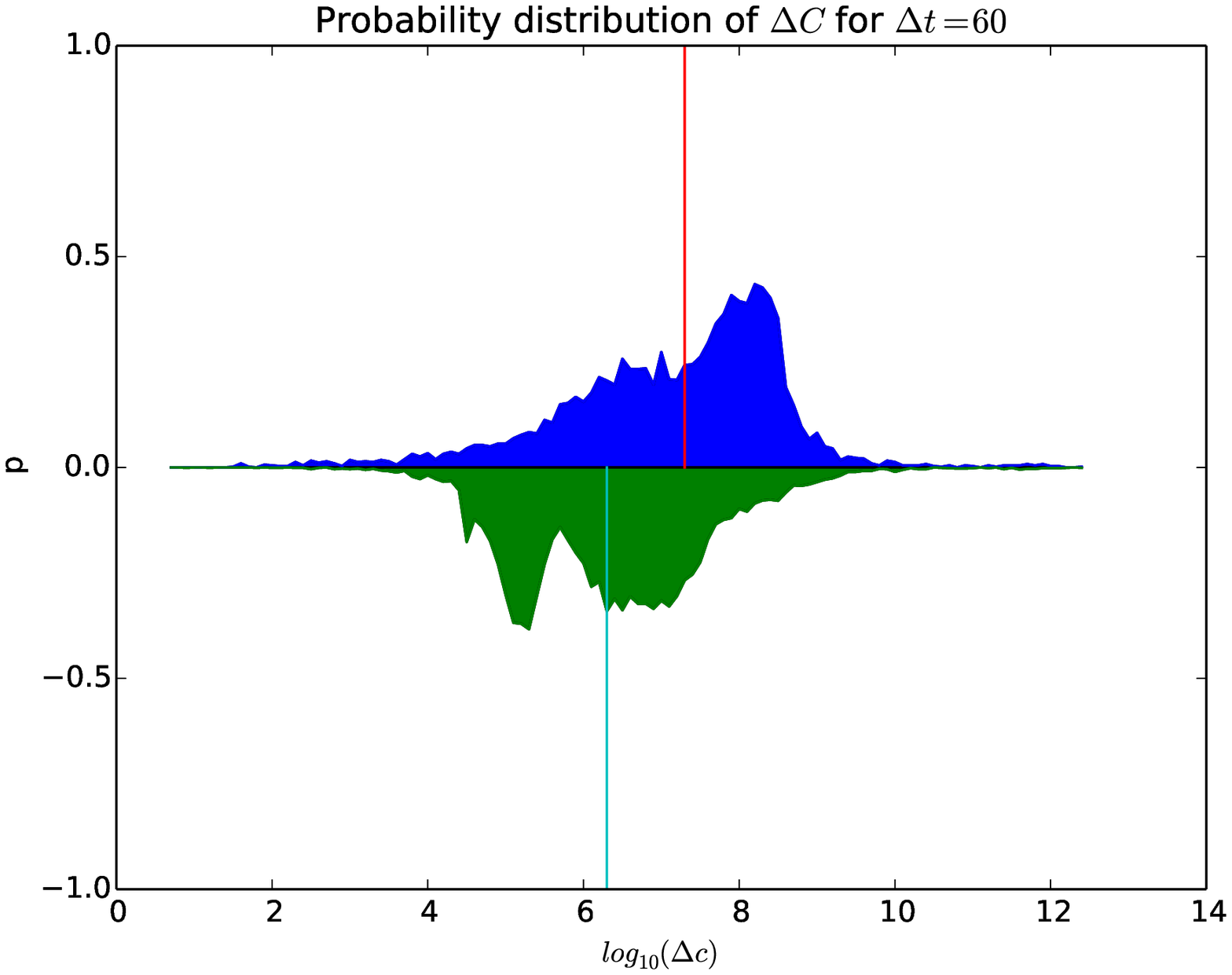}
}
\caption{In/out flow probability density}
\label{fig:FlowInOut1}
\end{figure}
The pictures in ~\ref{fig:FlowInOut1} on page ~\pageref{fig:FlowInOut1} finally show the problem with the disk cache misses. Note that for visual reasons the density function for outflow is plotted as negative density on the positive part of the inflow axes rather than as positive density negative inflow. This to more clearly show the contrast between inflow and outflow density. The diagrams show both the probability density of growth of the amount of \emph{active} data in the system, and on the negative axis the probability density of shrinkage in one minute intervals. If we remember that this density function is plotted on a logarithmic X axis, we can identify that the growth density on the bottom end of the graphs is higher than the shrinkage density on the upper end of the graph for most of the four investigations. This means that while there may be many minutes where there is significant shrinkage, over time the amount of \emph{active} data will grow as new data continues being submitted to the system. These results show clearly that for the OCFA system, for most of the investigations, there is an imbalance between the data input speed and the overall data processing speed. With knowledge about the dominant module in specific investigation, as discussed in the next subsection, we can further conclude that the significant production of new evidence data events by modules other than the Java based kickstart, significantly increases the visual noticeable imbalance in these graphs.  

%% file: ocfa-step5-results.tex
\subsection{Inter-module flows}
\begin{figure}
  \centering
  \includegraphics[width=130mm]{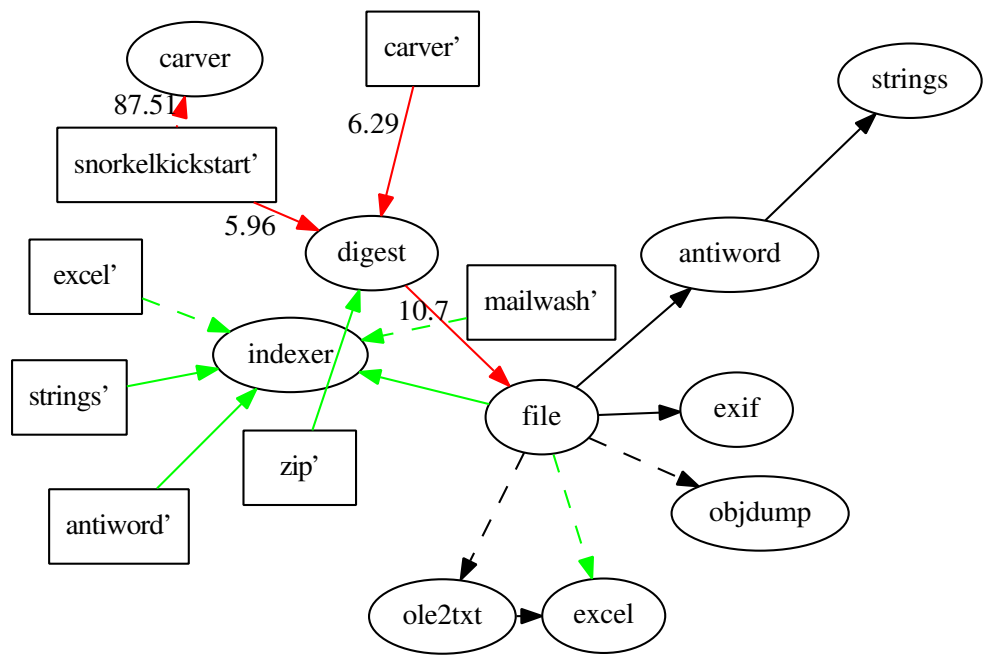}
  \caption{Case 1 inter-module flows}
  \label{fig:Case1Modules}
\end{figure}

\begin{figure}
  \centering
  \includegraphics[width=130mm]{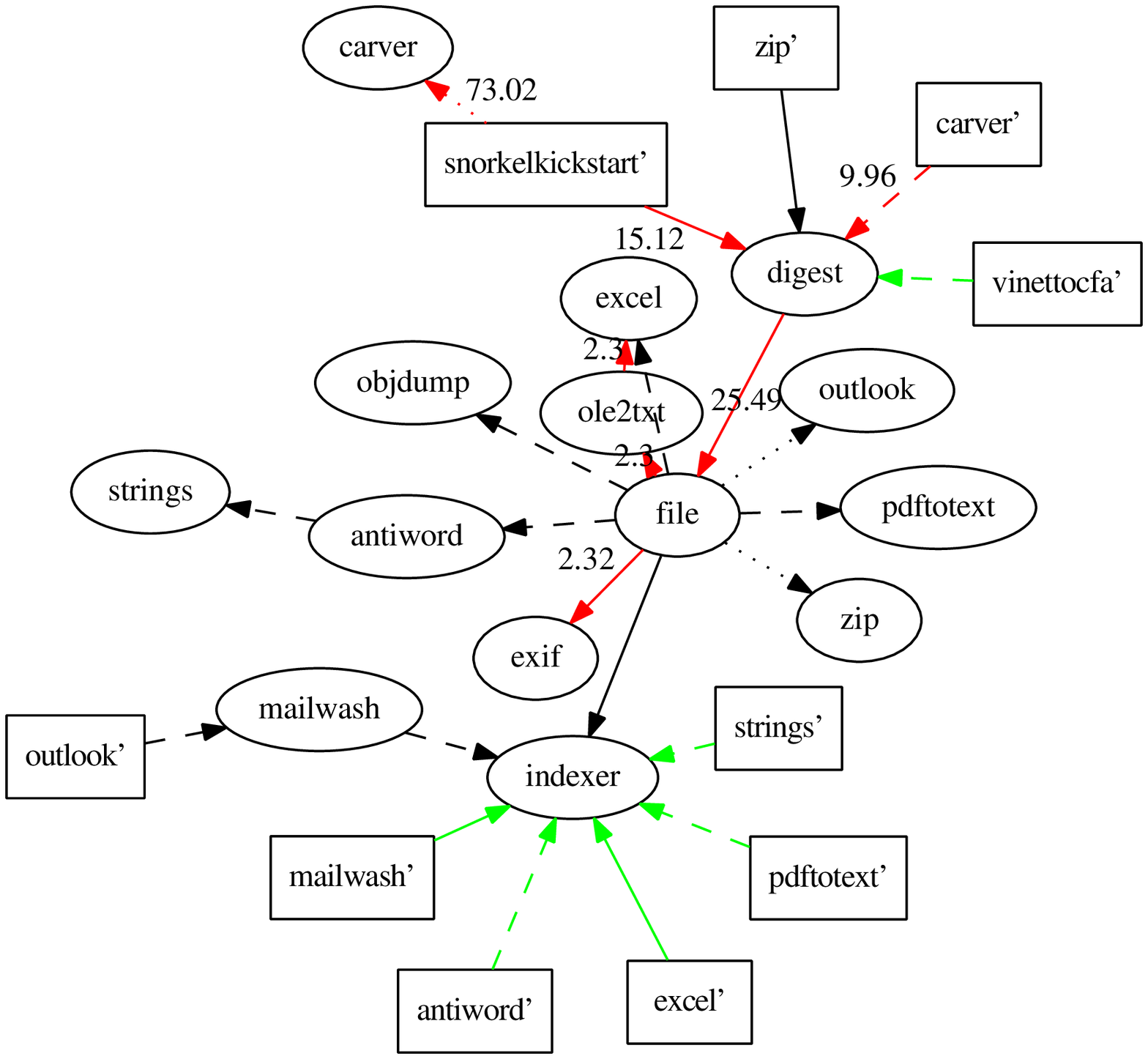}
  \caption{Case 2 inter-module flows}
  \label{fig:Case2Modules}
\end{figure}

\begin{figure}
  \centering
  \includegraphics[width=130mm]{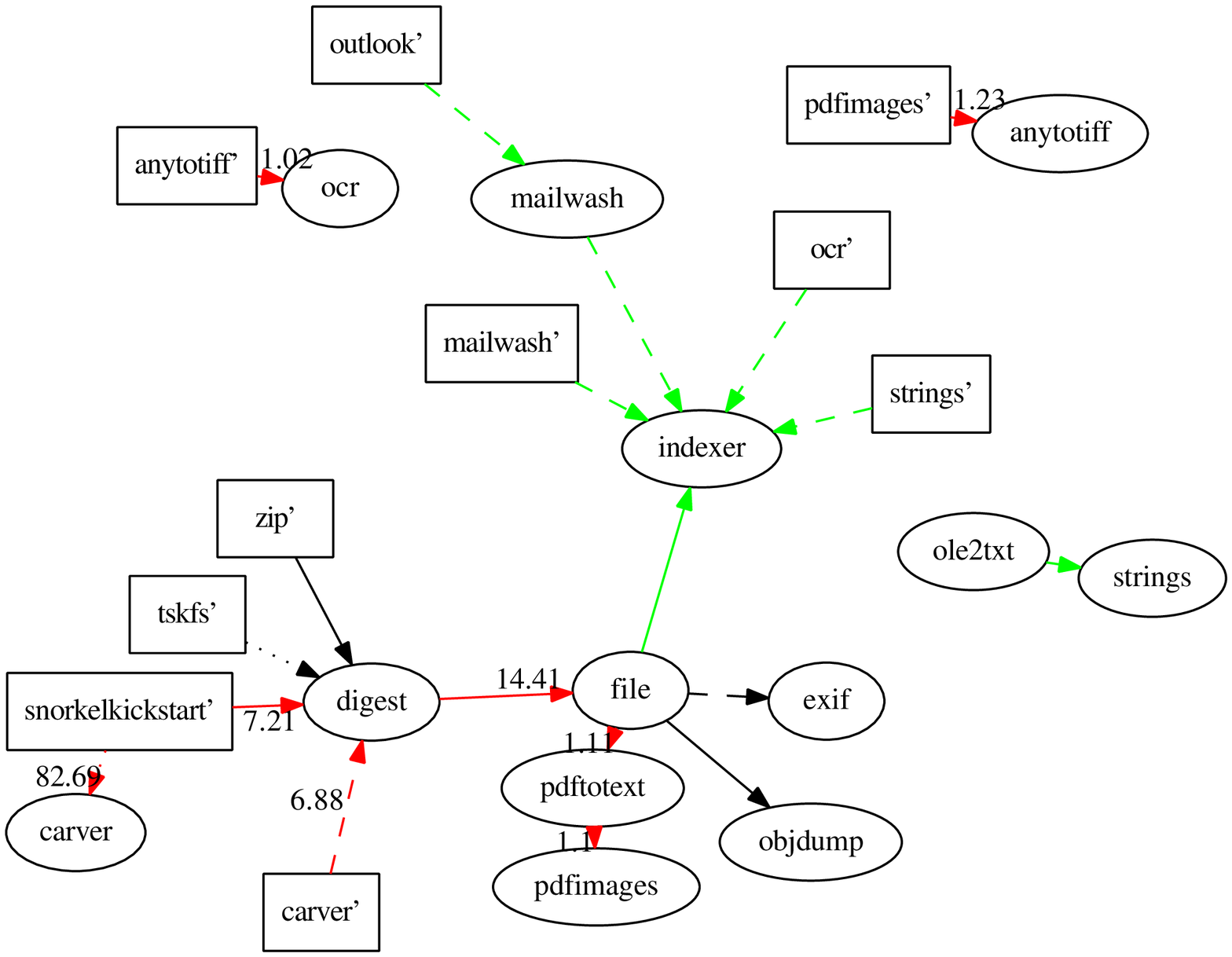}
  \caption{Case 3 inter-module flows}
  \label{fig:Case3Modules}
\end{figure}

\begin{figure}
  \centering
  \includegraphics[width=130mm]{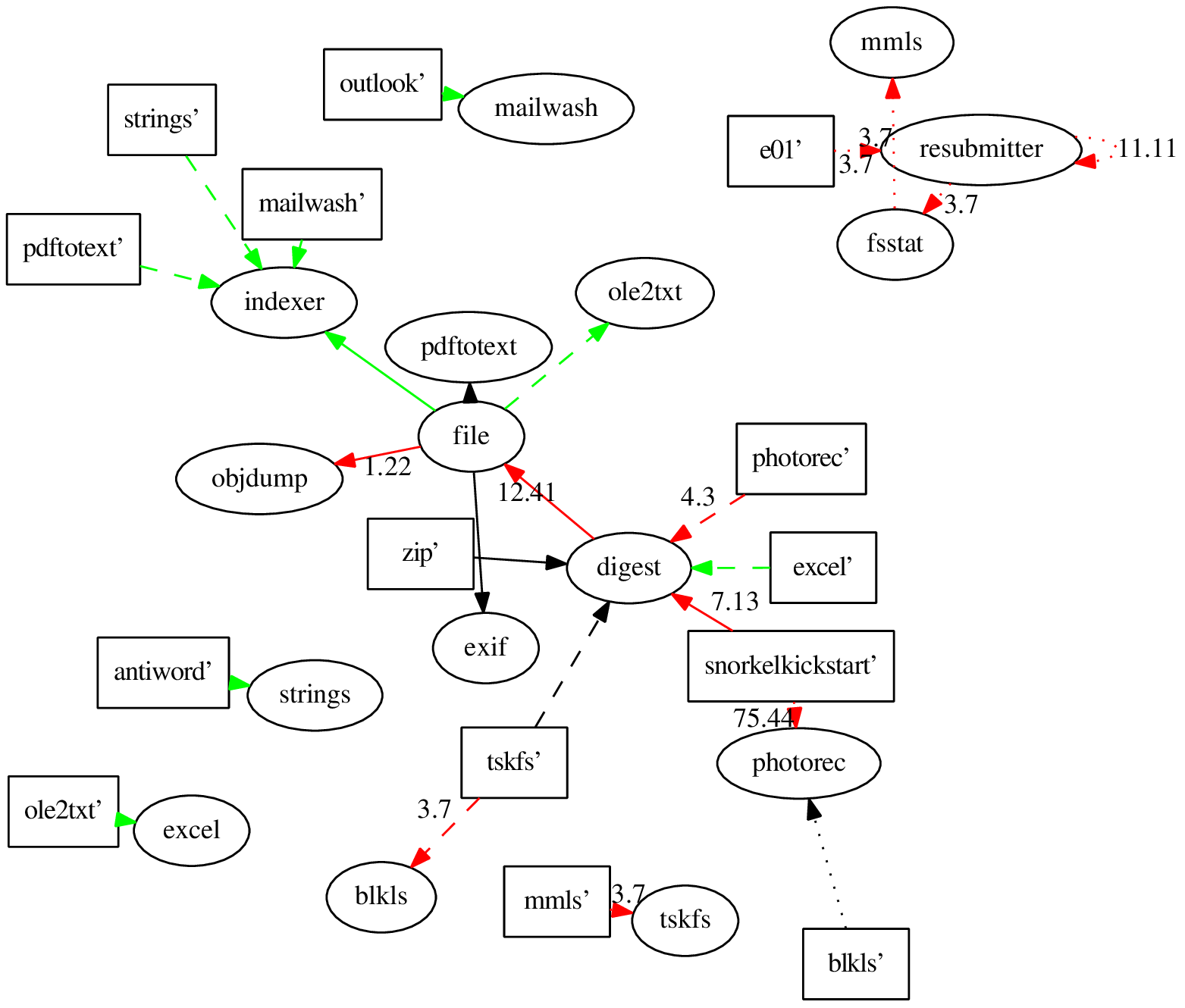}
  \caption{Case 4 inter-module flows}
  \label{fig:Case4Modules}
\end{figure}

\begin{figure}
\centering
\subfloat[case 1]{
  \includegraphics[width=50mm]{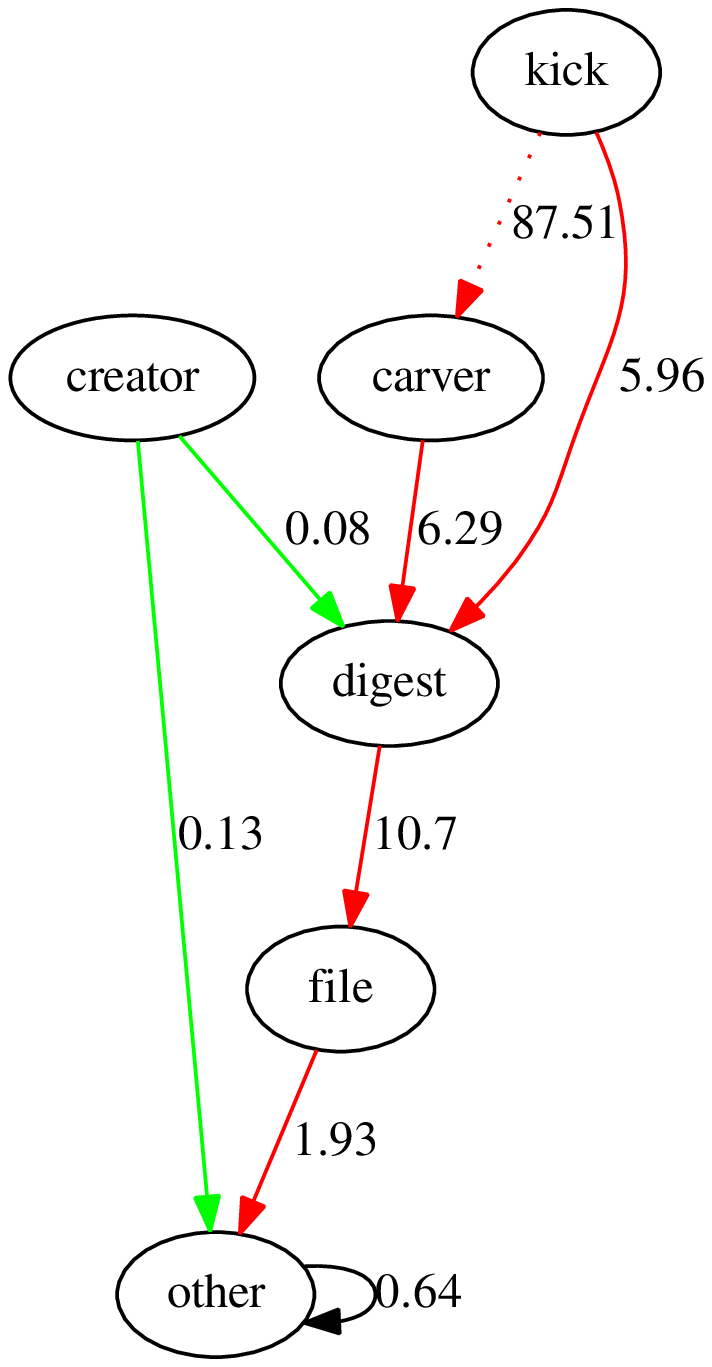}
}
\subfloat[case 2]{
  \includegraphics[width=50mm]{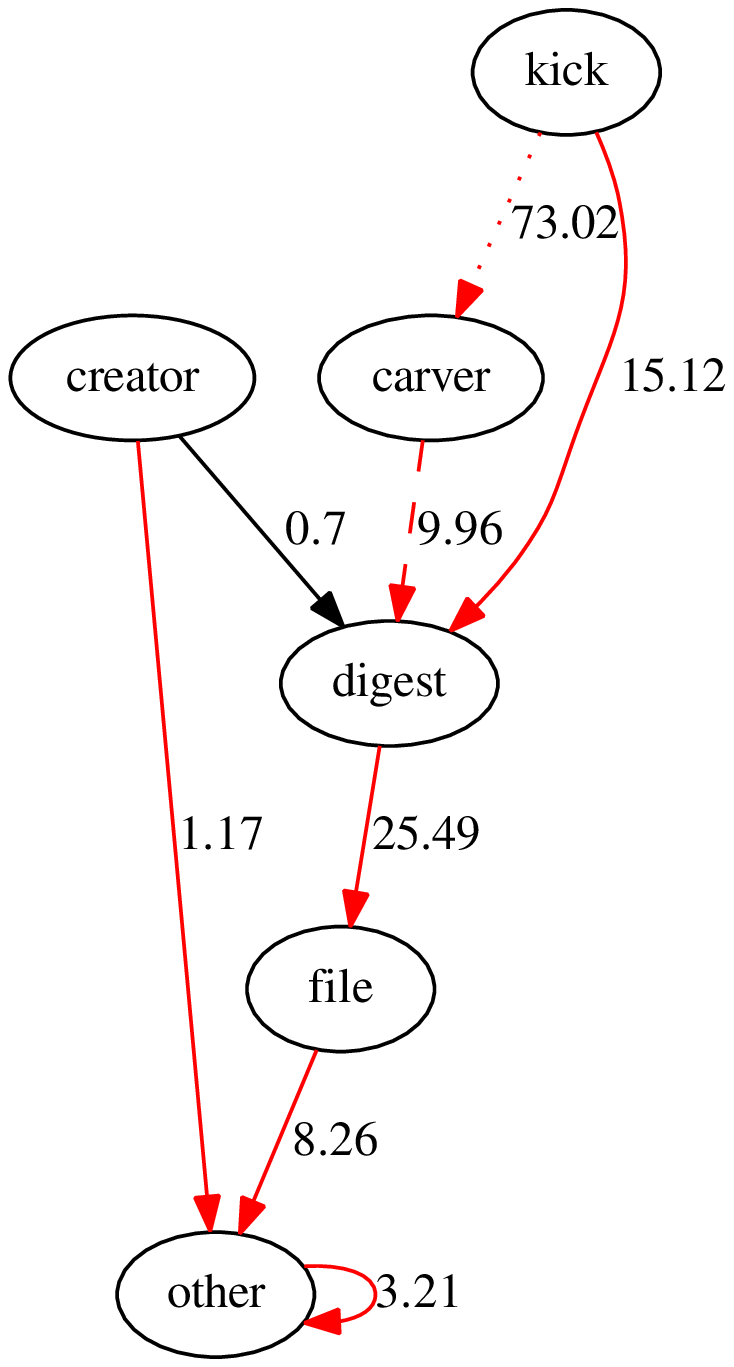}
}
\hspace{0mm}
\subfloat[case 3]{
  \includegraphics[width=50mm]{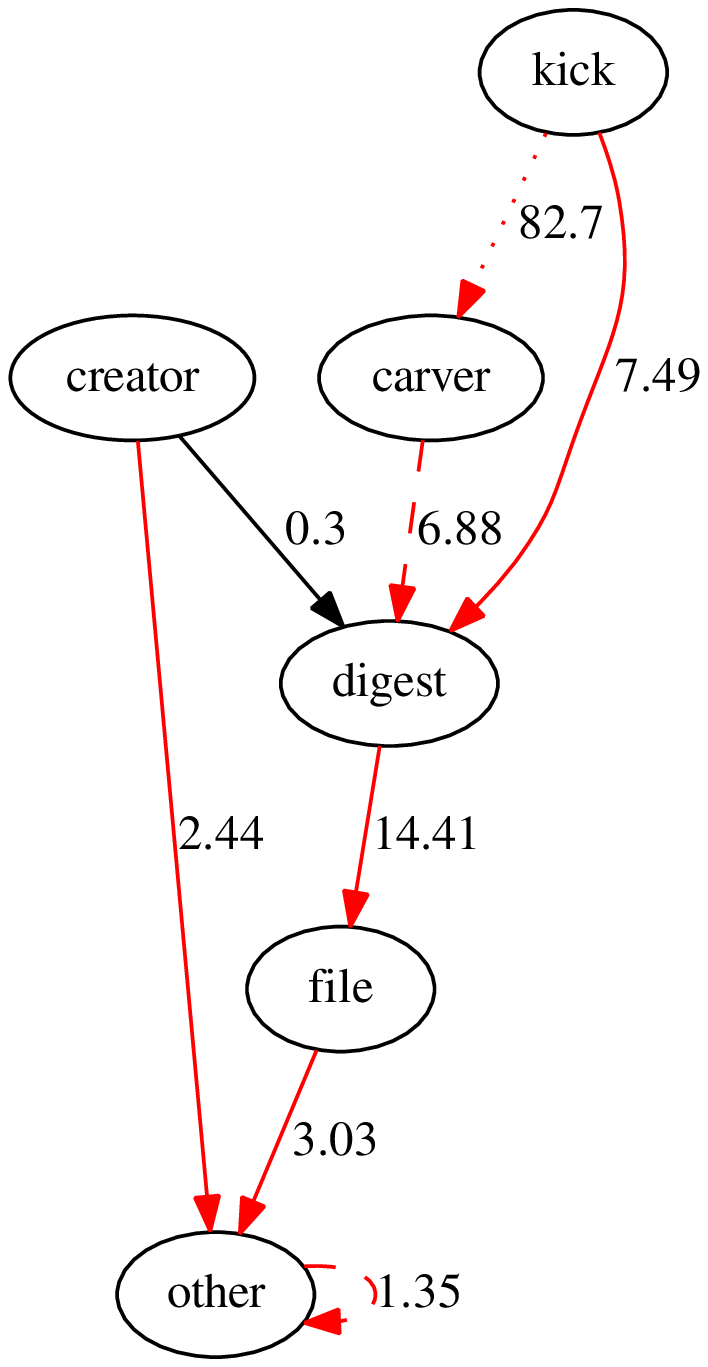}
}
\subfloat[case 4]{
  \includegraphics[width=50mm]{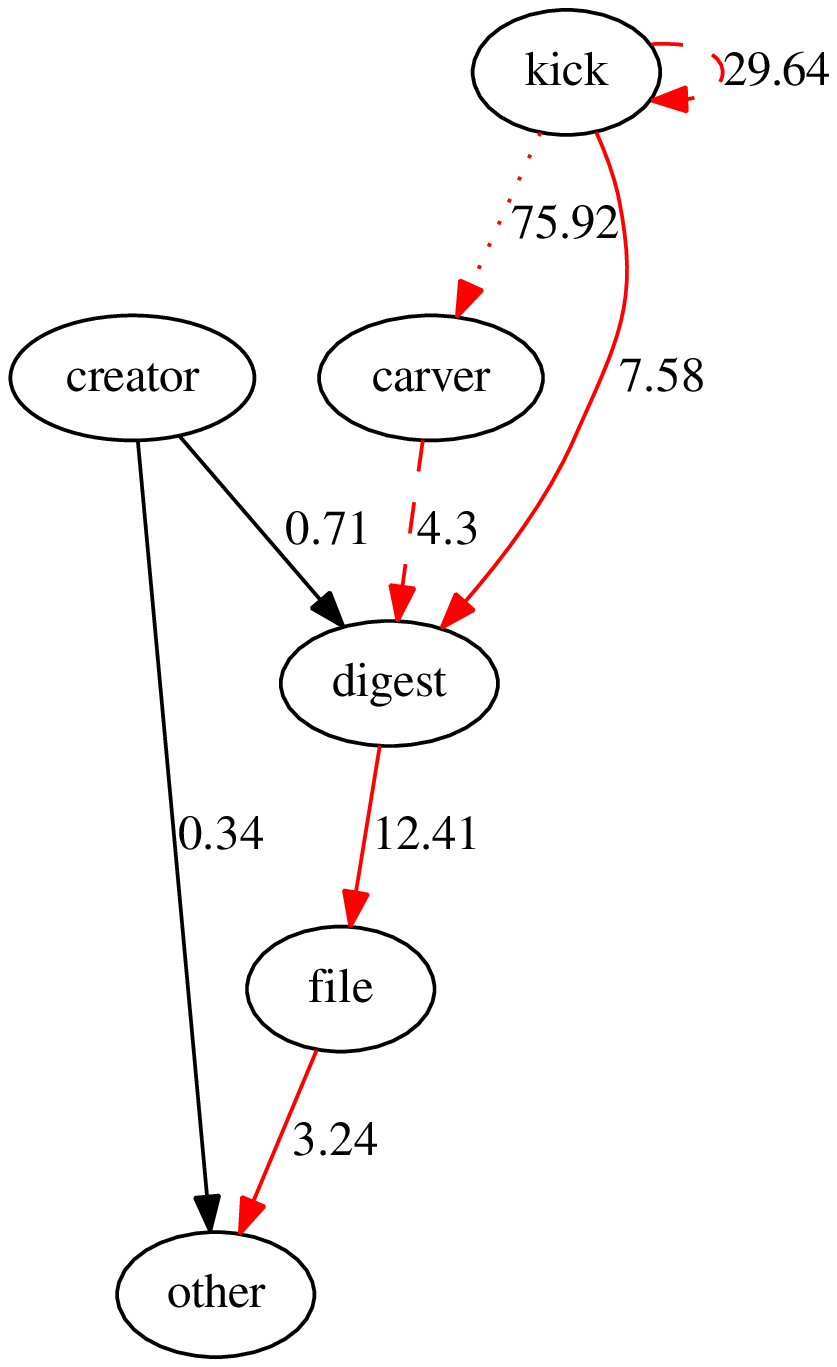}
}
\caption{Rough data flow}
\label{fig:RoughModules}
\end{figure}

The figures ~\ref{fig:Case1Modules}, ~\ref{fig:Case2Modules},~\ref{fig:Case3Modules} and ~\ref{fig:Case4Modules} show an overview of the data and event flow between modules in the four cases we have been examining. The most prominent flows data-wise are indicated in \emph{\color{red}red} arrows, followed by \emph{black} arrows for less prominent data flows and \emph{\color{green}green} for flows that are almost negligible in data flow but not in terms of event count. \emph{Solid} arrows indicate a prominence in the number of data events, followed by \emph{dashed} arrows for less prominent event flows and \emph{dotted} arrows for flows that are
negligible in event flow but not in data flow. Flows that are both almost negligible in data-flow and in event flow are not displayed to keep the figures from becoming too filled with irrelevant information. Note that modules in a role as as data producers are denoted using a \emph{square box} while (possibly the same) modules in their role as data consumer are denoted with an \emph{oval} shape. Even though there are important differences due to the differences in types of and size of investigations, we also see a few clear similarities in these flows. To emphasize these similarities, figure ~\ref{fig:RoughModules} on page ~\pageref{fig:RoughModules} shows these same information using a grouping for modules that play a less prominent role. Below we summarize the core information that we can conclude from this figure.
\begin{enumerate}
\item The majority of all data (roughly between 80\% and 95\%) is forwarded by the kickstart functionality to the carver functionality and never passes the digest checking module.
\item The majority of evidence data checked by the digest check module originates from either the carver or the kickstarting functionality.
\item About one third to half of the data sent to the digest check module originates from the carver, about 50\% to 60\% from the kickstart functionality.
\item Only roughly 1\% to 15\% of evidence data sent to the digest check module ends up matching. The remaining data is forwarded to the file-type check module.
\item Only roughly 15\% to 35\% of data forwarded to the file-type check module ends up being forwarded for processing any further.
\end{enumerate} 
This information, combined with what we have seen in our earlier analyses allows us to draw some important conclusions that we shall discuss in the following sections of this appendix.

%% file: ocfa-conclussions.tex
\section{Identifying the main bottlenecks within the OCFA architecture}
In this section we look at the information gathered from analyzing the OCFA timing info from four investigations and try to draw well founded conclusions from this data by combining it with knowledge about the OCFA architecture.
\subsection{Effectiveness of OCFA throughput measures}
As we have seen in the event based graphs looking at the cumulative probability density of per data entity timing, the priority queuing would seem to be effective, and would be quite effective if all data entities were similarly sized. Looking though at the data-volume-adjusted probability density graphs, we see that the effectiveness of priority queuing becomes questionable at best. We don't have sufficient proof to state that the measure may be anti productive from a cache hit rate point of view, but it's something that at this point in time seems like a viable hypothesis. A hypothesis though that we must leave untested as there is no comparative material available to test it against. It is safe to say though that from a cache-hit efficiency point of view, the priority queuing mechanism proves to be at best insufficient.
\subsection{The effects of meta-data messaging inefficiency and centralized router}
One of the known performance bottlenecks of OCFA is the implementation of meta-data access. Conceptually an evidence entity is forwarded between modules through the messaging system. Due to monitoring concerns though, an approach was taken where short messages were used instead with an internal reference to a BLOB inside of the Postgres database. The process for processing and forwarding an evidence entity to a module or router is the following:
\begin{itemize}
\item Module receives an incoming message from the Anycast relay and extracts the meta-data ID from the message
\item Module retrieves the XML blob containing the evidence trace from Postgres, validates and parses the XML into a DOM tree and then extracts the data ID from the XML
\item Module retrieves the evidence data filesystem path from Postgres
\item Module \emph{Processes the data}
\item Module adds additional meta-data to the DOM tree, serializes the DOM tree to a new version of the XML trace and updates the XML BLOB in the Postgres database.
\item Module sends a return message to the Anycast asking to forward it to the router.
\item Anycast marks the old message as processed and places the new message in the router queue. When router queue reaches message, Anycast sends message to the router
\item Router receives an incoming message from the Anycast relay, and extracts the meta-data ID from the message
\item Router retrieves the XML blob containing the evidence trace from Postgres, validates and parses the XML into a DOM tree, traverses the meta-data and determines the next module to process the evidence entity
\item Router adds additional meta-data to the DOM tree, serializes the DOM tree to a new version of the XML trace and updates the XML BLOB in the Postgres database
\item Router sends a return message to the Anycast asking to forward it to the specific next module.
\item Anycast marks old message as processed and places the new message in the router queue.
\item When module queue reaches message, Anycast sends message to the next module
\end{itemize}
It's easy to see that there is quite some overhead in messaging and routing.
There are a number of known bottlenecks in this process that would be suitable for addressing the speed of the OCFA framework. 

\begin{itemize}
\item The use of XML, XML-schema and related technology was very much current when OCFA was designed. In retrospect though, XML is a highly expensive serialization technique and other serialization possibilities like JSON or FlatBuffers
would likely have given significantly less overhead.
\item The use of a separate router process means that much of the overhead is \emph{repeated twice}. Moving the router functionality into the messaging library and thus into the module processes could half this overhead and make for a major overhead reduction. 
\end{itemize}
So to summarize, there is quite some room for improvement in the efficiency of the cross module timing. We can assume that any future OCFA successor that aims to use a similar message based concurrency model will implement such improvements. 
\pagebreak
\subsection{Bottlenecks and possible solutions}
From the analysis of the timing information from four real life OCFA runs, and our knowledge of the OCFA system we can draw some conclusions regarding the inability of the OCFA design to effectively minimize the amount of disk-cache misses. We can also draw conclusions with respect to what may be improved in the design of a new OCFA-like system that could improve both the disk-cache efficiency and the overall throughput of the system.
\begin{itemize}
\item The un-throttled input of new data into the system creates an imbalance between inflow and outflow that leads to far higher amounts of \emph{active} data than can be accounted for by disk cache, inevitably leading to massive amounts of disk cache misses. A form of \emph{throttling for new data submission} is a necessity for effectively addressing disk cache efficiency.
\item The percentage of data that is discarded after a file-type check is more significant when compared to the percentage of data that is discarded after a hash value check. It would seem logical to design a new system in such a way that \emph{file-type checking happens before} the file is actually fully read by the framework, as to remove unneeded overhead from reading the file data, either to copy it out or to determine the file hash. This seems one argument in favor of \emph{opportunistic hashing}. 
\item When looking at the chain of modules that processes particular data, it is not uncommon for the first full read to be the result of the on-creation hashing, while the second and often last full read happens in the last module before the (in OCFA) meta-data-only Data Store Module. This means that if hashing could be delayed to a later module, that while the entity processing time may stay approximately the same, the disk-cache hit could improve resulting from the reduced first full-read last full-read timing. 
\item The percentage of data that constitutes large entities without in anyway meaningful hash value (such as whole partitions or unallocated space cluster collections) makes up a majority of all data processed. Traversing such large data chunks for the sole purpose of calculating a set of hashes, as is done in the OCFA architecture is wasteful. This is a second argument in favor of opportunistic hashing.
\item The percentage of larger data that is processed that is a chunk of other data being processed is significant. This implies that the page cache size requirements could potentially be reduced by making extensive use of annotation based addressing such as used in CarvFS.
\end{itemize}
From a disk-cache efficiency viewpoint, we are proposing three distinct yet interdependent measures:
\begin{itemize}
\item \emph{Opportunistic hashing}: Calculate hashes opportunistically when an entity as a whole is either being written or read.
\item \emph{Annotation based data access}: Use of a CarvFS alike system for accessing data as chunks within a bigger whole.
\item \emph{New-data input throttling}: Keep track of \emph{active} data in the system and throttle input accordingly.
\end{itemize}
While not the core subject of this dissertation, the following improvements to an OCFA like  forensic framework design should be beneficial given that the above disk-cache related issues are solved first:
\begin{itemize}
\item Integration of router functionality into the module framework.
\item Integration of file-type module (libmagic) functionality into the module framework.
\item Meta-data separate from central database storage. No intermediate mutable result storage in database.
\item Use of more efficient serialization technology instead of XML.
\end{itemize}

%% file: mainpaper-fuse.tex
\chapter{CarvPath as core paradigm}
As discussed in the literature survey, CarvFS and ScalpelFS introduced the concept of the use of a flat secondary-storage model. In CarvFS this model was implemented through the use of so called CarvPath annotations. While ScalpelFS used the flat secondary-storage model purely for carving in the narrow sense, the use of CarvFS in the Sleuth-kit based modules of OCFA showed that a flat secondary-storage model can be applicable to a much wider range of embedded sub-entities within the computer-forensic process. For clarity, some examples of flat secondary-storage model annotation possibilities:
\begin{itemize}
\item A partition within a disk image.
\item A file within a file-system as identified by file-system analysis forensic tools.
\item An area of unallocated space within a file-system.
\item A carved file within an area of unallocated file-system space.
\item A file within an ISO file.
\item A file within a TAR archive.
\item An email within a mailbox.
\end{itemize}
In fact, it is conceivable that we could find an email within a mailbox within a TAR file within an ISO file carved within unallocated space within a file-system within a partition of a disk image. CarvPath annotations as introduced by CarvFS allow us to designate any such (possibly nested) piece of data within a flat secondary address space. CarvPath annotations build upon the concept of possibly nested structures like the one we just described. Each level of nesting is represented with a slash character in a way similar to the use of this character as separator in a file-system path. In fact, CarvFS uses this concept in order to map CarvPath annotations to pseudo files and pseudo directories within a user space file-system. Within a single nesting level, CarvPath defines two main types of tokens. \emph{Regular} CarvPath tokens and so called \emph{long-path} tokens. A regular CarvPath token consists of one or more fragment tokens, separated by an underscore character. As it is possible for a file to be highly fragmented, the length of a regular CarvPath token may end up exceeding the maximum size of a directory or file name, and to address this, CarvPath has the concept of the long-path token. Such a token consists of a secure hash of the regular CarvPath token that is then stored in some \emph{global lookup table} so that the original regular token can always be restored from the long-path token. When we look at the fragment tokens, we see two types of fragments. Regular fragments and sparse fragments. A regular fragment is annotated using an offset and a size that designate the location of the fragment within the parent-entity or nesting level. The offset and size are ASCII encoded decimal numbers separated by the plus character. Sometimes a file can be partially sparse. To accommodate that reality, CarvPath also allows for sparse fragments to be designated. A sparse fragment starts with the capital S character followed by the size of the sparse fragment. 
The following railroad diagram defines the CarvPath format:
\begin{rail}
((('D' digest) | ((('S'|(offset '+')) size) + '-') ) + '/') ".crv"
\end{rail}
An example of a CarvPath could be:

\emph{0+1056964608/32256+4096\_S24576\_36352+4096.crv}

If we take the use of a CarvFS alike system within a computer forensic framework as a given way to designate and use flat secondary storage address space, then the user-space file-system can become a cohesion providing central component for the framework that elevates flat secondary storage address space to a central paradigm that could help optimize for different page-cache related bottlenecks. With the proper design, the user-space file-system can keep track of the active data within the computer-forensic process, and can play a pivotal role in many page-cache efficiency scenario' s that we look at in the next few sections.

%% file: mainpaper-results2.tex
\chapter{MattockFS design considerations}
\noindent 
In this chapter, the results of the remaining steps of the adopted approach are discussed in detail. This is done by describing the outcome of the main phases of the research related to the design considerations for the MattockFS computer forensics file system as they relate to the solving of bottlenecks we identified from the OCFA provenance logs and bottlenecks as well as anti-forensic vulnerabilities identified in the analysis of the OCFA design. We identify the following main phases from our adopted approach:
\begin{enumerate}
\item Investigating page-cache friendly archive interaction.
\item Investigating page-cache friendly message-bus interaction.
\item Investigating the potential and implementation possibilities for opportunistic hashing.
\item Investigating the requirements and possibilities for satisfying forensic system-integrity needs.
\item Integration into a fully functional working proof of concept sub-system.
\end{enumerate}

%% file: mainpaper-archive.tex
\section{Page-cache friendly archive interaction}
Where CarvFS was initially designed to provide read-only access to individual image files using a forensic-disk-image-format loadable-module in order to allow access to EWF disk images, this concept ended up not scaling that well to larger investigations spanning a large multitude of disk images and demanding all these images to be simultaneously mounted. In later versions of OCFA and CarvFS, a hack was used to allow images to be appended to an already mounted \emph{archive}. This archive was a collection of huge sparse files, typically residing on SNFS volumes. The use of a single mount-point for an archive allowed better RAM resource usage and better IO performance than the use of a multitude of simultaneously mounted and on-the-fly decoded EWF images. We choose to build further on these insights and look at the possibilities for using such an archive in a page-cache friendly way.

In order to address page-cache friendliness, we should first look a bit at the Linux kernel and the interaction between open files and the page cache. Given that our archive shall consist of one or more huge sparse files that shall be kept open by our user-space file-system, and given that multiple processes shall be having pseudo-files opened within this user-space file-system, it is important to know how open files and page-cache usage are related. The Linux kernel will use part of its RAM as page-cache, where it will keep part of the data from files that are in use and that were recently interacted with  available for consecutive reads. Such consecutive reads shall not require additional disk IO when the data is still in the page-cache. Given that the available RAM will be much smaller than the amount of file-data, the kernel will need to pick what data to keep and what data to overwrite when other file-data is accessed. The kernel will try to be as smart about these strategies as is possible with the information it has available to it. With the \emph{fadvice} system call, a user space program such as our user-space file-system can communicate information about the file it is accessing in order to help the kernel in making wiser decisions with respect to the page-cache. As such, the file-system can mark pieces of the archive as \emph{expected to be read again} soon, or as \emph{no longer needed}. 

A core functionality thus needed to improve page-cache performance is keeping track of what parts of the archive are \emph{hot} and what parts of the archive are not. To determine this, we need to introduce a new concept, the \emph{tool-chain}. A piece of evidence data within the archive will typically pass through a collection of different tools or modules before processing is done. We say that \emph{hot} archive data is data that is part of an \emph{active} tool-chain. Combining this with our page-cache related kernel interaction, it suffices to say that any data that still is linked to an active tool-chain should be marked as \emph{hot} regardless of any tool currently having a pseudo-file open for that data. Data that does have pseudo-files open is \emph{hot} as well. It may not currently be getting processed, some process is accessing it still. This could be a User Interface process for example. Thus as long as either an active tool-chain exists for a CarvPath designated piece of archive, or if a pseudo file is currently opened for that piece of archive, than that data is \emph{hot} and should be marked as such with the kernel. If neither of these conditions is true, than the data is \emph{cold} and this should be marked as such as to allow the kernel to free up any page-cache. While the file-system should be perfectly capable of keeping track of open pseudo-files, it is essential that we find a way to also allow the file-system to know about tool-chains. With that information the kernel can than be informed in such a way as to make best use of the available page-cache assigned RAM.

If we know the amount of RAM available to the page-cache, and we know the amount of \emph{hot} data currently in the system, at some point we may reach the situation where adding more \emph{hot} data to the archive could be a bad idea. There is no need to limit the amount of \emph{hot} data to the available page-cache RAM, but beyond a certain level of over-commitment, adding more hot data to the archive becomes unwise and throttling of new archive data becomes the better option. Implementing throttling within the archive file-system itself would be a possible option, but putting too much functionality in the file-system may not be wise. We thus choose to let the file-system provide information to the framework in such a way that the framework becomes capable of implementing throttling based on the page-cache load numbers obtained from the archive file-system. 

\subsection{The CarvPath reference counting stack}
Both active tool-chains and open pseudo-files result in all the data designated by a given CarvPath to be considered \emph{hot}. In order to keep track of \emph{hot} CarvPaths, maintain indirect information about page-cache pressure, and map the high-level CarvPath entities to low level kernel interaction operations, a way must be devised to keep track of the status of the low level data chunks. To allow for this to be implemented a set-theory based combination of operations on two administrative data structures was devised. 
\begin{itemize}
\item For high level entity CarvPaths a reference counting map is used. Operations that increment the reference count for a given CarvPath from zero to one, and operations that decrement the reference count from one to zero result in an operation on the lower level structure.
\pagebreak
\item For lower level fragments, a stack of non-sparse CarvPath entities is used as a way to reference-count lower level fragments. An operation that increments the reference count for a given fragment from zero to one, and operations that decrement the reference count from one to zero result in a kernel interaction marking the fragment as active or inactive.
\end{itemize}
The reference counting stack works by combining CarvPath operations with set-theory operations. Each level in the stack contains a non-sparse CarvPath entity representing a number of fragment with a reference count of at least i+1, where i is the level on the stack. $S_{0}$ represents all fragments with a low-level reference count of at least one, $S_{1}$ represents all fragments with a low-level reference count of at least two and so on. We define two operations on the stack involving high level CarvPath entities:
\begin{itemize}
\item Merge up: When a new CarvPath entity is added to the stack, its fragments are \emph{merged up} into the stack.
\item Unmerge down: When a CarvPath entity is fully removed from the stack, its fragments are \emph{unmerged down} from the stack.
\end{itemize}
We look at both operations. Merge up starts with the full sparse-free version of our new CarvPath entity and with the stack in its pre-merge state. We shall describe our pre-merge stack in terms of $S_{0} .. S_{n-1}$ where $n$ is the pre-merge number of levels in the stack. Our post-merge stack shall be described in terms of $S'_{0} .. S'_{m-1}$ where $m$ is the post-merge number of levels in the stack. The initial sparse-free fragments in the CarvPath to be merged is designated by $C_{0}$. Consider $F(X)$ to be the set of all byte locations designated by CarvPath $X$, if we look at the actual \emph{merge up} operation, for each $i$, starting at zero, where : $F(C_{i}) \ne \emptyset$, we perform CarvPath operation so that:
\begin{itemize}
\item $F(S'_{i}) = F(C_{i}) \cup F(S_{i})$
\item $F(C_{i+1}) = F(C_{i}) \cap F(S_{i})$
\end{itemize}
And for i=0, M being the area that we shall be marking as \emph{hot}:
\begin{itemize}
\item $F(M) = F(C_{0}) \setminus F(S_{0})$
\end{itemize}

When merge-up results in $F(M) = \emptyset$, that implies all data designated by the CarvPath was already marked as \emph{hot}, and should be considered a sub entity of an active entity. Such a CarvPath entity is considered a candidate for \emph{secondary opportunistic hashing}, a subject covered in the next section of this paper. 

For the \emph{unmerge down} operation, we start at $i=m-1$, where: $F(C_{i}) \ne \emptyset$, we perform CarvPath operation so that:
\begin{itemize}
\item $F(S"_{i}) = F(S'_{i}) \setminus F(C_{i})$
\item $F(C_{i-1}) = F(C_{i}) \setminus F(S'_{i})$
\end{itemize}
And for i=0, U being the area that we shall be marking as \emph{cold}:
\begin{itemize}
\item $F(U) = F(C_{0})$
\end{itemize}
Finally, we can define the total size of all fragments of $S_{0}$ as a measure of the page-cache pressure for our archive. A number useful for throttling purposes. By implementing a reference counting stack like this within the archive implementation, the kernel is made aware of the intent of our framework with respect to the usage of fragments within our flat secondary address space archive and can use this information to allow for more efficient use of the page cache. By exposing the total size of the ref-count>0 fragments within the stack, the framework will have a useful marker to base a throttling strategy for new data on, again helping to further reduce the amount of page-cache misses within the system. In a next section we will look at more uses for the reference counting stack and it's potential role in optimizing performance in cooperation with a message-bus solution.

%% file: mainpaper-ohash.tex
\section{Opportunistic hashing}
In the previous section we discussed the low-level versus high-level archive interaction in terms of CarvPath fragments and page-cache interaction. In our evaluation of the OCFA bottlenecks, we discovered that many spurious reads were the result of early hashing, performed in order to facilitate the Content address Storage (CAS) needs of OCFA. Given that content hashing is such a common operation, and given that, using an efficient hashing algorithm in many current day settings, hashing operations are many times cheaper than spurious disk-IO. We look at the concept of introducing opportunistic hashing to our archive implementation. The primary goal for opportunistic hashing is reducing the time between first access and last access, and thus the total lifetime of the data within the system, by delaying the first read and combining it with hashing, thus potentially avoiding spurious IO operations. If we can hash the data as a side effect of data already being read for other purposes, then some spurious reads may be avoided.

\subsection{Using the right hashing algorithm}
Due to legacy and performance concerns, most forensic tools up until today use hashing algorithms that from a cryptographical point of view should be considered deprecated. Many other hashing functions, while being more secure are less suitable due to their software performance properties. As outlined in appendix C, we have investigated suitable available options and development with regards to collision resistance of commonly used algorithms. The conclusion was that both MD5 and SHA1 are currently effectively deprecated for use within the context of an automated computer forensic framework, and that SHA2 and SHA3 are sub-optimal replacements when it comes to performance and scalability. We found a very suitable replacement algorithm in BLAKE2 and as outlined in appendix C advocate the wide adoption of BLAKE2 into the computer-forensic process.

\subsection{Primary opportunistic hashing}
Primary opportunistic hashing aims to prevent premature reads that might unnecessarily increase page-cache pressure, and limit the amount of page-cache misses by eliminating hashing geared IO read operations. 
A second part of the opportunistic hashing function revolves around the idea of simultaneously having multiple nested hot CarvPath entities, and accessing these in a sequential manner with respect to location within the flat secondary-storage address space as defined with CarvPath annotations. That way, if a piece of data is read from or written to the archive file, that piece of data might contribute not only to the hashing of the high-level pseudo-file that is being accessed at that particular time, but also to the hashing of nested children of that entity, or of the parent entity in which the current entity is nested. This however is not the primary projected benefit of opportunistic hashing. The main benefit is that a CarvPath entity traversing its tool-chain will often end up having its content partially or fully hashed by preceding tools as a result of hashing opportunistically on low-level IO operations, before the moment that explicit hashing might end up being actually needed. Many small files for example may get hashed completely resulting from file type checking. To allow for opportunistic hashing, every \emph{hot} CarvPath is added to an archive wide opportunistic hash collection. This collection will hold active BLAKE2 hashing state and in-CarvPath hashing offset information for every \emph{hot} CarvPath in the collection. When low level IO takes place within the archive, the block of data is offered to each of the opportunistic hash objects in the archive. If the offset and size of the IO operation match the current hashing offset of the entity, the relevant part of the data is used to further the hashing process for the given CarvPath. 

\subsection{Secondary opportunistic hashing}
As mentioned in the previous section about page-cache friendly archive interaction, CarvPath entities that in a reference-counting-stack merge-up operation don't result in any new hot data being added, are candidates for \emph{secondary} opportunistic hashing. All the data in the CarvPath entity is already marked as \emph{hot}, and could potentially already reside in the page-cache. The Linux \emph{incore} system call allows checking if a file chunk fully resides in the page-cache, or does not. If a chunk of data resides in the page-cache, then reading it will fetch it from page-cache without any actual IO taking place. We define secondary opportunistic hashing as the process whereby all leading incore fragments of a candidate, as identified by a reference-counting-stack merge-up operation, are actively read by the archive implementation and are used in opportunistic hashing operations. The strategy of secondary opportunistic hashing could be particularly useful for carving-generated sub-entities. Carving is usually done using sequential reads on a parent entity. When a sub-entity is identified by a carving tool and submitted as sub entity to the messaging subsystem, the whole child CarvPath entity will have been read very recently and the likelihood of it still residing in page-cache would be very high. This means that the child entity could be hashed using read operations that don't translate to any actual IO operations on the underlying secondary storage technology.

%% file: mainpaper-bus.tex
\section{Page-cache friendly message-bus interaction}
Where the archive implementation allows the page-cache efficient use of active hot data and provides the framework with information that could be used for throttling purposes, neither of these measures will have any impact on the out-flow interaction. The OCFA timing analysis revealed that the priority queuing mechanism used by the OCFA messaging sub-system was ineffective at reducing the average volume normalized lifetime of data within the system, a parameter that is rather important for page-cache efficiency. As such, reevaluation of job prioritization should prove very important. Due to logistic dependencies, a full investigation of optimum job prioritization falls outside of the scope of the achievable goals of this research. We did however look at providing different prioritization and other performance and scalability geared facilities that should provide essential handles for future research and system development in this area. 
\subsection{Actor job picking policies}
Given that we have elevated the flat address-space secondary storage model and the use of CarvPath annotations to a central paradigm for computer forensic framework implementation. Given also the existence of both the reference counting stack and the opportunistic hash collection, it should be possible to integrate part of the knowledge from these facilities into an alternative to the failing static priority queue concept. We propose that replacing the priority queues in the Anycast message-bus system with sortable sets or sets that allow for selective picking according to a picking policy. Given that different modules in the tool-chains have different properties with respect to data access patterns, data IO versus CPU usage, etc, it is likely that there could be benefits to using different picking strategies for different modules. 
Another variable relevant to picking policies is page-cache pressure. If current page-cache usage is high, then other picking policies may be more appropriate than when page-cache usage is relatively low.
Given the potentially available CarvPath and opportunistic hashing related attributes, a set of possible picking policies can be defined for our Anycast message-bus sets:
\begin{itemize}
\item Highest reference count: The idea behind this policy is related to opportunistic hashing. Using this policy will prioritize entities that contain data fragments that are shared between the highest number of entities. Reading any such entity would have a high probability of contributing to the opportunistic hashing of multiple other entities. 
\item Lowest opportunistic hashing offset: Given that regular hashing happens sequentially, it can be beneficial if entities are processed in order of opportunistic hashing adjusted offset. This in order to minimize hashing mismatches resulting from data beyond the current hashing offset of a file.
\item Reference count of one: Prefer entities that contain fragments with a reference count of one. This policy is aimed at short-term outflow. After data with a reference count of one is processed by a module that doesn' t produce any new child entities, this data will be completely done after module completion, allowing it to be marked as \emph{cold}. This could make room in the page-cache to allow throttling to discontinue and new data to be submitted into the archive.
\item Highest density of ref-count one data: The idea of this policy is that this will give precedence to processing operations that will free up substantial amounts of page-cache relative to the amount of data processed.
\item Lowest density of maximum ref-count data: The idea of this policy is that entities containing large amounts of the highest ref-count data are less likely to free up much of the page-cache in the short run.
\item Lowest weighted average ref-count: An other policy aimed at cumulative effect. The idea is that using the weighted average ref-count may possibly produce a better overall outflow, rather than focusing purely on ref-count of one.
\item Smallest data-size: The idea behind processing small entities first is that small entities could potentially grow the size of the Anycast sets to the point where the sorting or picking overhead becomes a bottleneck in its own right. To avoid this, prioritizing small (fast) entities over larger ones could keep the picking process responding sufficiently fast.
\end{itemize}
We allow the above policies to be chained, and allow each module to set and update its own picking policy.
\pagebreak
\subsection{Message-bus/Archive integration}
\begin{figure}
  \centering
  \includegraphics[width=60mm]{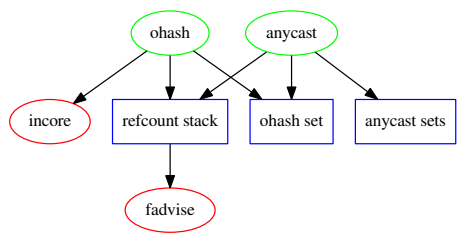}
  \caption{Dependencies}
  \label{fig:Dependencies}
\end{figure}
In the job picking policies we identified the need to use information about CarvPath designated entities that is maintained within the reference counting stack and opportunistic hashing layer of the data archive. After evaluating the required information exchange between the Anycast message bus and the archive with respect to the implementation of job picking, it became clear that information exchange would make up a level of overhead that would make performant and scalable messaging infeasible. Due to data-structure dependencies as shown in figure ~\ref{fig:Dependencies} on page ~\pageref{fig:Dependencies}, the Anycast message-bus, implementation of the job picking policies needs to be integrated with the archive in order for both to be able to work effectively and efficiently with both opportunistic hashing state and the reference counting stack. By integrating these two, the message-bus solution becomes part of the same user-space file-system that implements the data archive solution. By choosing this approach however, the message-bus in essence becomes a \emph{local} message bus and scalability concerns for the message bus become a reality. There are two possible approaches to address scalability concerns for the message bus. The suitability of these possibilities depends on aspects of the distributed setup for the archive solution. If we look at archive scalability to a multiple node setup, there are two main options for distributed access to a single archive:
\begin{itemize}
\item Running multiple instances of the file-system on top of a Storage Area Network (SAN) solution that provides for a shared-access file-system such as SNFS.
\item Running an NFS mesh-up with one local archive per node that is shared with all other nodes over NFS.
\end{itemize}
Analysis of the archive/message-bus integration setup has revealed that both solutions come with their own specific challenges with respect to scalability of the message-bus part of things.
\subsubsection{Message-bus in SAN setup}
In the SAN archive setup, each node will have its own user-space file-system running on top of the raw archive files. This has the advantage for the archive part of things that both the file-system and page-cache information on that one node are maintained on a per-node basis without the need for additional record keeping on the side of the file-system implementation. the page-cache is local, the file-system is local, all fits seamlessly with respect to the archive side of things and page-cache optimized file-system access. Direct messaging between nodes however is not possible and additional facilities are needed to allow for jobs to be exchanged between nodes when such a thing becomes desirable for scalability reasons.  We identify the need for a load-balancing facility in our message-bus implementation that could be used in the implementation of a cross-node mesh-up implementation.
\subsubsection{Message-bus in NFS setup}
As with the SAN solution, with NFS each node will have its own user-space file-system running on top of the raw archive files. Using an NFS mesh-up though, the file-system where the archive exists will be local for one node and remote for all other nodes. It is suggested that each node uses files on a local file-system for primary (non load-balancing) processing. NFS mounted archives should be used primarily for serving load-balancing needs. Again we identify the need for a load-balancing facility in our message-bus.
\subsubsection{Minimal data-move based scalability}
Whenever a chunk of data moves from a tool or module on one node, to a tool or module on another node, this move effectively equates a page-cache miss. As such, we consider a node-move to be a relatively expensive operation that should not occur unless the move is beneficial to the distributed processing to the extent that the price of the page-cache miss can be justified. Hot data should by default stick to being processed on the same node. When data migrates, the new node should become the new sticky node, minimizing the total amount of data moves and allowing the overloaded node to appropriately clean up page-cache after a data move. When there is imbalance between the load of different nodes, a move of data must be considered. Given the price of a page-cache miss implied by a data move, migration policies are very important for effective load balancing. We thus identify the need of job picking policies geared specifically at job migration for load balancing purposes.
\pagebreak
\subsection{Load balancing support}
We identified the need for a load balancing facility for multi=server setups. Given the integration of the message-bus and the availability of job picking policies, it is logical to implement this facility in terms of message-bus consumer and message-bus producer operations. The concept of load-balancing within our solution is that the user-space file-system allows for a tool-chain to be prematurely abandoned on one file-system instance to then be newly instantiated on an other file-system instance. Given the availability of per module job-picking, we choose to implement our load-balancing facility in terms of extended job picking. We identify the concept of \emph{module-picking policies}. The load balancing facility running on top of the user-space file-system should for example be able to pick a job from a module that is known to be CPU intensive. 
We define the following module-set attributes that the framework may set for load balancing purposes:
\begin{itemize} 
\item Module weight: This attribute is meant as a measure of how CPU intensive the operations of a module are.
\item Set overflow threshold: This attribute is meant to set a set-size threshold. Only if the set-size reaches this threshold will migration be considered. 
\end{itemize}
Based on these set attributes, the size of the set and volume of entities referred to in the set, a non-empty set will be picked using a given module set picking policies. We define the following policies:
\begin{itemize}
\item Weight: Pick the set with the highest module weight attribute value.
\item Count : Pick the set with the highest number of entities.
\item Volume: Pick the set with the highest total size of all entities combined. 
\item $\dfrac{W \times C}{V}$:  Pick a set with  small yet CPU intensive jobs.
\end{itemize} 
As with the job picking policies, these policies can be chained. They are also combined with the job picking policies. A policy for load balancing could be: \emph{Pick the job with the highest density of ref-count-one data from the module with highest job count from the module sets that share the highest weight}. 
\pagebreak
\subsection{Hooks for a distributed FIVES router}
In the analysis of tool-chain commonality in OCFA, we identified that many entities are discarded after digest and file-type checks.
In the analysis of OCFA bottlenecks, the central router had quite a dominant position in that its centralized nature doubled the amount of message-bus interactions. If every module process comes with both built-in file-type module functionality and a built-in OCFA-style router, then we would reap the following performance gains:
\begin{itemize}
\item Unsupported file-types would not need to be routed at all; No messages.
\item No need for doubling message. Each module sends message directly to the next module, halving the amount of Anycast messages.
\end{itemize}
In OCFA the router used a growing provenance-log blob to determine the next tool in the tool-chain to forward a data entity to.
The FIVES project introduced a replacement router for OCFA. the FIVES router used the OCFA provenance log to store rule-list state that would be used the next time the provenance-log was presented to the router. The concept of the FIVES router was a very powerful tool allowing for the expression of complex tool-chain rules. This is a concept that is important to preserve and accommodate in our message-bus solution.
The solution OCFA offered to the FIVES router, for performance reasons, must be considered not to be a viable one. The use of a large per-tool-chain mutable provenance log containing all module meta-data and being updated by each consecutive module has major IO performance issues. We advocate instead the separation of the core provenance log from meta-data collection, and the use of immutable storage for meta-data exchange. If we treat evidence data and evidence meta-data the same from an archive and messaging point of view, we can devise a minimal messaging format that does away with the need to interact with the large mutable meta-data provenance log blobs. Looking at the minimum needs for both modules and a potential new FIVES style router with statefull rule-list, we can define our core message format as being made up of five distinct and relatively small fields:
\begin{itemize}
\item CarvPath: The CarvPath that designated the immutable data or meta-data this message refers to.
\item Mime-Type : The mime-type of the data or meta-data this message refers to.
\item File extension: A file extension for the CarvPath to facilitate the next tool.
\item Next Module: The name of the tool or module functionality that this message is to be handled by.
\item Router state: A string representing all the rule-list traversal state that the next router functionality instance may need, in order to continue rule-list traversal for this tool-chain. 
\end{itemize}
It is up to the implementation of the new router functionality to minimize the size requirements of the router-state string without compromising the expressiveness of the rule-list. Ideally the expressiveness should be high enough for implementing the rule-list description as a domain specific language. It is important to note that in the proposed setup, non-provenance related meta-data extraction is to be submitted directly to some kind of data store module. As there should no longer be a need to use per tool-chain meta-data blobs, a dissecting module producing multiple child entities may choose to consolidate the meta-data for multiple child entities in a single meta-data storage entity, further reducing the amount of Anycast messages in the system. 

%% file: mainpaper-integrity.tex
\section{Forensic system-integrity needs}
As identified in the literature survey, in the light of overall system robustness, the reality of anti forensics and the provable correctness and repeatability of the provenance information, system integrity needs have changed quit a bit since OCFA was incepted. Part of these integrity requirements fitted within the scope of the research project and complemented the performance and scalability geared subject of this paper, while other parts did not. As our literature survey showed, the chain of custody requires guarding of provenance information and requires facilities that safeguard the integrity of the data. What the literature survey did not reveal is the realization that \emph{the chain of custody does not end when the data reaches the lab}. The forensic data in a computer forensics framework is processed with a wide range of tools from a multitude of authors. Given the reality of anti-forensic techniques, and given the reality of software bugs, we must consider the scenario that seized data might either crash a tool in a relatively benign way, or in the worst case might actually end up executing an exploit against one of the tools in the tool-chain. Even an non-exploit crash might in extreme cases end up corrupting the results of a framework such as OCFA. While stopping tool crashes and exploits from happening isn't a viable goal, minimizing the potential impact of such an event is. Some measures could be taken at a framework and system logistic level by means of Mandatory Access Control (MAC), IP firewall rules and the use of different user accounts for the different tools in the tool-chain. These fall outside of the scope of this research project. Other concerns however can be considered useful features for our integrated archive and message-bus solution.
\section{Frozen immutable CarvPath data}
We already established that mutable data had IO performance issues and the use of immutable write-once data could be beneficial to overall system performance. There is however also a major anti-forensic invulnerability advantage to technically implementing immutability. Data that is made immutable immediately after creation in a technically secure way can no longer be changed. If a tool, due to a bug, ends up trying to corrupt existing data, an immutability facility will stop this from occurring. Given that our archive is implemented as a user-space file-system, running this file-system as a different user than the modules should keep these modules from directly corrupting any raw archive data. The only access to the archive data is through the file-system. A file-system that we implement in such a way that only new data (kick-start or derived data) is temporarily mutable during creation. Once the new data is forwarded to a next tool in the tool-chain, the file-system will deny any further attempts at changing the data. This means that both data and meta-data, after creation, become completely invulnerable to any type of data corruption attempts. Only the creating module should be capable, when compromised, to create bogus output. The result: the impact of a bug on actual evidence (meta)data is reduced to an absolute minimum.
\section{Trusted provenance logs}
In OCFA, the modules themselves were responsible for provenance logging. The Provenance log was bundled with all the extracted meta-data in a mutable database BLOB that was updated by each consecutive module. By making all data storage per-module write-once, most of the integrity concerns are alleviated. Write-once, however, does not keep a module from either not reporting or lying about provenance information. The file-system message bus functionality has the ability to keep track of the actual provenance information. With the file-system running as a different user as the modules, this means that the file-system is fully capable of maintaining the provenance logs in an out of band way and writing them to a file or database inaccessible to regular modules. There should be no need to trust the different modules to be honest about their own provenance information. The file-system does provenance logging and no compromised tool will be able to intervene with the integrity of the file-system generated provenance logs.
\section{Use of sparse capabilities}
Combining the two facilities above, we have the core ingredients for providing a future computer forensics framework with the lab-side equivalent of a Sealed Digital Evidence Bag (SDEB). There are however still some access control issues that must be addressed. The base concept is that each module that derives either child data or meta data from an input entity, is able to create and seal a new virtual evidence bag by writing and  when doing so, implicitly linking the child/parent relation of new data to the old data. In order to implement this possibility, we shall be relying partially on the assumption that a future full solution will be combining MAC with the solution we are offering. The way we allow for SDEB equivalence, is by providing a special file-system based API to the framework modules. A file-system based API that builds upon the concept of sparse capabilities. Sparse capabilities are unguessable strings that both designate an object and allow access to that object. You can think of sparse capabilities as unguessable object names doubling as passwords for accessing those same objects. The following outlines the sparse capability based API that, when combined with MAC, the provenance log and immutability facilities should create a good line of defense against most forms of integrity compromising anti-forensic attacks or non-hostile yet potentially corrupting tool crash events.
\begin{itemize}
\item \emph{Instance cap} :A tool or module process can register itself as an instance or worker of a certain module/tool. When it does this ,it receives a sparse capability that links its operations to the provenance logs and provides an entry point to other file-system operations. The use of MAC to limit a module to only register as itself is strongly advised.
\item \emph{Job cap} Using the above instance cap, a tool can set some attributes such as job picking policy. After this it can ask the message-bus for a job to process. This job will be provided in the form of a second sparse capability. Accepting the new job will add an entry to the in-process provenance log for the tool-chain that this job is a part of. The job capability can be used to retrieve router state for the framework routing logic, it can be used to forward the current job to the a next tool in the tool-chain, and it can be used to create and submit child entities.   
\item \emph{New data cap} Using the above Job capability, a child storage entity can be created. Creating a child storage entity will yield a new-data capability. This capability allows read/write access to a new fixed-size area of archive storage space that can be used either for child data or extracted meta-data to be stored. After writing the data has completed, the new entity can be \emph{frozen} and submitted to a tool in the tool-chain.
\end{itemize}
The use of these three types of sparse capabilities provide a semi-secure file-system based API that provides for message-bus and data-entry capabilities. Given some fundamental cross-process insecurities with the Linux /proc filesystem, more MAC measures should be considered to make the use of these sparse capabilities secure against truly advanced anti-forensics attacks. This however falls outside the scope of this paper.

%% file: mainpaper-poc.tex
\section{MattockFS; a proof of concept}
Next to the research outcomes described in this chapter, an important piece of result from our research project is a fully functional proof of concept of the combined archive and message bus solution as a user space file-system. The FUSE file-system MattockFS combines almost all features described in this chapter. Due to the time constraints of this research project, this initial proof of concept was implemented in the programming language Python. We assume the performance to currently still be sub optimal due to language choice, job picking policy algorithms and the specific BLAKE2 algorithms available from Python, that all have not been optimized for speed. Functionally though, the proof of concept has most functionality that a production-ripe version will need to possess apart from some minor points of possible improvement. The proof of concept implements the full archive functionality and sparse-cap based file-system API, for what we also implemented a python wrapper API. The implementation of a lab-side SDEB equivalent is completely operational in the proof of concept. Most of the message bus functionality is operational, although job picking algorithms may not be optimally implemented with respect to processing efficiency. A feature that would allow messaging state to be restored from the provenance log would have been desirable, making the message-bus solution equivalent to the persistent priority queue solution that the OCFA Anycast relay used. Stopping and re-starting an incomplete run is not currently part yet of the proof of concept implementation. Primary opportunistic hashing has been implemented. Secondary opportunistic hashing, though, has not. Test code coverage for the current code-base, at over 90\% should be considered sufficient to have confidence in the correct workings of the current proof of concept implementation.

%% file: mainpaper-evaluation.tex
\chapter{Evaluation and discussion of results}
\noindent 
From a purely research oriented perspective, the viability of creating the foundation for an OCFA inspired next generation scalable message passing concurrency computer forensics framework suitable for both academic research and full scale criminal investigations that addresses the page-cache and system integrity concerns identified, has mostly been shown in this research project, and most of the research questions have been answered in a satisfactory way. Some questions about the practical interaction of different individually useful partial solutions however remain as challenges for future research and implementation. The proof of concept created in this research project shows the viability of the proposed facilities. There remain however scalability, performance and robustness concerns that shall need to be addressed in future work. Lacking other interacting framework components, not everything can fully be evaluated until a framework is created and realistic interaction can be objectively evaluated.
\newpage
\section{Evaluation of design and implementation}
In this section, part of the design and implementation of the MattockFS proof of concept is evaluated. It must be noted that the complete evaluation of MattockFS would require a fully functional computer forensic framework complete with a set of modules. Lacking these, we need to limit this evaluation to those part of the design and implementation that can be evaluated without these components being available. 
\subsection{Evaluation of data-input}
The MattockFS code-base includes a tool for importing forensic disk images in the EWF file format, This tool, ewf2mattock uses the python ewf library and the python MattockFS API library to copy the data contained in an EWF file into a MattockFS archive. To evaluate the MattockFS data-input functionality, we create a simple python script that basically converts an EWF file to a raw DD file using the python EWF library. Using a 10GB EWF image file representing a 40GB disk image, we compare the time required to run ewf2mattock with the time required for the simple script. We do this for systems with three different classes of CPU's in order to look for a trend with respect to CPU power available.

\begin{center}
  \begin{tabular}{ | c | c | c | c | c | }
    \hline
     \emph{CPU} & \emph{ewf2dd.py} & \emph{ewf2mattock} & \emph{overhead} & throughput \\
     \hline
     Core 2 Duo & 11:33 & 40:57 & 255\% & 133 Mbps \\
     \hline
     AMD A6 & 10:02 & 29:39 & 197\% & 184 Mbps \\
     \hline
     XEON E5 & 6:08 & 17:06 & 177\% & 320 Mbps \\
    \hline
  \end{tabular}
\end{center}
In this evaluation we notice the overhead of MattockFS with respect to data input is very high. We however also see that the overhead becomes significantly smaller as CPU power increases. This suggests that improvement of CPU efficiency of the MattockFS implementation, for example by reimplementing MattockFS in C++, might, for representable server hardware, reduce the overhead of data input to a level that would not end up making MattockFS a bottleneck on data input over, for example a 1Gbps network connection. We must conclude that for use on lower range CPUs, the current implementation data input overhead should be considered prohibitive, and that the CPU requirements for acceptable performance with the current code-base are relatively high. 
\subsection{Evaluation of opportunistic hashing}
\begin{figure}
  \centering
  \includegraphics[width=120mm]{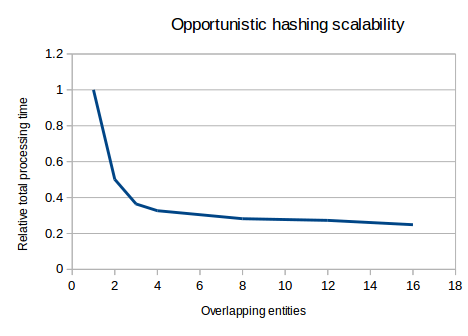}
  \caption{Opportunistic hashing overlap scaling}
  \label{fig:OppScaling}
\end{figure}

For evaluating opportunistic hashing in MattockFS, we again use three different computers with a different class of CPU. As OCFA used SHA1 and MattockFS makes use of BLAKE2 in Python, we look at the following:
\begin{itemize}
\item sha1sum on the raw archive file
\item b2sum on the raw archive file
\item blake2 on the raw archive file using the python API
\item reading from MattockFS
\end{itemize}
We notice that only on the low spec CPU system, there is any notable difference in the time needed to perform these tasks using the different tools. Both the increase in speed between sha1sum and b2sum, and the considerable overhead in the python implementation of BLAKE2 are reduced to insignificant numbers on the two more performant CPUs. On the XEON E5 based system, each of the above runs took 5:22 $\pm$ 0:02.
A second part of the evaluation of MattockFS opportunistic hashing lies in answering the question how the opportunistic hashing scales. Again, a full evaluation would require a complete working computer forensic framework that would run on top of MattockFS, but we can do a set of simple tests to determine how MattockFS opportunistic hashing scales to higher levels of data overlap. To do this, we open a number of files that span almost the complete size of the 41GB disk image and time the amount of time it takes to read the whole disk image.
Figure ~\ref{fig:OppScaling} on page ~\pageref{fig:OppScaling} shows the results of these tests.
 
\subsection{Evaluation of job picking policies}
The most performance sensitive part of MattockFS is the AnyCast message-bus implementation. While adding messages to the message-bus is straight forward, selecting the best job from a set uses one or more rather scalability sensitive job picking policies. The scalability is determined by the time required for picking a job from the set at different set sizes. To test the scalability properties of different job picking policies, we first fill an AnyCast set with a large number of messages and then fetch all the jobs in the set using the given policy recording the time needed for fetching a single message. In this evaluation we used the XEON E5 setup only. One particular policy stood out in such a negative way that it was needed to scale down the whole experiment for that specific policy. 
\begin{figure}
  \centering
  \includegraphics[width=120mm]{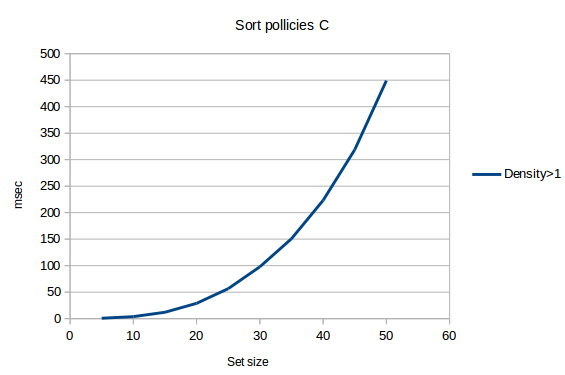}
  \caption{Message picking scaling for policy d}
  \label{fig:ScalingC}
\end{figure}
The current implementation of the \emph{d} policy, a policy that prefers jobs with the lowest possible density of chunks with a ref-count higher than one, performed so poorly that a different set-size range was needed to perform a useful performance and scalability test. The graph for this policy, shown in figure ~\ref{fig:ScalingC} on page ~\pageref{fig:ScalingC}, shows not only that performance degrades quickly at just a few dozen messages in the set, but also that message picking time follows an exponential curve. While it is likely that implementation in C++ would attenuate the poor performance of the implementation, the exponential nature of the performance curve is still a strong indicator that this would not increase the threshold of the set-size where this policy becomes unusable by much. 
\begin{figure}
  \centering
  \includegraphics[width=120mm]{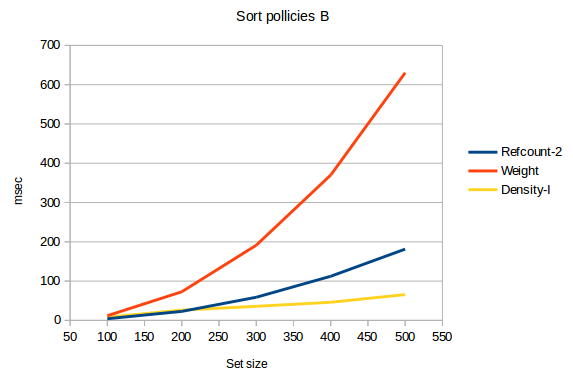}
  \caption{Message picking scaling for policy W/r/D}
  \label{fig:ScalingB}
\end{figure}
The \emph{W}, \emph{r} and \emph{D} policies, displayed in figure ~\ref{fig:ScalingB} on page ~\pageref{fig:ScalingB}, are not as poorly performant as the \emph{d} policy, and should be usable if throttling is used to keep the maximum number of messages in a set below a reasonable but possibly restrictive threshold.  each of these three policies yields a performance graph with a distinctly exponential shape. It is suggested that future research should look into the algorithms for the above four policies and investigate if better policy scalability can be achieved with alternative picking policies.
\begin{figure}
  \centering
  \includegraphics[width=120mm]{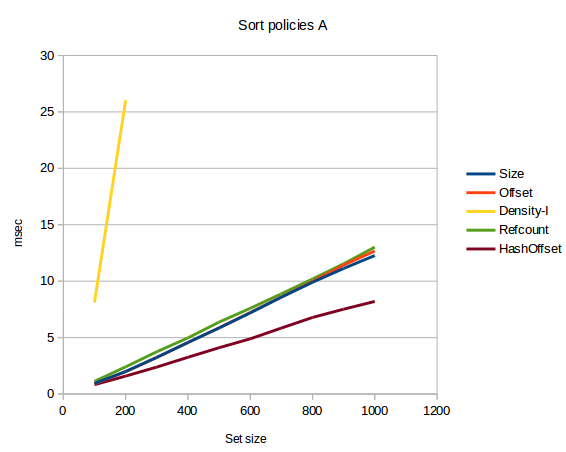}
  \caption{Message picking scaling for policies D/S/O/R/H}
  \label{fig:ScalingA}
\end{figure}
The remaining four policies, shown in figure ~\ref{fig:ScalingA} on page ~\pageref{fig:ScalingA}, have scalability properties that are good in two ways. Not only do they all show relatively decent performance for relatively large set sizes, they do so with what approximates a straight line. For these policies, it is likely that a C++ implementation should yield a truly high-performance local message-bus implementation.   
\subsection{Evaluation of hash/file-type switch}
In the final part of the MattockFS evaluation we look at the order of file-type determination and file hashing and evaluate what that could mean for one of the investigations from our OCFA timing analysis. While this analysis is purely a logical one based on a simplified model, the result should be indicative of expectable reduction in reads. If we look at investigation \emph{A} from our OCFA timing analysis, we can extract the file size of all files that:
\begin{itemize}
\item Pass through the digest module only.
\item Pass through both the digest and the file module but no other modules.
\item Pass through both digest and file module before being forwarded to an other module.
\end{itemize}
Depending on mount options and defaults, a file-type check will read either 32k, 64k or 128k of data. Applying this to the file sizes above, we can determine how many data blocks would be read by an initial file-type check, and how much will remain to be (un)read after that. Unfortunately we don't have the data to know if the files that match black/white-list matching would have matched a file-type check. As such we consider three scenarios:
\begin{itemize}
\item File-type split for non white/black-list files is representative for white/black-list files.
\item All non white/black-list files have a file type that needs further processing.
\item None of the non white/black-list files require any further file-type dependent processing.
\end{itemize} 
Finally, the upper and lower bound of blocks eventually red assuming that all or respectively none of the blocks of forwarded files are eventually read by the following modules. The following table shows the results of this analysis:
\begin{center}
  \begin{tabular}{ | c | c || c | c | c | }
    \hline
     \emph{Scenario} & \emph{Min/Max} & \emph{32k} & \emph{64k} & \emph{128k} \\
     \hline
     None & Min & 10\% & 18\% & 31\% \\
     \hline
     None & Max & 23\% & 30\% & 40\% \\
     \hline
     Representative & Min & 8\% & 15\% & 26\% \\
     \hline
     Representative & Max & 23\% & 29\% & 38\% \\
     \hline
     All & Min & 12\% & 23\% & 40\% \\
     \hline
     All & Max & 37\% & 45\% & 57\% \\
     \hline
  \end{tabular}
\end{center}
It is important to note that reducing the block size for FUSE based read operation, while significantly reducing spurious reads in these scenario's, doing so will also increase the overall number of user space kernel space switches for the FUSE file-system, greatly increasing FUSE file-system overhead. As such, we can conservatively conclude that the reduction of spurious reads should roughly be between 43\% and 69\%. It might be possible to reduce it further, possibly to a figure between 63\% and 90\%, but at substantially increased numbers of kernel/user-space switches. Determining if this tradeoff is worth it should be considered subject of further study.
\newpage
\section{answers to the research questions}
The historic page-cache related bottlenecks within the OCFA have been located and different ways of addressing these bottlenecks in a future framework have been identified and evaluated individually. The usefulness of the flat secondary-storage address-space model in respect to solving page-cache related bottlenecks has been shown, as has the possibility of an opportunistic form of hashing. The anti-forensic weaknesses of the OCFA architecture have been identified and to a great extent addressed and demonstrated in our proof of concept. It was shown that the use of sparse capabilities, combined with a file-system based API, can provide an important, yet not a complete, solution for addressing these issues. We have shown with our proof of concept implementation that each of the concerns mentioned can be addressed and implemented within a computer-forensics framework component. Lacking other components for a forensic framework, spurious read reduction could not be reliablly quantifies. 

\section{From proof of concept to operational system}
While all of the research questions have been addressed, the current proof of concept is not yet fully suitable as foundational component of a future next generation computer-forensic framework. We propose that the following is needed in order to turn the current proof of concept into a production ready implementation:
\begin{itemize}
\item \emph{Restore state from journal}: While the current implementation maintains a journal log file that in theory could be used to allow the messaging subsystem to pick up where it left off after a restart, restore from journal isn't currently implemented. Implementing this would effectively give MattockFS the same atomic persistence that the OCFA Anycast had.
\item \emph{Implement temporary-quarantine facility}: In OCFA, the Anycast implemented a quarantine facility for data that would crash a module. This facility, that OCFA implemented in the form of the \emph{never} priority allowed the investigation to keep running while maintenance programmers would fix the module. Research is needed into the possibility of implementing a similar data quarantine functionality in MattockFS.
\item \emph{Port MattockFS to C++}: While Python is a good programming language for prototyping, our evaluation has shown that the implementation is relatively slow with respect to import and possibly messaging. This slowness could potentially partially nullify part of the performance benefits from the page-cache and opportunistic hashing. Rewriting it in C++ is likely to lead to a much faster implementation. 
\item \emph{Move to BLAKE2bp}: The current implementation uses a Python module without support for the parallel multi-core BLAKE2bp implementation. With just one file-system on a node for a given archive, this would mean at most one core could be working on opportunistic hashing for the whole node, possibly creating a hashing bottleneck for the system. Implementing  BLAKE2bp, either by patching the BLAKE2 python module code, or by moving to C++ and appropriate library could remove that potential bottleneck.
\item \emph{Re-factor sorting to picking}: While the picking policies are suitable for a proof of concept, some policies have shown to have poor scalability properties in their current implementation. It is likely that moving from a sorting type algorithm to a more picking oriented algorithm would lead to a more efficient picking process for some of these policies, especially for larger sets.
\item \emph{Implement and evaluate secondary opportunistic hashing} : Secondary \emph{incore} based opportunistic hashing could potentially greatly improve the interaction between carving and opportunistic hashing. 
\item \emph{Write ports of the base API to C++} Next to Python, C++ should be considered an important language for writing modules in. The current Python wrapper API for access to the file-system based archive and messaging API should also be ported to C++.
\item \emph{Redis and privilege separation}: The current implementation of MattockFS uses REDIS as in-memory distributed storage for CarvPath long-path entries. The use of a Redis server for distributed CarvPath long-path storage is currently an Achilles' heel in the access control model of MattockFS. It may be necessary to rethink this part of MattockFS to remove this potential vulnerability.
\item \emph{Mandatory access control}: Without limiting access to /proc in an effective way, the sparse capability based API might be compromised through capabilities that could be snooped from other processes through pseudo files in the /proc hierarchy that give access to process internals including open files. Using AppArmor or SELinux, access to these pseudo files could be effectively prohibited. Further, in a directed anti-forensic attack against MattockFS, a running module might spawn a child process that could register under the name of an other module. Again a threat that could be removed by using AppArmor or SELinux for MAC. 
\item \emph{Framework fadvise hooks}: Different file-types can have file-type specific processing and expected access patterns. It is likely that providing hooks for allowing the framework to predict such patterns could allow the file-system to set smarter fadvise values that might further reduce spurious reads and improve page-cache efficiency.
\item \emph{Fuse read-block size}: As we identified in our evaluation of file/digest order, reducing the read block-size for FUSE could potentially greatly reduce the amount of spurious reads, but at the expense of file-system overhead. This should be evaluated and if worth the price, implemented into MattockFS.
\end{itemize}
\section{Towards a complete framework}
In appendix D, an outline of a possible computer forensic framework on top of MattockFS is described. The base idea would be that asynchronous-framework based processes will be used as module instances or workers. These processes would combine shared framework logic with module specific code. The framework logic should, next to functionality for serialization, evidence tree walking and logic for communicating with MattockFS, also contain FIVES like routing logic and file-type checking functionality (libmagic). The combination of MattockFS and these processes should make for a single node forensic framework implementation. Secondly a load balancing mesh-up should be added. That mesh-up would be working on load-balancing of initial disk-image kick-starts, but also work in the limited migration of jobs between different nodes for load balancing purposes.
\section{Other future work}
While the job picking policies, the reference counting stack, and the use of opportunistic hashing have each been examined in isolation, the policies and their interaction with both opportunistic hashing and page-cache efficiency can only truly be evaluated when a full framework implementation is available and a real set of disk images is processed by a real set of modules. As such, only \emph{after} the completion of a complete framework together with the most essential modules, the results from the currently concluded research could fully be evaluated. Especially the interaction between opportunistic hashing and page-cache hit oriented features and policies.It is suggested that a profiling and evaluation run, such as done on OCFA within this research project, should be conducted after completion of a full framework with its base modules. This profiling should look at the different available policies and evaluate each for its merits and performance characteristics. It is likely that such research would yield new insights on the interaction of opportunistic hashing and page-cache strategy possibilities that could combine to arrive at strategies aimed at optimum overall IO performance and a minimum of spurious reads.
Part off the effective interaction between the module framework and MattockFS relies on the throttling implementation in the framework. A suitable throttling policy is yet to be devised and devising such a policy in a way that neither under- nor over-commits is considered the most crucial piece of algorithmic design that is yet to be conducted. 
Currently MattockFS only uses two states to mark archive-file-chunks. The \emph{fadvise} function defines additional states that might be useful. For example a state that marks a \emph{hot} section for sequential access. Exploring the use of other state markers could potentially further improve page-cache efficiency. Further, MattockFS currently only knows two values for the status of a CarvPath fragment. Either it is \emph{hot} or it is cold. Hot could currently mean two things:
\begin{itemize}
\item The data has been read before, likely in page-cache and will be accessed again. (\emph{hot})
\item The data is part of designated area that has not actually been read yet but will likely be read in the near future. (\emph{warm})
\end{itemize}
Exploring the possibilities of differentiating between these two could constitute an area of useful future research. Finally there is the subject of non sequential access. Non-sequential access to the parts of a data entity is undesirable from an opportunistic hashing perspective, but will not always be avoidable. There might be some merit in the use of Merkle-tree based file-hashing that would allow opportunistic hashing to take place even with non-sequential access patterns. It is unknown, due to the greater amount of required state and the unknown access patterns for different file-types, if the effect of Merkle-tree hashing could yield a beneficial result. This is a potential subject for further research.  
\section{Recommendation}
The recommendation of this author is to continue work and research on the creation of a future open-source computer-forensic framework. The work done for the MattockFS proof of concept could act as a first step towards such a framework and the insights gained with respect to page-cache efficiency and access control could also benefit other forensic framework efforts. While the proof of concept lacks the persistence and, due to language and algorithm choices, part of the performance characteristics needed for a production system, the base design should be usable for use in a C++ port that addresses each of these issues. The existing code is stable enough to be used as foundation during the development of a framework on top of it. this means that work on MattockFS improvements and work on the framework could be done in parallel.

%% file: mainpaper-references.tex
\noindent 
\renewcommand{\bibname}{List of references}

%% file: ocfa-scriptsreference.tex
\noindent 

\renewcommand{\bibname}{Scripts created for OCFA timing analysis}

%% file: linux-linux.tex
\chapter{Page-cache in Linux}
In this appendix we examine the core properties of the Linux page-cache and the possibilities for interacting with the page-cache and related memory layout parameters in a way that could limit the disk-cache miss rate of a multi-process system running on this OS. First we will look at the main memory usage strategy used by Linux and at the way that page-cache competes for RAM with other RAM consuming Linux subsystems. We will look at information we can extract from the virtual /proc filesystem and at POSIX API's that provide a way to interact with the way the page-cache works. We then move on to project this knowledge to our fictitious computer-forensic framework and discuss what measures can be taken to effectively interact with the Linux system in order to limit page-cache misses on a Linux system that is hosting such a framework. We close this appendix by reasoning about possible strategies that could be used by a computer-forensic framework in an attempt to optimize disk-cache hit percentages while at the same time not introducing other sources of stagnating throughput to the resulting setup.
\section{Allocation and use of page-cache}
RAM in a computer is a shared resource. It is a scarce resource that is managed by the operating system. The OS kernel itself runs in RAM as do the user processes. Both kernel/process code, heap and stack take up pieces of RAM and on a busy system some RAM pages may end up getting swapped out of RAM to an on-disk swap facility. Not all RAM is kept available to kernel and processes. Part of RAM is allocated to hold a copy of on-disk data and to act as a cache for file data that the Linux kernel assumes is likely to be read again in the foreseeable future, allowing for more efficient use of the, relatively slow, reading from hard disk media. An other part is allocated to hold data pages that should eventually end up on disk and aren't synced to disk yet to improve the overall system performance related to issues with frequent short writes. The Linux kernel needs to balance the desire to avoid wasting disk-IO on swapping out process memory with the desire to not waste disk-IO on data that could have been cached. It's as if there is a rubber band between the process oriented RAM usage and the file-IO oriented RAM usage. As such, memory available to page-cache is not static. It depends on the behavior of the different processes on the system. Our main interest in RAM for the purpose of this research is an interest in the part of the RAM used for the read-part of the page cache. While there is a lot to say about the use and tuning of write oriented caches, this falls outside of the scope of this paper and thus shall not be discussed here. 
\section{/proc/meminfo}
In the /proc filesystem, the pseudo file /proc/meminfo can be used to access kernel information about the size and use of the system RAM. Some potentially interesting variables that can be retrieved from this pseudo file are:
\begin{itemize}
\item \emph{Cached} : In-memory cache for files read from disk (the page-cache).
\item \emph{MemFree} : The total of memory that is free to be used for anything.
\item \emph{Active} : Recently used memory that ideally isn't reclaimed.
\item \emph{MemTotal} : The total usable system RAM.
\item \emph{SwapTotal} : Total amount of swap space available.
\item \emph{SwapFree} : The currently unused portion of the swap space.
\end{itemize}
The information from this pseudo file can be used to initialize or later tune our efforts at throttling data input into our system. 
\section{/sys/block/\$DEV/stat}
Other than directly looking at the memory usage divisions, the actual disk activity should be a good indication of how well the page-cache is performing across a long running operation. The system pseudo filesystem defines a directory under \emph{/sys/block} for every block device. In any such directory resides a pseudo file named \emph{stat} that contains information that should be useful in this sense. The pseudo file gives access to I/O statistics of a given block device. Each line contains the current disk stat variables for a single device. 
\begin{enumerate}
\item Number of \emph{reads completed}.
\item Number of \emph{reads merged}.
\item Number of \emph{sectors read}.
\item Number of \emph{milliseconds spend reading}.
\item Number of writes completed.
\item Number of writes merged.
\item Number of sectors written.
\item Number of milliseconds spent writing.
\item Number of I/Os currently in progress.
\item Number of milliseconds spent doing I/O
\item Weighted number of milliseconds doing I/O
\end{enumerate} 
For our research we are interested in the first four numbers. The reading behavior. If we are to compare strategies for disk-cache efficiency, a test run with a lower number of reads, a higher number of reads merged and a smaller number of milliseconds spent reading would be the preferred result. As disk-cache misses would result in undesirable additional reads and increased amount of time spent reading, a good page-cache efficient solution should result in those numbers going down. 
\section{Starting off with a clean slate}
If we want to repeatedly run tests on the page-cache behavior of the system, it will be important that we can reinitialize our page-cache so that a first test does not mess with a second or third test. So how do we reinitialize the page cache so we can start our test with a clean slate? There are two things we need to do.  First make sure all page cache \emph{can} be initialized cleanly by invoking \emph{sync} as root. After that we run the following command (as root):
\begin{itemize}
\item \emph{echo 3 > /proc/sys/vm/drop\_caches}
\end{itemize}
This command will write the number \emph{3} to the pseudo file /proc/sys/vm/drop\_caches. Doing so will prompt the kernel to free the cache and slab objects. Finally we should save a copy of both \emph{/proc/meminfo} and the relevant \emph{/sys/block/\$DEV/stat} files before we can run our test. After running the test, a second copy can be made of these pseudo files. The differences between the two should yield information related to the page-cache mis/hit-rates.
\section{API interaction with the page-cache part of the Linux kernel}
\subsection{posix\_fadvice}
The Posix API call \emph{posix\_fadvise} or \emph{posix\_fadvice64} advises the kernel about the expected behavior of the application with respect to a region of data as associated with an open file descriptor. This advice helps the kernel in its page-cache behavior, as it will allow the kernel to determine if specific pages should get precedence over others when deciding what old page-cache data to keep and what to drop. A few important values for the advice are:
\begin{itemize}
\item \emph{POSIX\_FADV\_WILLNEED} : The data is expected to be accessed in the near future
\item \emph{POSIX\_FADV\_DONTNEED} : The data is NOT expected to be accessed in the near future
\item \emph{POSIX\_FADV\_NOREUSE} : The data is expected to only be accessed once.
\end{itemize}
\subsection{mincore}
It is possible for the application itself to find out if data it wants to access is currently present in the page cache. To do this, the application can invoke the \emph{mincore} API call.  
\subsection{readahead}
If the application knows that a file-descriptor is about to be used to read a specific range of data, it may populate the page-cache explicitly by invoking the readahead Linux specific API call.
\section{Interaction between a computer-forensic framework and the Linux kernel}
When we assume a CarvFS-like user space file-system and an OCFA-like routing and relaying architecture will be used, and if we look at the capabilities and knowledge that these and other subsystems may have, we get to the following list of assertions:
\begin{itemize}
\item The CarvFS-like user space file-system, assuming that it keeps a file descriptor opened to the underlying data archive file, is the only place where API based interaction with the page-cache kernel subsystem is effective.
\item Knowledge about \emph{active} CarvPaths within the framework context, is possibly not directly available to the CarvFS-like user space file-system and could come from explicit interaction of either the modules involved or the routing and messaging infrastructure. Alternatively this interaction can be avoided by integrating other functionalities into the file-system component.
\item With this knowledge, the CarvFS-like user space file-system can use invocation of the \emph{posix\_fadvice} API in a more structural way, properly marking the designated areas according to their projected remaining cross-module paths and usage. 
\end{itemize} 
\section{Possible strategies}
If we assume that a future computer-forensic framework will take the advice from the main paper at heart and will integrate the routing functionality in the libraries used by the different modules, we suggest that a strategy for page-cache interaction would be definable inside of the routing rule definitions. That is, we will want to allow the shared library code to set specific strategies either on a static module based manner or dynamically dependant on the specific routing configuration rules triggered by evidence meta data. 
While in a routing library based approach the use of \emph{readahead} may be warranted, within the scope of our research we shall focus on \emph{posix\_fadvise} and throttling. Given that some modules will look only at the file header. As a carvpath may be a sub entity of an other bigger entity, the current state of the parent region should also be relevant to our strategy for the sub entity. 

Other than the \emph{posix\_fadvice} policies, we also need to look at throttling. 
There are two types of throttling that are relevant. Throttling based on processing chunks of a large CarvPath entity that isn't marked  with posix\_fadvise in order to remain stored in the page-cache, and throttling with regards to the entering of fresh data into the growing archive of underlying data. Based on the information from \emph{/proc/meminfo} combined with the size of the currently mapped-as-active data chunks, throttle levels can be calculated. 
\section{Conclusions}
The use of the \emph{posix\_fadvise} API, especially the use of \emph{DONTNEED} to indicate a chunk of content no longer needs to reside in the page-cache, seems to be a pivotal concept within our project. 
The communication with the kernel gives us many hooks to implement a forensic framework in a way that allows us to reduce the amount of page-cache misses in the system. As a result of multiple factors the interaction between the framework and the kernel page-cache mechanism will go through a CarvFS-like user space file-system and shared library code used by all individual modules within the framework. Our first proof of concept will be less configurable than would be possible, yet our projected file-system should implement all the needed interaction points to allow a future implementation of an advanced router library to implement more advanced tuning through interaction with the file-system. The file-system will need to implement a type of CarvPath state graph that allows it to identify and eliminate data chunks that are no longer required to stay in the page-cache. It also needs to implement a way for a library to interact with it and set a policy or update the state for a specific CarvPath. Implementing throttling as a library poll loop seems reasonable for our proof of concept. We shall need to define a type of file-system based API, using either extended attributes or ioctl. As this will need to be combined with our desire to implement opportunistic hashing, the definition of this API will be the subject of appendix E.

%% file: mattock-hash.tex
\chapter{Secure hashing algorithm selection}
Traditionally digital forensics has always used secure hashes for multiple purposes. With new cryptographic advances in the subfield of secure hashes, the specific algorithms that are commonly used in digital forensics have come to be considered cryptographically deprecated. This appendix looks into these deprecation issues with commonly used algorithms, compatibility issues with moving to other algorithms and performance and scalability considerations with alternative secure hashing functions.
It also addresses how opportunistic-hashing geared algorithms in MattockFS further underline the need for specific hash function properties. 
\section{Hashing algorithm choice is essential} 
Traditionally MD5, and later also SHA1 have been used as hashing algorithms in digital forensics. From a cryptographic point of view, MD5 is now deprecated while SHA1 is in the process of being deprecated. Many public and law-enforcement-only hash collections like the Virusshare hash set or national child pornography hash sets are still distributed only with MD5 and/or SHA1 hashes. Others like the NIST NSRL contain MD5 and SHA1 hashes but are now supplemented with SHA256 hashes. While the last may sound like good news, there is another important issue with SHA256 for use in a forensic framework; SHA256 may be significantly more secure as a hashing algorithm compared to SHA1 or MD5, it is also significantly more CPU intensive and not parallelizable. These properties may undermine the whole concept of opportunistic hashing as presented in this paper. While today NIST still considers the use of SHA1 for purposes as defined in this paper as \emph{acceptable use}, we must consider the retroactive impact that a cryptography expert testifying on behalf of the defense and questioning the use of SHA1 in the forensic process may have a few years from now if SHA1 reaches the same level of deprecation that MD5 has today. Signs that SHA1s practical deprecation is near are plentiful. Recently a freestart collision has cast serious doubt on the remaining validity of the NIST \emph{acceptable use} stand and has shown that the use of GPUs could be leveraged to significantly reduce the potential cost of generating a SHA1 collision. 
It thus can be argued that moving forward to a non-deprecated secure hashing algorithm should be considered a priority for digital forensics.  Given the CPU resource issues with SHA256, it is also of paramount significance that the algorithm we move towards should have at least reasonable resource requirements in its software implementation. A quick study into available secure hashing algorithms reveals a small family of secure hashing algorithms share properties that would make each of these secure hashing algorithms prime candidates for supplementing and eventually replacing SHA1 as primary hashing algorithm for computer forensics.

\section{Deprecation}
In this section we will look at why MD5 and SHA1 should be considered cryptographically deprecated and while this has limited impact on the technological and economic realities of digital forensics, it may pose a serious problem in the legal arena.
\subsection{Collision resistance}
When looking into the deprecation of SHA1 and MD5, the sole reason for practical deprecation of these two algorithms stems from attacks that reduce the collision resistance. Collision resistance comes in multiple forms, but the base property lies in the difficulty of finding two data entities with the exact same hash in a situation where both data entities may be manipulated. In its simplest form such collision would be fixed size relatively small chunks of data, but in a more complex form collisions may be generated by manipulating just the last part of larger data entities. Given the use of white-lists with hashes of files that are part of known software packages, a collision like this might be used as anti-forensic technique by for example generating a collision between a modified version of a clip-art image and an image containing illegal content. If the clip-art can than be made part of some whitelistable collection, the illegal content shall remain under the radar indefinitely. A similar attack may be possible without poisoning the white-lists. A computer forensic framework may opt not to process the same entity twice and may use a secure-hash to stop itself from doing so. If a subject can poison the \emph{already-processed} hash collection, than a second file with different content may be omitted from being processed later on. These attacks are technically possible with MD5 and could be theoretically achievable with SHA1 in the not too distant future. From a practical point of view though, the use of cryptographic containers that support the concept of plausibly deniable encryption would seem a much more economical way to achieve the same data hiding goals. There currently are no concrete indications that collision attacks might be in use as anti-forensic data-hiding technique. 
\subsection{First pre-image resistance}
Other than collision resistance, pre-image resistance lies in the difficulty of finding a data entity that, when hashed, yields a given hash. If a hash function is collision resistance, than it should be pre-image resistant as well. An economical pre-image attack could allow simpler anti-forensic data-hiding. Collisions could be created with any hash from the commonly used white lists in order to hide data. Alternatively a denial of service against known bad could be possible. Imagine a large set of photo's and other images being modified to match hashes from a child pornography related database. A first pre-image attack against a hashing function would have a significant impact on the use in forensic frameworks. One possibly even worse problem with a first pre-image attack possibility would be the the legal ramifications of the possibilities of investigators to inject fabricated evidence into a disk image while maintaining the validity of the recorded hash. Digital forensic good-practice dictates that a secure hash is generated and recorded out-of-band when a disk image is created. This practice ensures that lab-investigators can't accidentally or out of malice change the content of the disk image. A pre-image attack could allow an investigator to inject false evidence into a disk image file without changing the validity of the secure hash that was generated during acquisition. 
\subsection{First pre-image resistance and hash sets}
Combined hash-sets used in digital forensics can grow relatively big. As a result an anti-forensic pre-image attack could become somewhat more viable resulting from the fact that a typical hash collection for combined black and white listing will typically have \(2^{26} .. 2^{28} \) hashes. This fact makes it feasible that a first pre-image attack that isn't practically feasible against a single target hash, may be feasible when implemented as \(2^{28}\) parallel attacks. 
\subsection{Use in forensic frameworks}
Given that SHA1 and MD5 due to their diminished collision resistance should be considered deprecated from a collision resistance point of view does not imply that from a technical point of view they are now unusable for all computer forensic purposes. The combination of diminished collision resistance with the properties of the hash set sizes used in the forensic process should however be sufficient to cast doubt on the use of these deprecated algorithms in a forensic framework. While today anti-forensic attacks may be unseen and impractical, the continued use of these algorithms in combination with sizeable list of process-flow-influencing hash lists seems a fundamentally flawed approach waiting for a serious attack factor to happen.
\subsection{Use for guarding the forensic process}
While there seems no technical ground \emph{yet} for discontinuing the use of first pre-image resistant hashes for guarding the forensic process as done by the out-of-band recording of hashes at acquisition time, we must remember that a judge is bound to lack the deep level of technical insight needed to distinguish between collision resistance related deprecation of a hash, possible large-collection first-pre-image related issues, and first-pre-image related issues. Given this reality, it would be quite easy for a judge to get confused between the responsible use of for example SHA1 for safeguarding the integrity of a full disk image (a use that should only suffer when first pre-image resistance is significantly reduced), the use where a \(2^{28}\) entry large set of hashes would make a less significantly reduced first pre-image resistance a serious problem and the use where even the reduced collision resistance could pose a problem. Given this likely confusion, it should be considered a real possibility that a judge, using a cryptographic expert to gain some understanding of the weaknesses of a hashing algorithm might for example dismiss digital forensic findings based on the fact that the digital forensic investigators knowingly used deprecated hashing algorithms and compromised the essential integrity of the forensic process by not using any of the available modern secure hashes. One additional problem with a judge coming to such a conclusion is that legal rulings can have ramifications that stretch well beyond the confines of just the specific legal case. While on technical and economical grounds the continued use of cryptographically broken hashes may be defensible, from a legal risk perspective for the prosecution, moving forward seems even more urgent as it does from a technical perspective. 
\subsection{MD5}
MD5 is a secure hashing algorithm that has been in wide use in digital forensics and is still used as sole hashing algorithm for for example the Virusshare hash set. In 2004 Xiaoyun Wang and Hongbo Yu found that MD5 is not collision resistant. Today MD5 collisions are not just possible but practical to the extent that using commodity hardware, generating multiple collisions per second has become a reality. In 2009 Yu Sasaki, Kazumaro Aoki found a pre-image attack against MD5 with a complexity of \(2^{123.4}\). It is generally accepted that MD5 is deprecated as a secure hashing algorithm and should not be used, even though the pre-image attack may not be of practically usable complexity.
\subsection{SHA1}
When MD5 started showing major problems, digital forensics started to largely move towards the use of SHA1. The performance of SHA1 does not deviate significantly from that of MD5. Today, in 2015 Stevens, Karpman \& Peyrin demonstrated a freestart collision using ten days and a 64 GPU cluster. In cryptographic circles SHA-1 is widely considered as deprecated and Stevens, Karpman \& Peyrin's work on freestart collisions supports this position. At the moment of writing NIST still addresses the use of SHA1 for the computer forensic purposes of known-good/known-bad matching as acceptable use. These statements however predate the freestart collision. Accelerated deprecation of SHA1 in the wake of this freestart collision would be in line with expectations.
\section{Alternatives}
While there is sufficient reason to avoid the use of MD5, and plan for the eminent wider deprecation of SHA-1, the path forward is not directly evident. NIST includes SHA256 hashes in their hash sets, yet SHA256 may not be the most suitable for use with digital forensic processes. We look at different alternative hashes.
\subsection{SHA256}
SHA256 is one of the hash functions from the SHA2 family. While NIST has added SHA256 to its NSRL hash lists, there are reasons why SHA256 may not be the most logical step forward for extensive use in a forensic framework. With NIST's choice for SHA256 and the lack of concrete plans for adding other hashes to their datasets, SHA256 could be a good choice from a compatibility point of view. Performance and system resource usage, when compared to MD5 and SHA1 however, raises serious questions about its usability. While in the past hashing overhead may have been negligible due to IO overhead and throughput speeds, in modern system setups, using SSD, IO latency and throughput have changed the balance. In situations where multi-layered opportunistic hashing is used, even when using non-SSD technology, the practice of opportunistic hashing and page-cache efficiency enhancing techniques shifts the IO versus hashing operation ratios away from IO towards hashing operations.  Considering these two realities, picking a hashing algorithm that triples hashing overhead, today has become more of an obstacle than it would have been in the past.   
\subsection{SKEIN, BLAKE \& BLAKE2}
While SHA2 was meant to succeed SHA1, SHA2 itself has now been succeeded by SHA3. The SHA3 standard was the result of a competition by NIST to find a successor for SHA2, as it was feared that SHA2 might soon proof to be broken in a way similar to SHA1. Where SHA3 aimed to provide a general purpose hashing function, the needs for digital forensics are more specifically geared toward software implementations and variable length data. If we look at the SHA3 candidates and focus purely on the selection criteria that would make these candidates suitable for digital forensics, then the final winner (Keccak) would not have been the first candidate for a secure hash replacement for SHA1 in the field of digital forensics. There were however two SHA3 candidates, SKEIN and BLAKE that showed software implementation properties that make these algorithms more likely candidates for an acceptable SHA1 replacement. Further, there has been some development on an algorithm that must be seen as part of the same family as SKEIN and BLAKE. BLAKE2 is a continuation of the BLAKE efforts that is significantly faster than SHA1 on the modern server architectures that will be typical in computer forensic investigations.
\section{Choice of algorithm}
If we discard the interesting and important efforts in the fields of partial, Merkle-tree based and fuzzy hashing, and focus on full-entity hashing, there are four main ways how hashes are used in computer forensic processing.
\begin{itemize}
\item Check against known good. This includes abandoning further processing for for example Windows OS system files.
\item Check against known bad. This includes marking for example known child pornography image files.
\item Check against \emph{seen before}. This can keep a forensic framework from for example unzipping a large archive that is found in many places for every place where it was found.
\item Applying set-theory to groups of files from different sources within a single investigation. This includes things like allowing an investigator to select all office documents unique to the intersection of office files found on the systems of two suspects.
\end{itemize}
For the first two purposes, compatibility of the hashing algorithms used in the known file hashes data set with the hashing algorithm used by the framework is important. For the other two purposes there exists only one algorithm to be compatible with; the one chosen to be used by the framework. It is with the first two types of usage that we run into an issue with choosing a hashing algorithm for MattockFS. The problem is that data-set producers seem to be behind the curve with regards to secure hashing. Many known-file data-set providers only provide SHA1 hashes. According to NIST, SHA1 is deprecated for the \emph{generation} of secure hashes (as was MD5 before it). Even NIST though still distributes their \emph{NSRL} known-file hash data-set with (next to SHA256 hashes) SHA1 hashes. As use of hashes in computer forensics involves the hashing of large amounts of data, and as access speeds to that data are increasing as the use of solid state disks in small scale computer forensics setups gains more traction, the poor performance of the more secure SHA256 becomes a serious performance consideration. There are several alternative hashing algorithms that combine a strong and secure hash with a decent or good performance on a software platform. Most notably two (non-winning) candidates for SHA-3: SKEIN and BLAKE. Those however come with a lack of compatibility with our dataset providers. Finally, a successor to BLAKE, BLAKE2bp comes in forms optimized to take advantage of 64 bit multi core systems. 
A short comparison against SHA1 :
\begin{table}[]
\centering
\begin{tabular}{llllr}
Algorithm & Secure  & Compatible & Speed \\ \hline
SHA1 & NO  & YES & 100\% \\
SHA256 & YES  & YES & 30\% \\
SKEIN & YES & NO & 75\% \\
BLAKE2b & YES & NO & 135\% \\
BLAKE2bp & YES & NO & 420\% \\
SHA1+BLAKE2bp & YES & YES & 80\% \\
\end{tabular}
\end{table}
When we look at the numbers above, we can conclude that combining SHA1 with the multi-core 64 bit optimized BLAKE2bp algorithm will give us at least the security of SHA256 combined with the compatibility of SHA1 at a performance hit of only 20\% when compared to only using SHA1.
Given the current compatibility needs and the expectation that fast solid-state storage will continue to shift the balance between IO and hashing performance, we propose that digital forensics as a field consider the following migration path away from the usage of SHA1. 
\begin{itemize}
\item \emph{legacy} : SHA1 only (100\% speed; compatible/insecure)
\item \emph{transitional} Dual SHA1/BLAKE2bp hashing (80\% speed; compatible/secure)
\item \emph{performance} BLAKE2bp only (420\% speed; incompatible/secure)
\end{itemize}
For the use with opportunistic hashing, the performance aspect is significantly more dominant given that a single read may result in a multitude of hashing progress operations, the more performant BLAKE2bp justifies the choice to not provide for a migrational path in MattockFS.
\pagebreak
\section{Conclusion}
As recent development regarding traditionally used secure hashing algorithms seem to accelerate, the need for their deprecation, the reality that lack of understanding of the impact of this deprecation may have in a court of law further underlines the urgency for finding a migration path away from MD5 and SHA1. The multi-core variant of the BLAKE2 algorithm appears to allow for a migration path that at first includes the sustained use of SHA1 during a short transitional period while allowing for hash-set providers to catch up. The migrational path also allows the digital forensic process to become prepared for a time when the balance between IO and hashing performance shifts more towards hashing being the primary bottleneck.  We shall, due to opportunistic hashing specific performance concerns not be using this migration path in our implementation of the MattockFS. Instead MattockFS  is built using BLAKE2 only.
The findings in this appendix also justify us advocating the application of a migration path to the whole field of digital forensics. 

%% file: mattock-mattock.tex
\chapter{A Full Framework: An outline}
This appendix tries to show the rough outlines of a possible future computer forensics framework that would built on the good parts and lessons learned from the Open Computer Forensics Architecture (OCFA), current day technological advancements, modern insights from the field of computer forensics and high-volume data processing, access control technologies and the concepts and sub-system introduced in the MattockFS dissertation that this document was written as an appendix for. 
The idea of this appendix is to draw a rough outline of a possible, partially OCFA inspired framework using modern day information technology components and insights gained from years of OCFA usage and the analysis done on OCFA timing that was described in Appendix-A. A framework that, if completed, could fill the gap that in an academic sense was left by the discontinuation of OCFA development by the Dutch police. While this gap has been filled successfully for Dutch law enforcement by the Xiraf framework developed by the Netherlands Forensic Institute, and while on a smaller scale the framework provided by PyFlag and possibly the Sleuthkit Hadoop Framework have taken interesting steps, PyFlag, has not moved significantly beyond any of the same legacy implementation choices that OCFA made.
The Sleuthkit Hadoop Framework efforts on the other hand, like OCFA seem to have been abandoned. Neither architectures have addressed the basics of disk-cache efficiency as addressed in this paper. As far as a literature study has revealed, no open source frameworks that exists today has managed to provide the academic community with a framework suitable for both small scale experiments and for experiments that work with a more sizeable lifelike data corpus and server cluster setup. Computer forensic research projects of the scale of for example the FIVES project that used OCFA at its core would hardly be possible today. 
This while OCFA in itself was actually never designed with an academic setting in mind. The framework described in this appendix
aims to first and foremost address the needs of computer forensic academic research, while keeping use as an alternate tool in cybercrime and other real world computer forensics investigations a serious option. 
While implementation of a full open source academic forensic framework with the required properties falls far outside of the scope of a single M.Sc research project, this appendix will provide a rough architectural outline of such a framework and will explain how MattockFS could play a pivotal role in the realization of such a framework. We shall address the base architectural design proposed in this appendix as \emph{The Mattock Computer Forensic Framework}. The idea of the naming stems from the naming of the carving tool scalpel, metaphorically referring to the scalpel used in forensic pathology. The idea is that while a tool the size of and with the precision of a scalpel is indispensable at any scale, if you want to scale up your investigations to the scale of a rogue burial site, you will also need bigger tools like a mattock for excavating the bodies first. Its just a working title for the prospect architecture. Anyone picking up on this paper is invited to come up with a more suitable name for the resulting framework. 
\section{The OCFA architecture}
\begin{figure}
\centering
\includegraphics[width=100mm]{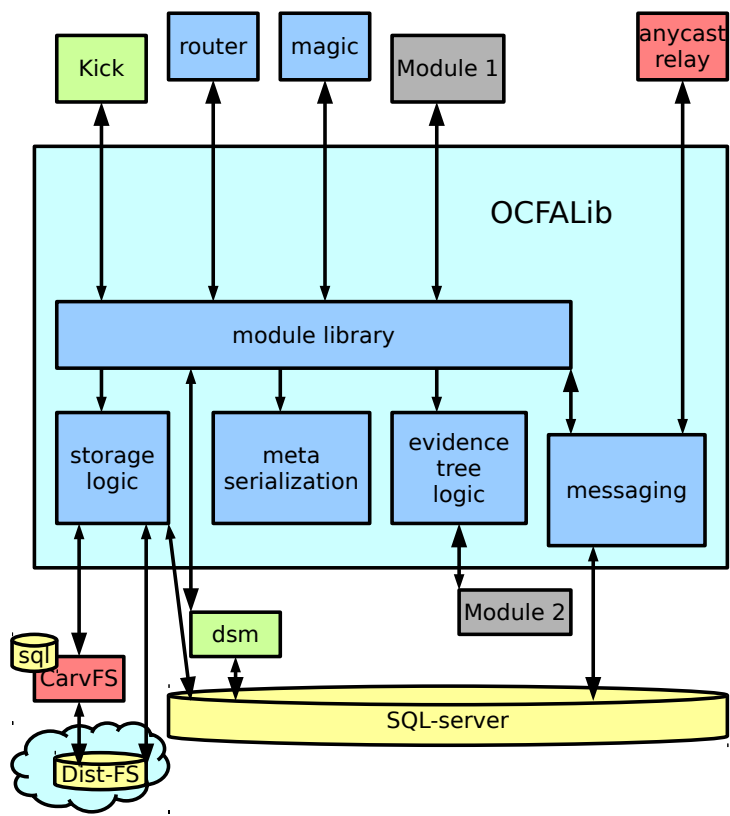}
\caption{The base OCFA architecture}
\label{fig:OCFA}
\end{figure}
When we look at the OCFA architecture, at its core we can clearly identify a fully custom build asynchronous messaging framework for processing modules that communicate with a small set of networking components that act to combine the modules into dynamic chains of tools that are applied to parts of the forensic evidence data. The use of commodity software components is mostly limited to the pervasive use of XML technology and a central relational database. We shall now have a look at some of the core components of the OCFA framework.
\subsection{The AnyCast Relay and PPQ}
At its basis, OCFA was a message passing concurrency based system. One known issue with message passing concurrency is the use of buffers. While message passing environment like the Erlang programming language and platform opt to put producing processes to sleep when buffers fill up, OCFA opted for a different approach. In OCFA, the message passing buffers were managed by on-disk persistent priority queues. These queues only contained references to in database large text objects. The persistent queues were meant and designed to be fully crash resistant. The priority queues had a special \emph{'never'} priority to hold messages that were observed to crash specific modules. This allowed modules to be restarted and to skip problematic data until a maintenance programmer would look at the problematic data and buggy module to fix the problem and re-submit the messages in the never queue for further processing. The AnyCast relay was built on top of the persistent priority queue. Every module connected to the AnyCast relay and registered as a consumer of a certain type (a module \emph{instance}) and would go into a message processing event loop asking the AnyCast relay for new jobs. When a module was done with a piece of evidence data, or when a module derived a sub entity from such data (for example an attachment as child entity of an e-mail message), the module would send a message to the AnyCast Relay addressed at a special process named the \emph{router}. The AnyCast would keep track of irresponsive and broken network connections and would play an important role in having stale or crashed modules restarted in a way not unlike what is common practice in Erlang based architectures. On such a detected crash, messages that were still pending a response would be put aside in the never queue to be looked at by a technician at a later point in time. In OCFA the AnyCast relay served as a single server for all modules, independent of the server these modules would run on.
\subsection{OcfaLib, a domain specific asynchronous framework}
While today NodeJS has mainstreamed the purpose of a generic asynchronous framework, and while in other programming languages generic asynchronous frameworks such as Twisted for Python or Boost::asio for C++ have been available for quite a while, OCFA was first built long before such systems became mainstream. As a result, OCFA basically ended up building its own asynchronous processing framework. We could say that OcfaLib, the C++ OCFA library was a domain specific asynchronous framework for use with the AnyCast server. 
\subsection{The legacy Module API}
OCFA came with two quite distinct module Application Programming Interfaces (APIs). This fact was the result of chronology of development. The first version of OCFA came with a module API not much unlike that of the current day Sleuthkit framework. A module would get a file to process and could add meta data to that file, or, when it wanted to for example mark an extracted e-mail attachment as child entity, would submit that information to the framework. The API consisted of a module initialization part and a single method called 'processEvidence' that a module was supposed to overload. From within processEvidence the module could either add meta or submit a child entity with added meta date.
\subsection{The Tree-graph API}
After new modules got added to OCFA, the legacy module API was found to be lacking in the meta-data area. The problem was that a module deriving a tree of children would not be able to set meta-data for deeper child entities. Only level zero and level one meta data was possible. As a result, the more powerful tree-graph API was added. As porting old modules to the new API was considered a waste of precious development time, the old API was also still continued after the introduction of the new API.
\subsection{The legacy CAS storage}
OCFA in its initial release came with a Content Addressed Storage system for storing data entities. Data was created or, lacking CarvPath facilities, first copied to a temporary file and hashed during copy. Once the hash was fully calculated, the temporary file was either moved to a location derived directly from the hash of its content, or discarded if an entity with the same hash was already present in the repository. 
\subsection{CarvFS}
Later releases of OCFA were made compatible with the use of CarvFS for parts of the storage needs. CarvFS integration has however remained a bit of a hack. The storage sub-system of OCFA used physical symbolic links to CarvFS CarvPaths inside of its primarily CAS based storage system. This meant that for example when using a CarvPath aware Sleuthkit MMLS module, storage of a partition in the OCFA CAS storage system required the full partition to be read for hashing purposes before it could be symlinked in the storage subsystem.
\subsection{The meta-data based message router}
At the core of the OCFA architecture was the central meta-data based router. This XML technology based router would parse the meta-data that modules had gathered regarding a data entity, and would based on an, also XML-based, rule-list determine the next hop in the tool-chain for that data entity. 
\subsection{The use of an SQL server}
There were two technologies that were pervasively used within OCFA. One is the XML we already discussed. The second one was a relational database. It was used by the storage subsystem, by the messaging subsystem, and thirdly by the meta-data storing data-store-module. In each of these uses, the use of an SQL database turned out to be a sub-optimal choice for a number of reasons. Some related to the creation of run-time performance bottlenecks and others related to the nature of the data structure and the nature of useful analysis-time queries on this data. More on this when we discuss the alternatives for Mattock.
\section{PyFlag}
\begin{figure}
\centering
\includegraphics[width=50mm]{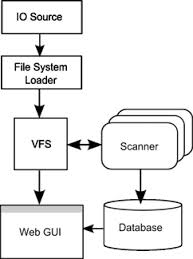}
\caption{The base PyFlag architecture}
\label{fig:Pyflag}
\end{figure}
Next to OCFA, the PyFlag framework deserves mention in this paper. While PyFlag has quite a different scope than OCFA it shares a lot of similarities too. Where OCFA is meant purely as a framework for computer forensics, PyFlag is a hybrid system that also addresses the field of network forensics. This hybrid approach makes that this framework would in a fundamental way be much more difficult to apply disk-cache related optimizations to. It thus would be unfair to look at PyFlag purely from a large scale computer forensic data processing point of view in a comparative way. 
\section{The Sleuthkit Hadoop framework}
A more promising development was the Sleuthkit Hadoop framework. It aimed to combine The Sleuthkit with specific distributed technologies. The technology proposed was a combination of the HDFS distributed file-system and the distributed NoSQL database HBASE, both part of the Hadoop technology stack. That is, a distributed file-system and NoSQL technology. We shall look in more detail at suitable NoSQL technology. While HBASE isn't a bad choice at the processing end, there are data structure and analysis concerns that would point to different NoSQL technology as potentially more suitable. Further, the \emph{initial import} performance of \emph{redundant} distributed storage might make looking closely at the performance/redundancy aspects for any distributed storage system a crucial selection step. Today the concept of erasure encoding based distributed storage might provide an interesting middle ground between the two. Yet, taking a more NUMA or striping approach to scaling the storage for distributed computer forensic data processing might also proof an interesting alternative. 
\section{Non-open frameworks}
It is important to note that due to the closed (non open-source) nature of the frameworks involved, this paper does not look into some major and widely successful non-open forensic frameworks such as Xiraf or FTK Distributed Install. The scope of this appendix is limited to open tools and publications.
\section{Digital Evidence Bags}
So far we have been looking purely at scalability and performance of forensic architectures without considering other important new computer forensic insights. One conceptual idea that all current frameworks seem to discard are the key concept introduced with so called Sealed Digital Evidence Bags (S-DABs) by Bradley Schatz and Andrew Clark in 2006. While the details of S-DEBs fall outside of the scope of this appendix, one key aspect deserves special attention: The concept that both data and meta-data require a form of tamper-proofness. That is, once a piece of evidence data or meta-data is entered into the system, this (meta-)data should be considered to be logically immutable. We shall take this concept with us in our outlines for a next generation scalable framework.
\section{New insights}
Looking at OCFA and other vintage forensic frameworks, there are in retrospect important suboptimal choices that would be made differently if a framework like that was to be developed today. In this section we summarize a few important insights that arose from many years of using OCFA, a survey of other open frameworks, a literature survey and the results of the timing analysis in the main paper.
\begin{itemize}
\item \emph{A tree-graph API is essential} : While simpler API's can be useful for some trivial modules, all such modules could also work with a more generic tree graph API. Having just one generic tree-graph oriented API could facilitate a much wider range of modules and if the API is defined in asynchronous terms, the API could be portable to alternative architectures.
\item \emph{Leveraging asynchronous frameworks is promising} : OCFA implemented its own custom asynchronous framework and the part of the OCFA code-base involved with implementing that functionality was substantial. With current day asynchronous frameworks such as Twisted, Boost::asio or NodeJS, the need for a custom built asynchronous framework has disappeared.
\item \emph{Evidence sealing, privilege separation and access control facilities are a must} : It is essential to limit the mutability time-span and scope to an absolute minimum. Both from an anti-forensics point of view, and from the point of view of the legal credibility of the integrity of the implemented forensic process.
\item \emph{Reducing disk-cache misses and other spurious reads is essential for performance}: While many papers focus on CPU cycles being wasted by inefficient forensic processes, the truth is that much of the forensic process is IO rather than CPU constrained. As such, the fact that the same data \emph{will} get read multiple times should make clear that a disk-cache miss will impact throughput and that a forensic framework such as OCFA with a design that does not effectively mitigate disk-cache miss rates will suffer from disk-cache-miss related IO bottlenecks.
\item \emph{Zero-storage carving is essential} : The process of locating, carving and validating files on disk images is complex and will either result in high false positive or high false negative counts. In the case of high false-positives, copy-out will result in unreasonable requirements for forensic archive storage of derived entities. In the case of high false negatives essential data may be missed. Applying zero-storage carving facilities such as CarvFS will allow for a relatively low cost of false positives while minimizing the amount of additional storage required for processing.
\item \emph{Simple priorities don't really work} : While our research has shown that priority queuing as used in OCFA would be effective for homogeneously sized chunks of evidence data, not taking into account the size of the evidence entities and their likely disk-cache status in prioritizing has been shown to yield such poor results that the usefulness of priority queuing in such a way must be seriously questioned. 
\item \emph{Most current day forensic disk image formats are poorly suited for large-scale processing archives} : In large scale investigations encompassing hundreds of full-size disk images, the in-lab usage of the common computer forensic disk image storage formats (EFF \& AFF) have shown to be rather poorly suited due mostly to scalability issues of keeping hundreds of opened disk-image state-object open. Lacking a forensic-lab storage format for large archives of disk images, the use of simple sparse dd images currently seems to be better suited in a scalable lab environment than the direct usage of these formats. It would be good if future research would investigate the possibilities of archive friendly low system-footprint storage of a multitude of computer forensic data. 
\item \emph{Data migration should not be taken lightly} : While distributed file-systems or storage systems such as SNFS can facilitate in making evidence data available on many nodes, it is important to realize that accessing data from a different node will by definition result in a disk-cache miss on that node. It is suggested that data migration should prefer either the early migration of data that has not yet been fully cached by the originating module, or the migration of relatively small chunks of data targeted for relatively high-CPU processing on the other node. 
\item \emph{Relational (SQL) databases are a poor fit on most fronts} : In OCFA the Postgress SQL database was used for many things. In retrospect, certainly given the current technological landscape, these things would today all have better alternatives. First of all, the database usage for \emph{mutable} meta-data and for the storage and messaging subsystem together was a major performance bottleneck. Apart from the fact that in retrospect in-process mutable meta-data does not fit in with the S-DEB view of things, a relational database is a poor choice of technology for implementing either such meta-data \emph{document} storage or the extra indirections implemented within the storage and messaging subsystems. More than that though, the usage of an SQL database by the data storage module and the user interface have shown that many of the more advanced analysis's have had such shape and form that the database and queries would have been much better of having a more \emph{graph} oriented infrastructure. If we look at OCFA and than look at modern NoSQL technology, we see that parts of the SQL functionality could better haver been implemented without a database, some could have been implemented with a distributed \emph{key/value store} database, some with a \emph{document} database and some with a \emph{graph} database.  One thing all aspects of OCFA have in common is that SQL technology in todays technology landscape would be a sub-optimal choice. 
\item \emph{Kick-starting may not start at a server node} : In OCFA kick-starting took place from a server node. Before this could be done however, the EWF files had to be placed on an SNFS partition accessible to the server from a client through an SNFS connected file-server. Combine this with the need for converting EWF to dd or an other lab friendly format, the lack of a client-based EWF submitter led to massive disk write inefficiency. A kick-starting network client could be an essential component in a modern forensic framework.
\item \emph{Not all modules are alike} : Todays open source computer forensic frameworks treat modules as equal citizens. The reality however is that some modules such as a file-type module are so common in data processing that framework embedding would be justified, some modules such as simple carvers are used mostly on huge data files and are IO intensive while others like OCR work on relatively small data chunks and take up significant CPU resources. Treating all modules and all data as similar will inevitably lead to poor overall framework performance. 
\item \emph{Globally valid designations are the key} : As each server in a forensic data processing cluster has its own page-cache, the concept of dividing the load between different nodes by distributing and redistributing to different nodes should be positively influenceable by allowing nodes to have a common communicable notion of the portions of the global investigation data they and other nodes likely still have in their cache. For this, a globally valid annotation for data chunks is essential. CarvPath annotations could play a major role in this.
\item \emph{Meta-data serialization technology matters} : At the time that OCFA was devised, XML was the only logical choice for meta-data serialization. The XML technology stack though is far from being the most efficient serialization form for forensic meta data. While there are multiple papers proposing standardized XML formats for forensic meta-data exchange, the inefficiency and resource requirements of XML processing forms a major bottleneck in event-rich high throughput processing environments such as within a computer forensic framework. More efficient serialization options such as JSON, Protocol Buffers or Cap'n Proto should be seriously considered as alternative to XML. 
\item \emph{Hashing algorithm choice is essential} :  Traditionally MD5 and later also SHA1 have been used as hashing algorithms in digital forensics. From a cryptographic point of view, MD5 is now deprecated while SHA1 is in the process of soon becoming deprecated. Many public and law-enforcement-only hash collections such as for example the Virusshare hash set or national child pornography hash sets are still distributed only with MD5 and/or SHA1 hashes. Others like the NIST NSRL are now supplemented with SHA256 hashes. While the later may sound like good news, there is an other important issue with SHA256 for use in a forensic framework: SHA256 may be significantly more secure as a hashing algorithm than SHA1 or MD5, it is also significantly more CPU intensive. So much so that it may undermine the whole concept of opportunistic hashing as presented in this paper. 
A quick study into available secure hashing algorithms reveals a small family of secure hashing algorithms. SKEIN, BLAKE and BLAKE2 share properties that would make each of these secure hashing algorithms prime candidates for supplementing and eventually replacing SHA1 as primary hashing algorithm for computer forensics. In the previous appendix we made the case for one of these. The important insight here however is the notion that while SHA1 is close to deprecation, SHA256 puts to much strains on resources in a high performance computer forensics setup and we need to pick a more suitable replacement.     
\end{itemize}
\section{A modernized OCFA inspired open-source architecture}
With the good parts from the old OCFA architecture combined with the new insights above, we can sketch the base outlines of a next generation message passing concurrency open computer forensic framework. 
It is important to note that this section is not a comprehensive architectural design document for a complete framework. It is just a preliminary outline of a possible base setup for such an architecture, primarily focussed on describing the position of MattockFS within such a framework.
Let us start out with a rough description of the changes from the OCFA architecture to our new architecture outline. The most obvious change can be seen in the dependencies. We no longer have a central SQL server as dependency.  We see that the custom asynchronous framework gets replaced with a standard asynchronous framework. This could be boost::asio for C++ or for example twisted for the Python language. We see that file-type logic and meta-data router are no longer a separate module and network service but are now integrated in the base functionality for each module process. As in OCFA, a module instance runs in its own process as part of the asynchronous framework that it runs under. One module instance per async-framework process and possibly more module instances per module type if the server architecture and module type warrant multiple instances of the same module to be run on the same machine.  These would typically be CPU intensive modules on a server with a high amount of cores. 
We see that the number of module APIs is reduced to only the tree-graph API.  The most notable change however is the fact that there is no longer a direct dependency between generic modules and database technology. The SQL server has disappeared and has been replaced by other technology. The Data Store Module (DSM) maps the meta-data into some NoSQL database. This will most likely be a distributed graph-database or a distributed document-database (or possibly a hybrid combination of the two such as ArangoDb). Both data and meta-data are stored in a write-once way within MattockFS, that implements a subset of the S-DEB concept by means of privilege separation, immutability and trusted provenance logging. The long-path SQL database as used in CarvFS is replaced with a distributed key/value NoSQL database. While MattockFS could initially still run on something like SNFS or an NFS mesh-up, the use of a distributed file-system might be a potential way forward. The AnyCast Relay functionality has mostly moved to a system bus like facility provided by the MattockFS filesystem, combined with a monitoring and load balancing mesh-up that connects the AnyCast functionality of all Mattock server nodes and will migrate jobs to other nodes when and only when the CPU needs for a job are expected to outweigh the cost of the page-cache miss on the peer node. The kick-starting process is revised. EWF processing is moved to the client side of things. The client connects to a central kick balancer that will query all load balancers to find the most appropriate kick server to redirect the client to. 
\begin{figure}
\centering
\includegraphics[width=100mm]{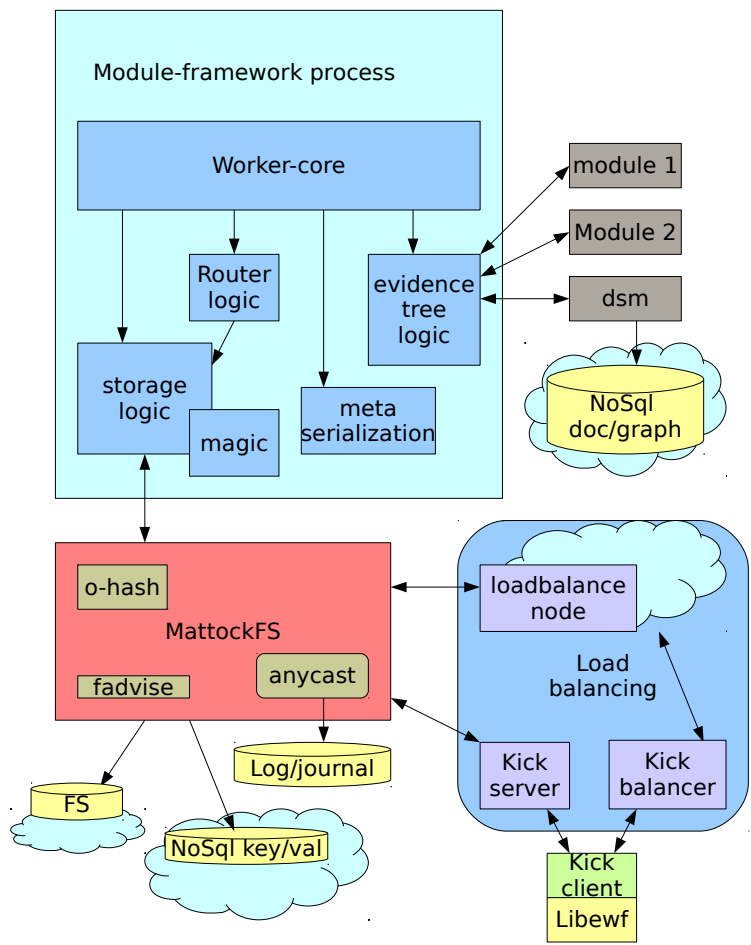}
\caption{The base Mattock architecture}
\label{fig:Mattock}
\end{figure}
\subsection{The distributed long-path store}
The use of CarvPath annotations works very well for most data entities originating from digital media. In some cases however, such as for log files on busy servers or large files downloaded in parallel over slow P2P connections, the files can be so fragmented that the CarvPath ends up being longer than what would be usable as a file or directory name in a file-system. To overcome that problem, long CarvPath annotations are replaced with a secure hash of the actual CarvPath. A storage tokens should be the same on each server node, knowledge to resolve these short notations back to their full CarvPath needs to be distributed. Given that the data stored are just simple key/value pairs, we asses that a NoSQL key/value store database technology could be most suitable to fill this task. A solution like Redis could be a viable implementation, but consideration regarding the speed versus persistence trade-off, and system integrity concerns are important.
\subsection{MattockFS}
MattockFS is discussed in full detail in the main paper. It is a user-space file-system that is centered around CarvPath annotations and entities that consist of fragments and/or sparse regions.  The file-system provides the different async-modules with a file-system based API to storage and messaging. Apart from the CarvPath based data access, the MattockFS file-system based API provides:
\begin{itemize}
\item Immutability: The file-system provides facilities to create entities that are mutable only on creation.
\item CarvPath based batches that aim to provide for opportunistic hashes. A CarvPath can be marked as part of an active tool-chain traversal or batch, and as it (and possibly its ancestors or descendants) is handled by multiple modules before the batch is decommissioned, a hash may be calculated opportunistically.
\item Batch based fadvice: as batches get created and decommissioned, MattockFS can communicate with the kernel about chunks of repository data being potentially needed or will never get read again. The kernel should be able to use this data to become smarter with respect to page-cache usage, thus reducing the potential for page-cache miss performance issues.
\item Throttling: As the file-system keeps track of all active tool-chains and notifies the kernel about data chunks no longer needed in the page-cache, MattockFS can provide statistics to async-modules that act as throttling advice for these modules. It provides total batches-size information and provides a \emph{what-if} facility allowing a module to query how much a new CarvPath would grow the active batches volume. An async-module should use this info to temporarily restrain from entering new (derived) data or uncached CarvPaths into the system until the point when the total amount of active tool-chain space has gone down. 
\item AnyCast inter-process data exchange: Using extended attributes, symlinks and special directories, the file-system provides the possibility to mark data with a \emph{next} marker, placing it in an AnyCast like queue. It also allows to accept data for processing as a module.
\item Sparse capabilities: For processing a job and for one-time mutable access to a newly created data entity, the file-system provides sparse capabilities. These are basically temporary passwords, granting access to small chunks of authority. Passwords that can however be communicated between processes so that tool invocations by modules remains a possibility.
\item Total potential page-cache pressure: Provides the load balancer with info needed to potentially migrate jobs to other nodes.
\end{itemize}
\pagebreak
\subsection{The load-balance mesh-up}
Migrating a batch to an other node is expensive. Any opportunistic hashing will need to start from nothing, data won't exist in page cache on the other node, and unless the storage was highly redundant, there is the extra networking overhead of making the data available at all on the other node. For CPU intensive modules however, migrating a job might be well worth the cost. A load balancer process tries to keep track of things. The amount of page-cache potentially used up by active batches, the overall system load, the CPU and RAM usage of different worker modules, etc. This info is then all shared with load balancers on other server nodes. If a large imbalance is detected between node's, a load balancer may act as one or more modules, accept jobs and close their corresponding batches on the originating server node. These jobs are than recreated by a load balancer on a different node where they should live to complete the rest of the tool-chain.
\subsection{Client/Server kick-starting}
With processing being done on a server or a cluster of load-balance connected nodes, and with forensic disk images normally coming in on external media, entering the evidence data into the lab-cluster is of main concern. Avoiding spurious image copying and performing load-balancing at this point will seriously reduce the total processing time on a distributed processing environment. To avoid spurious disk image copying, we define a per MattockFS-file-system instance network kick-start-server that accept data from a client side forensic disk image processing client. This means the whole of the image data will only be copied once, directly into a MattockFS repository. In order to avoid spurious load-balancing actions being needed, getting the client to send its disk image data to one of the the least loaded of the processing cluster is important. To allow this to happen, a per-cluster kick-start balancing server is defined. When a kick-start server starts up, it registers with the kick-start balancing server. The kick-start balancing server is part of the AnyCast balancing mesh-up network and is kept informed on system and page-cache load for each of the nodes. When a client wants to upload an image, it communicates this with the kick-start balancing server and is sent the network address and port of the kick-start server where it is expected to deliver its data. With such a construct, kick-starting will be relatively evenly spread amongst the server nodes.
\subsection{Module framework embedded lib-magic functionality}
In OCFA the libmagic file-type testing functionality was placed in a regular module. As we identified in the main paper, this module would in most cases where the mime-type of the data was unknown be invoked as first and often last module for a batch. We identify that libmagic functionality is so common that it is warranted to make libmagic functionality part of the module framework.
\subsection{Distributed state-full routing logic}
In OCFA, the router module tended to be a processing bottleneck. For OCFA, lacking proper throttling facilities, this bottleneck turned out to actually mitigate the effects of the lack of throttling, but in a new architecture with solid throttling support, such bottleneck is undesirable. We pose that routing functionality can easily be distributed amongst the modules and be made an intrinsic part of the module framework. The FIVES project produced an alternate OCFA router that, other than the OCFA router was state-full in its rule-list processing. The use of such statefullness, combined with the availability of module generated meta-data, makes that the amount of meta-data exchanged between individual modules, in most cases can be limited to just three values:
\begin{itemize}
\item A router state string, allowing router logic inside of one module to continue the rule list where the previous module left of.
\item The data mime-type, allowing a module that is able to process multiple mime-types to distinguish between its types of input.
\item The CarvPath of the (meta-)data.
\end{itemize}
This choice makes it easy to quickly forward any other module generated meta-data blob to a default data-store module (DSM). So other than in OCFA where a mutable meta-data trace file was forwarded between all modules in a batch, in the Mattock architecture the per-module meta-data is stored as a derived-data MattockFS blob (with mime-type x-mattock/ENC; where ENC could be any serialization format identification) that is then made immutable and is immediately forwarded to the DSM.  
\pagebreak
\subsection{From SQL to graph/document based NoSQL}
We have already eliminated the use of SQL-server dependencies from most of the architecture. The module framework only has MattockFS as dependency. MattockFS uses an in-memory \emph{distributed} key/value store NoSQL database for a small part of its annotations. The data storage module however needs some place to store all meta-data in such a way that it becomes suitable both for use in a front-end system by investigators and for advanced querying by investigative-analyst/digital-investigator duos. Experience with such duos and the use of OCFA databases has shown that most advanced queries tend to be exactly those queries that would translate to much more efficient and easy queries if the data had been modeled into a graph of vertexes and edges. Next to that, OCFA used XML blobs combined with specific indexed fields for much of its basic front-end related data storage. A task much more suitable for document databases. We suggest that taking these two distinct use-cases, a database technology used by Mattock should most likely opt to choose either graph or document oriented NoSQL technology over traditional SQL servers. Further, the chosen database technology preferably should be distributed. A quick market scan has revealed ArangoDb to possess all three desired qualities. ArangoDb is a distributed hybrid-model database that supports document and graph operations alike. More research is however needed to pick the most suitable database technology for Mattock.
\pagebreak
\subsection{Language-native asynchronous frameworks}
When we look at the sizeable code-base of the OCFA framework, we can identify that a major part of the code-base is made up by the implementation of a custom asynchronous framework and a custom messaging bus. Where MattockFS mostly does away with the need for a custom messaging bus as it abstracts away the message bus with a file-system layer, the need for a custom asynchronous framework has fully disappeared. NodeJS has mainstreamed language-specific asynchronous frameworks, and choosing for example Twisted for a Python based implementation of the module framework would take away much of the implementation time of such a module framework. The remaining per-language module framework would only need to implement:
\begin{itemize}
\item Communication with MattockFS according to the filesystem  based API that that system provides.
\item Throttling.
\item Use of a libmagic library for the given language as to determine mime-types of unidentified data.
\item Distributed state-full rule-list processing to set the next module for a batch.
\item An evidence tree API for implementing modules in a given language and the throttled evidence tree tree-walking logic.
\end{itemize}
\subsection{A single tree-graph, lambda and asynchronous operation oriented API}
OCFA had multiple APIs for making a module. The historic API was quite simple, but less suitable for deriving deep hierarchies from a single source. The newer tree-graph API was a bit more complex but also powerful enough to allow meta-data to be added to entities at each level of the tree. Given that a tree-graph API can be used to express each type of module, even simple ones, it is suggested that the Mattock module framework API be inspired by the OCFA tree-graph API. Given modern developments in programming languages that support a more functional higher order approach to programming. A modern API should probably leverage these developments and move away from a mostly object oriented approach to a more lambda oriented API. The creation of such an API should be subject of further study.
\section{Conclusion}
While MattockFS provides proof of the possibility to leverage the power of CarvPath entities as basis for a disk-cache efficient privilege separated forensic framework, the implementation of a whole framework that is usable in both an academic setting and for full scale investigations will require quite some additional research. This appendix sketched the outlines of such a framework.

%% file: MattockFS-fsapi.tex
\chapter{File-system structure as API}
\section{Base file-system structure} 
\begin{itemize}
  \item \emph{mattockfs.ctl}
  \item \emph{carvpath/} 
  \begin{itemize}
    \item \emph{<cp-entity>.<ext>} 
    \item \emph{<cp-entity>/} 
    \begin{itemize}
      \item \emph{<cp-entity>.<ext>}
      \item \emph{<cp-entity>} 
    \end{itemize}
  \end{itemize}
  \item \emph{actor/}
  \begin{itemize}
    \item \emph{<actor-name>.ctl}
    \item \emph{<actor-name>.inf}
  \end{itemize}
  \item \emph{worker}
  \begin{itemize}
    \item \emph{<worker-handle>.ctl}
  \end{itemize}
  \item \emph{job}
  \begin{itemize}
    \item \emph{<job-handle>.ctl}
  \end{itemize}
  \item \emph{mutable}
  \begin{itemize}
    \item \emph{<mutable-handle>.dat}
  \end{itemize} 
\end{itemize}
In this appendix we describe the technical interface to the MattockFS user-space file-system. This interface consists of three main parts. 
\begin{itemize}
\item The classical CarvFS-style interface of the \emph{carvpath} based read-only self-flattening data access
\item The framework geared interface centered around the core concepts of \emph{actors}, \emph{workers} and \emph{jobs}
\item The interface for the write-once approach to \emph{mutable} data and meta-data storage
\end{itemize}
The second and third part aim to be an \emph{API implemented as file-system} interface. To understand the API as file-system interface, we first need to discuss the core concepts of actors,workers and jobs, and the concept of sparse capabilities and their use as handle to in-file-system state.
\section{Actors, workers \& jobs}
On a system running MattockFS there can be multiple worker processes running. A worker process handles evidence data or evidence meta-data of a particular type, and the code for processing such data is called an actor. The process running that code is seen as a worker. There can be multiple workers of the same actor running on one machine. This can be particularly useful for single-threaded worker with a high CPU-load/data-volume ratio on a server with multiple dozens of processing cores. Most actors will need only one worker running on a single MattockFS instance. Most actors will not initiate any work, but will work with evidence-data or meta-data that is given to its instance as \emph{job}. 
During the processing of a job, the evidence data can be used to extract either sub-entities (that can be designated by a carvpath annotation), extracted entities that introduce new content to the repository, or evidence meta-data that represents meta information regarding the evidence data. Once a job is done, the worker needs to submit the evidence-data for further processing by an other actor or for chain termination by the data-store-module. Sub-entities , extracted data and meta-data also constitute jobs that must be submitted to other actors.
\section{Sparse capabilities as handles}
MattockFS aims to provide a least-authority cross-tool laboratory variant of a sealed digital evidence bag (S-DEB). To do this, MattockFS runs under a special uid different from that used by any actor. It defines all data in the repository as file-system enforced immutable data after the initial initialization. Further, the file-system is made responsible for provenance meta-data regarding the lab tool-chain. The workers provide the normal evidence meta-data as storage entities, but the file-system keeps track of provenance as part as its role as \emph{trusted} sub-system. In order for the file-system to truly function as anti-forensics resistant trusted code-base, running it as a different user is not sufficient. The system will need to be hardened to keep workers from interfering with each others processes and provenance chains, and at the same time, allow actors to be written that can invoke tools that still have access to essential data or sub-functionality. This is needed to keep the system flexible enough for reactive usage.
In order to facilitate this, we look at the concept of sparse capabilities. Sparse capabilities are authority tokens or handles that are represented as unguessable strings. The same way that in a programming language an object gets passed by reference to an other object, so can a sparse capability be handed to an other process. An example where this technique of using sparse capabilities together with a user-space file-system and hardening setups was shown to be usable is the MinorFS least-authority file-system.
\section{mattockfs.ctl}
This file represents the archive as a whole. This pseudo file defines two readable extended attributes:
\subsection{mattockfs.ctl::full\_archive}
This extended attribute returns a relative CarvPath representing the whole current archive.
\subsection{mattockfs.ctl::fadvise\_status}
This extended attribute is meant as a hint to the framework for throttling purposes. The value is a semicolon separated list with core file-system figures that may be used to determine if input throttling should be considered.
\begin{itemize}
\item \emph{fadv\_normal\_size} : The total amount of the MattockFS archive currently marked as normal.
\item \emph{fadv\_dontneed\_size} : The total amount of the MattockFS archive currently marked as dontneed.
\end{itemize}
\section{\emph{carvpath}/}
The carvpath directory is the core of the MattockFS interface. The directory itself has its access mask set to \emph{x} only (0x111). This means that no directory listing is allowed, nor are file or directory creation actions. Files that fall within the size range of the repository don't need to be created though, they just are and can be accessed.
\section{carvpath/<cp-entity>.<ext>}
Any designation of a carvpath that is valid within the bounds of the repository size can be used to get a read-only pseudo file that can be used by any forensic or non-forensic tool that expect to work on a regular file. These CarvPath files have a set of read only extended attributes to be used either by human operators, the Mattock framework or a possible future user-interface.
The file may be designated with any file extension as to accommodate processing tools. Some tools for processing a specific type of file won't run unless the file name has an expected file extension. For this reason, any file-extension is treated the same.
\subsection{carvpath/<cp-entity>.<ext>::opportunistic\_hash}
This read-only extended attribute represents the current opportunistic hashing state of a carvpath. Its content consists of two semicolon separated fields:
\begin{itemize}
\item \emph{hash} : If opportunistic hashing has completed, this field is non empty and contains the BLAKE2 hash of the carvpath entity.
\item \emph{hash\_offset} : If opportunistic hashing has not yet completed, this field contains the offset of the first file-data yet to be included in hash calculation.
\end{itemize}
\subsection{carvpath/<cp-entity>.<ext>::fadvise\_status}
The value of this read only extended attribute is a semicolon separated list with information of the overall fadvise related state of the carvpath: 
\begin{itemize}
\item \emph{fadv\_normal\_size} : The total amount of this carvpath currently marked as normal.
\item \emph{fadv\_dontneed\_size} : The total amount of this carvpath currently marked as dontneed.
\end{itemize}
\section{carvpath/<cp-entity>/}
Without an extension, the CarvPath annotation in the \emph{carvpath} directory refers to a sub directory. Like the \emph{carvpath} directory, this directory has mode \emph{0x111} and thus can't be listed. The directory provides the possibility to work with nested carvpath entities by means of flattening symbolic links, thus allowing CarvPath aware forensic tools to produce new valid CarvPath entities by designating a relative path and dereferencing the symbolic link.
\section[carvpath/<cp-entity/<cp-entity> and carvpath/<cp-entity>/<cp-entity>.<ext>]{carvpath/<cp-entity/<cp-entity> and \\ carvpath/<cp-entity>/<cp-entity>.<ext>}
If an entry itself is a valid CarvPath, with or without an extension, than the entry is represented by a symbolic link back into the \emph{data} directory, carrying the exact same extension. The file-system will flatten the carvpath into representing the proper data entity. For example: \emph{data/3145728+786432/ 1048576+65536.gif} will be a symlink to \emph{../4194304+65536.gif}.
\section{actor/}
The \emph{actor} directory is meant as an entry point to the API into the MatockFS AnyCast functionality. It allows a module to register as a worker for a given actor. Like the \emph{carvpath} directory, the \emph{actor} directory has its access mask set to \emph{x} only (0x111). This means that no directory listing is allowed, nor are file or directory creation actions.
\section{actor/<actor-name>.ctl}
This per-actor file is meant to be used by the workers for the specific named actor themselves. It is advised that in an operational setup, access to this pseudo-file is restricted to the specific named actor implementation using MAC facilities such as SELinux or AppArmor in order to prevent anti-forensic attacks against MatockFS.
\subsection{actor/<actor-name>.ctl::register\_worker}
If a process wants to register as an worker for a particular actor, it should fetch a worker sparse-cap by reading this extended attribute. On each attribute read invocation a new instance representation is generated within MattockFS. The extended attribute should thus only be retrieved once by every module instance. The link generated refers to a unique per worker control node where the worker can start accepting jobs and cooperating with the rest of the Mattock framework actors.
\subsection{actor/<actor-name>.ctl::weight}
This mutable extended attribute defines the weight that is assigned to the AnyCast-set of this actor for load-balancing purposes. The default value is 100. The weight is meant to represent the CPU/data-volume ratio in such a way that 100 represents the 100\% value of a default module. A very CPU intensive module may have a weight multiple orders of magnitude higher. The idea is that jobs for high weight values are good candidates for migration to other nodes in load-balancing scenarios.
\subsection{actor/<actor-name>.ctl::overflow}
This mutable extended attribute defines the number of entities in the actor's AnyCast-set that should not be considered candidates for migration. The default value is 10. By changing this value, the eagerness of the load-balancing process is tunable. We consider migration to be expensive and premature migration undesirable. It may be desirable to decrease the value of this attribute for high weight modules or to increase it for modules that are low on both IO operations and CPU load, for example modules that only extract limited meta-data from large archives.
\section{actor/<actor-name>.inf}
This per-actor file is meant to be used by workers of other modules that need to route messages to the given actor. No MAC protection is required.
\subsection{actor/<actor-name>.inf::worker\_count}
This read-only extended attribute returns the count of the number of workers that are currently registered for the named actor.
\subsection{actor/<actor-name>.inf::anycast\_status}
This extended attributes provides information about the anycast-set for the designated module. The content of this attribute is a semicolon separated list with the following meta-data:
\begin{itemize}
\item set\_size : The current number of entities in the anycast set for this actor.
\item set\_volume : The current amount of archive data the current anycast set represents.
\end{itemize}
It is up to the envisioned module framework code-base to make sure the information is used for effective throttling, balancing the need for effective system resource and processing power usage with the mitigation of disk-cache-miss levels.
\section{worker/}
Like the \emph{carvpath} and \emph{actor} directory, the \emph{worker} directory  has its access mask set to \emph{x} only (0x111). This means that no directory listing is allowed, nor are file or directory creation actions. The directory is meant as holding point for instance handle pseudo directories. While protection of this directory with MAC is not needed, the Linux /proc file-system might allow leaking of sparse-capabilities used here, thus MAC limitations on /proc are advised in order to protect access to this directory.
\section{worker/<worker-handle>.ctl}
Each worker has its own worker-handle node for interacting with the MattockFS anycast functionality.
\subsection{worker/<worker-handle>.ctl::job\_select\_policy}
In our fact-finding study on OCFA timing we discovered that the priority queuing used by OCFA was largely ineffective. In MattockFS we no longer have priority queues, we have sortable sets instead. So rather than asking an Anycast functionality for the next highest priority entity in the queue, we ask for the first entity in the set according to a given sorting policy. MattockFS implements a number of choices to the framework layer in the form of a policy string. The policy string consists of a number of consecutive letters that represent set-member properties that can be used for sorting. The next letter from such a string is only used in case of equality according to previous letters. In the end, the entity to be returned will be the one at the start of the resulting sorted set. The sort-policy string can consist of the following letters;
\begin{itemize}
\item \emph{R} : Prefer entities that contain fragment chunks that have the highest currently existing reference-count. The idea would be that processing these entities might have the greatest opportunistic-hashing impact if processing this entity could make the opportunistic hashing of more entities progress.
\item \emph{r} : Prefer entities that contain fragment chunks that have a reference-count of one. The idea would be that processing this entity will allow the ref-count=1 chunks to be marked as no longer needed, thus reducing page-cache pressure.
\item \emph{O} : Prefer entities that contain a fragment chunk with the lowest possible fragment offset. The idea would be that opportunistic hashing is sequential and not processing this entity first might intervene with its opportunistic hashing later on.
\item \emph{H} : Prefer entities that contain a fragment chunk with the lowest possible hashing offset parent reference. The idea would be that opportunistic hashing is sequential and not processing this entity first might intervene with its opportunistic hashing later on. The difference with the O lies in the fact that here the data that has already been opportunistically hashed is ignored.
\item \emph{D} : Prefer entities with the lowest density of chunks with the highest currently existing reference-count. The idea is that processing entities containing much highest-ref-count fragments are less likely to contribute much to freeing up page-cache in the short run.
\item \emph{d} : Prefer entities with the lowest density of chunks with  a reference-count other than one. The idea would be that processing these first will free up page-cache.
\item \emph{W} : Prefer entities with the lowest weighted average reference count. The idea would be that processing these first would contribute to freeing page-cache in the least amount of steps.
\item \emph{S} : Prefer entities with the lowest total size. The idea is that small entities take up relatively much in-process state for the file-system and can be handled quite quickly.
\item \emph{K} : Ignore the anycast set for the module, create a new kick-starting job instead.
\end{itemize}
Note that different sorting policies may be beneficial depending on the current page-cache pressure caused by MattockFS\@.
It might also be different depending on the specific module access patterns and/or system load characteristics. Further research is needed to determine the most appropriate policy. It is up to the framework, not the file-system to choose the proper sorting policy 
\subsection{worker/<worker-handle>.ctl::actor\_select\_policy}
For normal modules the concept of anycast-set selection is not relevant as those modules only access the set bound to their module name. For the load-balancer however, the next job may come from any module and a set selection policy is needed. This policy is partially related to the \emph{weight} and \emph{overflow} extended attributes of \emph{actor/<actor-name>.ctl}, settable attributes that plays a part in selecting the set. Like the sort-policy string, the set-selection-policy string consists of a number of ordered characters used in sorting, but now it's the sets themselves that are sorted. Other than for the sort-policy string, the sorting will favor the largest value rather than the lowest. It is important to note that the \emph{overflow} extended attribute defines a minimum value of entities in a module's set that is not to result in any migrational activity and thus modules staying below this threshold will not be part of the sorting and selecting process.
\begin{itemize}
\item \emph{S} : Prefer the set holding the highest number of entities. This policy should be preferred for CPU related load-balancing.
\item \emph{V} : Prefer the set with the highest total entity volume. This policy might be preferred for page-cache over-pressure related load-balancing.
\item \emph{D} : Prefer the set with the highest entity-count/volume ratio, what is the same as the lowest average entity size. 
\item \emph{W} : Prefer the the set with the highest weight module (as set with the \emph{weight} extended attribute.
\item \emph{C} : Prefer the set with the highest $\dfrac{weight \times entitycount}{volume}$ number. This should represent small yet CPU intensive jobs that should be very much suitable for migration. 
\end{itemize} 
We suggest that the \emph{C} should be a sane default, and given that this string is ignored for regular modules may be uses for those as well.  
\subsection{worker/<worker-handle>.ctl::unregister}
This attribute will be set to the value \emph{0} until it is explicitly set to \emph{1} indicating the de-registration of the module instance.
\subsection{worker/<worker-handle>.ctl::accept\_job}
Explicitly fetching this extended attribute will return a new sparse capability for a \emph{job}. The worker accepting this job is expected to assume responsibility for it and handle the job appropriately.
\section{job/}
This directory is meant to provide the JOB related API functionality for module instances. Like the \emph{data} and \emph{module} directory, the \emph{job} directory has its access mask set to \emph{x} only (0x111). So again no directory listing is allowed, nor are any file or directory creation actions. As with the \emph{worker} subdir, no MAC is needed to protect this directory, yet MAC restriction for /proc is strongly advised in an operation setting.
\section{job/<job-handle>.ctl}
This pseudo file is meant as main handle for job processing by a module instance.
\subsection{job/<job-handle>.ctl::routing\_info}
This extended attribute is made-up out of two parts separated by a semicolon. The first part is the name of the module that was to receive this entity as job. The second part is meant to contain a state variable that could be used by a distributed FIVES-OCFA-router style routing facility within the computer-forensic framework. The idea of the FIVES-OCFA-router was that an entity traversed a routing rule-list based on meta-data artifacts. When after being processed by an other module the entity returned at the router functionality, the router state string was used to find the location in the rule-list to continue the processing.
A module may update this extended attribute to a valid module name and router state. Doing so should 
close the job for the current worker and forward it to a next actor in the projected tool-chain.
\subsection{job/<job-handle>.ctl::job\_carvpath}
This readable attribute returns the carvpath for the data or meta-data that belongs with this job. This carvpath designates the data that the module should process in this job.
\subsection{job/<job-handle>.ctl::submit\_child}
A module instance may create new child entities using a carvpath designation relative to the parent entity. Submitting a child is done by writing a set of semicolon separated fields to this extended attribute:
\begin{itemize}
\item carvpath of the child entity
\item actor name for next hop in tool-chain
\item a router-state string
\item mime-type of the data
\item file extension suitable for the data
\end{itemize} 
\subsection{job/<job-handle>.ctl::allocate\_mutable}
If instead of marking a designatable carvpath, a module instance needs new storage to store \emph{extracted} data or meta-data from the input evidence-data, then the module-instance can ask for a new mutable file to store the derived data in. The value of this attribute should be set to the desired data size. After setting this attribute, the extended attribute \emph{current\_mutable} will be assigned a value.
\subsection{job/<job-handle>.ctl::current\_mutable}
This extended attribute returns a mutable data sparse capability for the last mutable allocated for this job. This is a fixed size sparse file that should be filled by the module.
\subsection{job/<job-handle>.ctl::frozen\_mutable}
While a mutable sparse-capability can be used within the processing of a job, it should not be forwarded to the next actor. To forward the new data to a next actor, it should first be frozen. Retrieving this extended attribute will yield a carvpath for a frozen version of the mutable data. After this attribute is first read, the mutable entity gets invalidated and ceases to be mutable.
\pagebreak
\section{mutable/}
Again a directory with mode 0x111. This directory is meant as holder for fixed-size mutable file entities. As with the previous two directories, this directory is used with sparse capabilities and as such it's security hinges on appropriately secured access to the /proc file-system on Linux.
\section{mutable/<new-data-handle>.dat}
A file accessible through a new-data-handle is a fixed size, initially sparse, mutable data entity that remains valid and mutable within a small time-window determined by the use of job-level attributes. The file may be used for extracted data, for example from a compressed file, or for storage of evidence meta-data. We should look at this time-window as the filling of a digital evidence bag just prior to sealing the bag. All later operations on the data can only take place in a read-only manner, as the data is considered to be sealed after initialization.
\section{Python API}
The file-system as an API described in this appendix is not meant to be used directly by the module framework or load balancer. Instead, a convenience API is included in the MattockFS code-base. This API consists of two parts:
\begin{itemize}
\item A MattockFS Python API
\item A Carvpath Python API
\end{itemize}
\subsection{Base module operations}
The following snippet of code shows how the MattockFS API is imported, a mount-point is bound to and an actor context is acquired.
\begin{lstlisting}
#!/usr/bin/python
from mattock.api import MountPoint
..
mp = MountPoint("/var/mattock/mnt/0")
context = mp.register_worker("ewf2mattock","K")
..
\end{lstlisting}
A full description of the Python API falls outside of the scope of this appendix. Important is that the API provides all functionality that the file-system as an API does. There should be no need for a module framework or load balancer written in Python to directly use the file-system as API. The two APIs are to be found in the files \emph{api.py} and \emph{carvpath.py} in the MattockFS code-base.